\documentclass[%
 reprint,
superscriptaddress,
nofootinbib,
 amsmath,amssymb,
 aps,
]{revtex4-2}

\usepackage{graphicx}% Include figure files
\usepackage{dcolumn}% Align table columns on decimal point 
\usepackage{listings}
\usepackage{tikz}
\usetikzlibrary{arrows.meta,decorations.markings}
\usepackage{verbatim}
\usepackage{bigints}

\newcommand{\RR}{\mathbb{R}}

\newcommand{\ZZ}{\mathbb{Z}}

\newcommand{\ket}[1]{|#1\rangle}
\newcommand{\bra}[1]{\langle #1 |}

\newcommand{\pr}[1]{\operatorname{Pr}\{#1\}}
\newcommand{\OO}{\mathcal{O}}
\newcommand{\ex}[1]{\langle #1 \rangle}
\newcommand{\re}{\operatorname{Re}}
\newcommand{\im}{\operatorname{Im}}
\usepackage{bm}% bold math

\usepackage{xcolor}

\usepackage{hyperref}% add hypertext capabilities

\newcommand{\edit}[1]{{\color{black} #1\color{black}}}
\begin{document}

\preprint{APS/123-QED}

\title{Birth-death-suppression Markov process and wildfires}% Force line breaks with \\

\author{George Hulsey}
 \email{hulsey@physics.ucsb.edu}%Lines break automatically or can be forced with \\
 \affiliation{%
Department of Physics, \\
UC Santa Barbara, Santa Barbara, CA 93106
}%

\author{David L.~Alderson}

\affiliation{
 Operations Research Department, \\
 Naval Postgraduate School, Monterey, CA 93943% with \\
}%x 

\author{Jean Carlson}%
 \email{carlson@ucsb.edu}
 \affiliation{%
Department of Physics, \\
UC Santa Barbara, Santa Barbara, CA 93106
}%

\date{July 21, 2023}% It is always \today, today,

\begin{abstract}
Birth and death Markov processes can model stochastic physical systems from percolation to disease spread and, in particular, wildfires. We introduce and analyze a birth-death-suppression Markov process as a model of \edit{controlled culling of an abstract, dynamic population}. Using analytic techniques, we characterize the probabilities and timescales of \edit{outcomes} like \edit{absorption at zero} (extinguishment) and the probability of the \edit{cumulative population} (burned area) reaching a given size. The latter requires control over the embedded Markov chain: this discrete process is solved using the Pollazcek orthogonal polynomials, a deformation of the Gegenbauer/ultraspherical polynomials. This allows analysis of processes with bounded cumulative population, corresponding to finite burnable substrate in the wildfire interpretation, with probabilities represented as spectral integrals. This technology is developed in order to \edit{lay the foundations for} a dynamic decision support framework. We devise real-time risk metrics \edit{and suggest future directions for determining optimal suppression strategies, including} multi-event resource allocation problems and potential applications for reinforcement learning. 
\end{abstract}

\maketitle

% \tableofcontents

\section{Introduction}
Markov processes provide simple but robust models of many real-world stochastic phenomena. The birth and death process is a continuous-time Markov process describing the dynamics of an abstract population including its possible extinction. 
Birth-death processes have been applied to queueing theory, epidemiology, mathematical biology, and even thermodynamic diffusion;  see \cite{anderson,feller1967introduction,kendall1949stochastic,Crawford2012} for general review and applications.
In recent work \cite{petrovic}, the authors introduce a modified birth-death process as a model of the dynamics of a wildfire under suppression, i.e. firefighting. 
Here, we extend this work and establish the analytical solution of the specific Markov process.

\edit{The} technical focus of this work is on the solution of the birth-death-suppression process, \edit{but a} guiding motivation is wildfire modeling. Dealing with wildfires at a regional level requires allocation of a finite amount of suppression resources---firefighters, trucks, aircraft---across potentially many active wildfire events of varying intensity and risk. There is an ongoing need for decision support tools that incorporate new conditions and allow for optimization-based assessment of trade-offs in suppression strategies. Markov processes have previously been applied to multiple facets of wildfire management, including fire spread, risk to infrastructure, and fire mitigation, among others \cite{CATCHPOLE1989101,largeScaleMitigation,somanath2014controlling,thompson2013modeling,abdelmalak2021markov,meng2015mapping,diaz2022modeling,diao2020uncertainty,griffith2017automated,soderlund2018markovian}. 

\edit{We stress that the introduction of the present model is not meant to replace or even compare to the high-fidelity, physics-based fire modeling tools employed by agencies today. Instead, the model is meant to capture important trade-offs in time-dependent resource management in a simple environment, where simulation is cheap and analytical results are available. As it contains no spatial information about the spread of a wildfire, the birth-death-suppression model is effectively a `mean-field' approach to describing wildfire dynamics.}
\begin{figure}[t]
    \centering
    \includegraphics[width = \columnwidth]{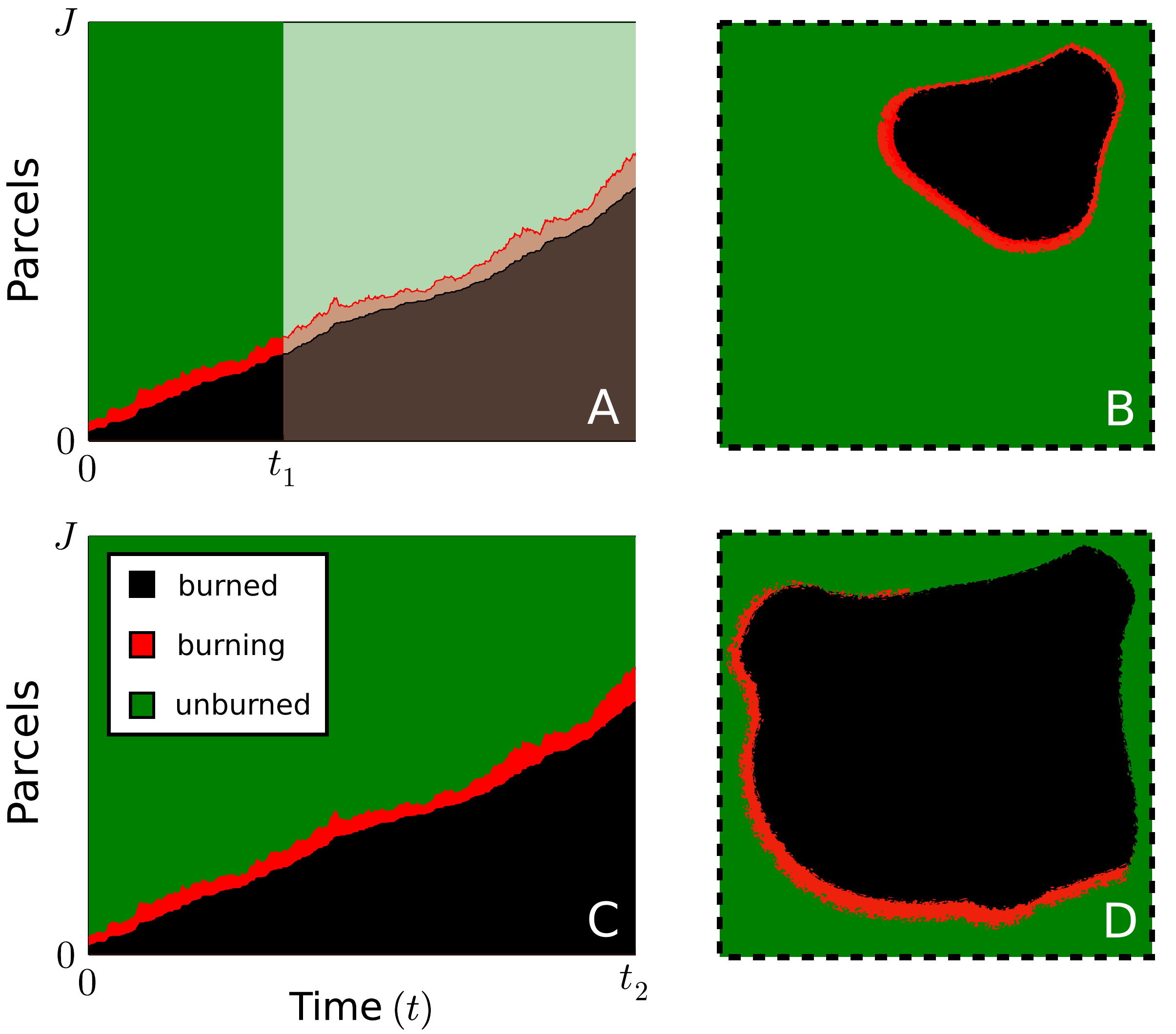}
    \caption{A simulated birth-death process (left) alongside the schematic representation as a wildfire (right). In A; the state of the simulated process at the time $t_1$, with all $J$ parcels in one of the three states: unburned, burning, burned.  In B; the corresponding spatial representation: a small fire at an early time $t_1$, the active population shown in red. In C, most parcels of the simulated process have entered the burned state. In D; a depiction of the same fire at the later time $t_2$: the footprint (black) has grown to almost fill the total burnable region (green). }
    \label{fig:cartoon}
\end{figure}
\edit{More generally, the birth-death-suppression process can model any population with a natural and an artificial source of eradication: the rate of the former being proportional to the size of the population and the latter independent of it. For example, one could interpret the process as modeling the size of an infected population in the presence of treatment efforts. Indeed, the dynamical equations of the susceptible-infectious-recovered (SIR) model of an epidemic are known to reduce to those of a birth-death process in certain regimes \cite{kendall1956deterministic}. The model we consider describes populations in a linear growth stage, where no dynamical limit to their growth exists, and where there is no chance of resurrection after complete eradication. }
\begin{table*}[t]
    \centering
    \setlength{\tabcolsep}{12pt}
    \begin{tabular}{c c c c c }
    \hline
    \hline
    Quantity & Symbol & Definitions & Wildfires & Epidemics\\
    \hline 
       birth rate & $\beta$ & 
        $\beta > 0$ & spread rate & infection rate \\
        death rate & $\delta$ & set $\delta = 1$ & (natural) extinguish rate & recovery rate\\ 
        suppression rate & $\gamma$ & $\gamma \geq 0$ & firefighting rate & treatment rate \\ 
        initial population & $N$ & $j(0) = F(0) = N$ & seed ignition size & initial outbreak size\\   
        threshold & $J$ & $F \leq J$ & burnable area & susceptible population\\
         & & & & \\
        population & $j(t)$ & $j(t) \geq 0$ & actively burning area & infected population\\
        
        footprint & $F(t)$ & $F(t) \geq j(t)$ & burned area & total epidemic size\\
        
        lifetime & $T$ & $j(t) = 0,\ t\geq T$ & extinguishment time & eradication time\\
        \hline
        \hline
    \end{tabular} 
    \caption{Reference of the parameters ($\beta, \delta,\gamma,N,J$) and variables ($j,F,T$) of the birth-death-suppression model.  Included are their interpretations as describing wildfire, as in \cite{petrovic}; and epidemic dynamics, as in \cite{kendall1956deterministic}. Setting $\delta = 1$ fixes the units of time. The lifetime $T$ is defined as the exact time of absorption (reaching the state $j(T) = 0$). The process pictured in Fig. \ref{fig:cartoon} has parameters $N = 10,\ J = 500$, $\beta/\delta = 1.1$, and $\gamma = 0$. }
    \label{tab:vars}
\end{table*}

The main subject of our model is a \edit{discrete} stochastic variable $j(t)$ we call the \textit{population}, \edit{representing in the wildfire interpretation} the number of actively burning (spatial) parcels of fire, or \textit{firelets}; one may take a single firelet to represent e.g., a burning acre of land. 
% The terminology `parcel' emphasizes that the population $j(t)$ takes discrete values. 
In Fig. \ref{fig:cartoon}, the population $j(t)$ is shown in red, both as a stochastic variable and schematically as the actively burning area of a wildfire.

As time progresses, the variable $j$ undergoes stochastic nearest-neighbor transitions $j \to j\pm 1$ with probabilities dependent only on the previous state. The transitions $j \to j\pm1$ represent \edit{births and deaths in the population}. There is an aggregate birth rate $\beta \cdot j$ of parcels per unit time which may change to reflect \edit{local conditions: fuel and weather in the fire interpretation or virulence in an epidemic interpretation}. There is also a suppression parameter $\gamma$ which reduces the size of the overall population, representing the application of 
\edit{eradication resources; firefighting or treatment}. A reference table of these parameters, their definitions and interpretations is included in Table \ref{tab:vars}.

If at some time the process reaches the state $j = 0$, the population has died out \edit{and the process is terminated}. Estimating the probability of this event, and the time at which it happens, is a primary concern. However, fires propagate, and their associated risk is related to how much total area they burn, not just their active size. A fire could maintain a small active size but still propagate and burn a very large area before going out. \edit{Similarly, the severity of an epidemic is generally defined as the cumulative number of cases, not the number of currently infected individuals.}  The \edit{cumulative population} is a new stochastic variable $F(t)$, \edit{which we refer to as} the \textit{footprint}. This variable shares all the positive transitions of the population $j(t)$: if $j \to j + 1$, then $F \to F + 1$, but if $j \to j-1$, $F$ is unchanged. The footprint, as the area burned by a wildfire, is shown in black in Fig. \ref{fig:cartoon}, where one can see its monotonic increase in size as the process (left) or fire (right) progresses in time.

To be sensitive to the cost associated with a large burned area, one can ask whether the footprint $F$ likely remains below some threshold value $J$. \edit{Asking whether the footprint satisfies} $F \leq J$ \edit{is the type of question} relevant for fires in the wildland-urban interface. \edit{When the footprint reaches $F \geq J$, one interprets the fire as having reached the built infrastructure.} While the model is purely temporal, the \edit{threshold} $J$ incorporates the spatial notion of a burned area: the correspondence between this implicit spatial representation and the continuous-time Markov process is the subject of Fig. \ref{fig:cartoon}, with snapshots of the `fire' juxtaposed with a simulated birth-death process.  \edit{The purpose of suppression is to keep $F \leq J$, thereby containing the cumulative size of the fire or epidemic. To this end we aim to compute the \textit{escape} probability that $F \geq J$ at the end of the process, for any finite $J$. }

% While our focus is on the wildfire interpretation and\edit{ subsequent applications}, the birth-death-suppression process is one less-studied corner of a family of linear birth-death processes, deserving mathematical treatment in its own right. 

Some immediate credence for the introduction of the birth-death-suppression model in the context of fire dynamics is supplied by the known statistics of fire footprints: if $P(F \geq J)$ is the probability of a given fire having footprint $F$ greater than or equal to $J$, a power law  $P(F \geq J) \sim J^{-\alpha}$ is observed. In particular, measurements of the area burned by real fires suggests that the exponent $\alpha$ of the empirical footprint distribution is close to $\alpha \approx 1/2$ \cite{maxjeanWildfireHOT}. This power law distribution is reproduced in our birth-death model for fire close to the critical threshold where the birth and death rates are equal. 

Power law distributions have heavy tails, which means that large or rare events dominate the total cost of such events. The presence of a power law distribution is common to many types of natural phenomena and various mechanisms exist which explain the origin of such scaling \cite{heavytails,aspectsriskassesment,generative,morenormal,bak2013nature,stumpf2012critical}. These notably include the idea of `highly optimized tolerance,' a model of systems which have evolved or been organized in some fashion \cite{hot2,hot1}. Intuitively, highly optimized systems are tolerant to common events but unprepared for rare, catastrophic events. This results in the `cost' or `loss' associated to such rare events comprising the majority of the total loss experienced. The suppression-free birth-death model recreates the desired footprint distribution, and the suppression parameter, once introduced, allows one to explicitly compute the effect of policy choices---allocation of suppression resources---on the distribution of possible outcomes of a fire event. 

Birth-death processes were initially studied by Kendall \cite{kendallGeneral}, who introduced the cumulative population; and Karlin and McGregor \cite{karlinClass,karlinDiff}, who developed the method of orthogonal polynomials as a means of solution. Linear birth-death processes, which include the birth-death-suppression process, were further studied by Karlin and McGregor in \cite{karlinear,linearkilling}, where the notion of the cumulative population (footprint) was addressed. However, they did not determine the dynamics of the cumulative population away from the critical point. The analytical tools for doing so were applied by Askey and Ismail \cite{ismail} and involve analyzing the embedded discrete birth-death process, a random walk. However, the solution of the birth-death-suppression process---including the cumulative population---has not been fully synthesized in the literature. For other work on birth-death processes and their properties, see also \cite{byron1979,jointBDintegrals,crawford2014birthdeath,Crawford2018}.

The birth-death-suppression process is notably similar to the birth-death-immigration (BDI) process \cite{anderson,bd_migration,ismail}. In the latter one has an immigration rate parameter which represents the growth of a population due to migration rather than births. This plays an antipodal role to the suppression rate in our model. The analytical methods of this work closely follow the approach of Askey and Ismail to solving the BDI process, and many potential applications of the birth-death-suppression have equally strong analogs that have been studied for the BDI process. 

\edit{In fact, the birth-death-suppression process is the dual, in a technical sense, of the BDI process \cite{anderson}. The duality mapping takes ergodic/recurrent processes (which do not terminate), like the BDI process, to transient processes (which may terminate in finite time) such as the birth-death-suppression process. The BDI process contains as special cases many other familiar continuous-time Markov processes: Poisson, pure birth, pure death, and the $M/M/\infty$ queueing model \cite{crawford2014birthdeath}. While all the aforementioned have been well-studied in the literature, the birth-death-suppression model---their image under the duality mapping---has garnered less direct attention. This is surely related to the fact that the presence of suppression complicates the standard approaches to solving the process.}
\subsubsection*{\edit{Outline of paper}}
The general goal of this work is to gain a more analytic understanding of the birth-death-suppression process as a simple fire model, an interpretation suggested in \cite{petrovic} to which the reader should refer for more background. \edit{Throughout the paper, we will use the language of wildfire to describe the process, though we emphasize its generality in describing branching processes in phases of linear growth.}

The paper is intended to be readable with different focuses. Section \ref{sec:BDprocesses} reviews the general theory of birth-death processes and defines precisely the model under consideration. In Section \ref{sec:populationdynamics} we review the solution of the continuous-time process\edit{: this includes the characterization of the absorption times and the explicit formulae for transition probabilities.} In Section \ref{sec:predfootprint} we solve the embedded random walk\edit{; it is here that the novel results of the work are developed, including the escape probability. After solving the model in question, in Section V we show some basic applications and sketch future directions for the technology developed in this work.} The mathematically inclined reader may focus on Sections \ref{sec:populationdynamics}, \ref{sec:predfootprint} \edit{which contain the bulk of the material on the specific Markov process itself. In the appendices we include a thorough review on background material, a majority of the technical details, and review of techniques applied in the paper such that this work is relatively self-contained. }

\section{Birth and death processes}
\label{sec:BDprocesses}

Defining the stochastic birth-death process begins with the state space, which is composed of discrete parcels. \edit{These parcels represent individual elements of the population; e.g., acres of land in the wildfire interpretation.} Each parcel can be in one of three states: unburned, burning, or burnt. A firelet is a burning parcel; $j(t)$ therefore counts the number of firelets at time $t$, and $F(t)$ represents the total burned area: the sum of burning and burnt parcels.  We take the convention that $N \equiv F(0) = j(0)$, so that $F(t) \geq j(t)$ in general.

\begin{figure}[!h]
    \centering
    \includegraphics[width = \columnwidth]{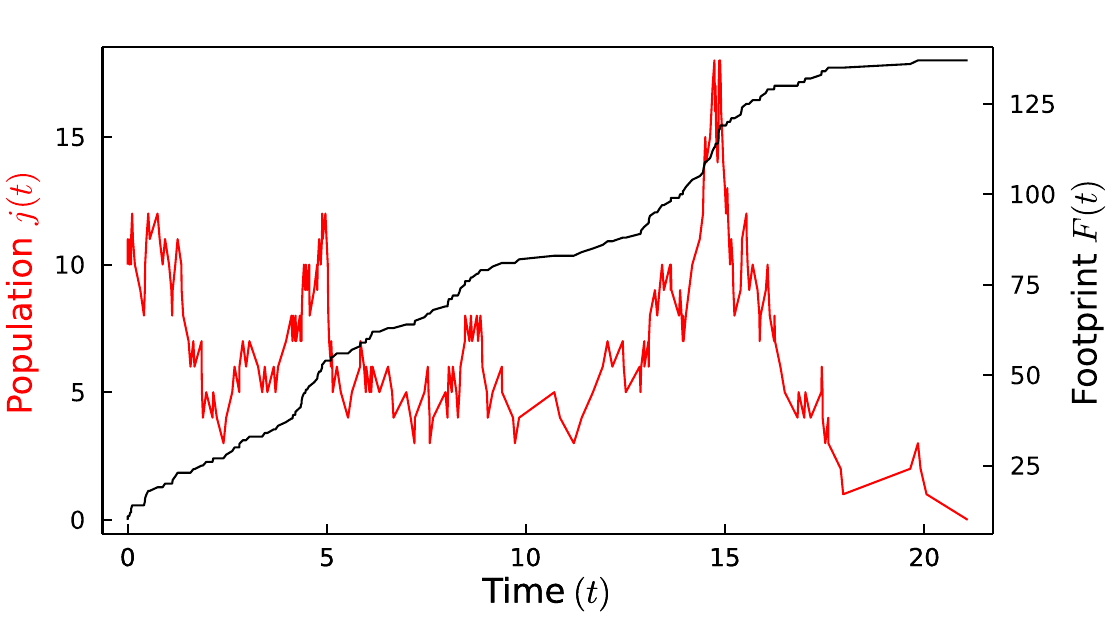}
    \caption{An example birth-death fire process. Here, the \edit{population is initially of size 10}, and hence $j(0) = F(0) = 10$. The process pictured has equal birth and death rates $\beta = \delta$ and reaches $j = 0$, ending in finite time.}
    \label{fig:exampleFire}
\end{figure}

As time progresses, unburned parcels can become firelets ($j \to j+1, F \to F+1$) and burning parcels can become burnt ($j \to j -1, F\to F$). Both the variables $j,F$ are non-negative integers, with $F \geq N \geq 1$. Fig. \ref{fig:exampleFire} shows a example simulation of the process, with $j(t),F(t)$ displayed in red, black respectively. 

Recall the \edit{idea of a threshold $J$} on the total number of available parcels. Any process has $F \leq J$ initially; reaching the point $F = J$ means the fire has consumed a burnable area of size $J$.  Beginning such a process with $j(0) = F(0) = N$ active firelets, \edit{one of two possible outcomes will occur first}: the process extinguishes, reaching $j = 0$; or \edit{it remains active and grows its footprint} until it reaches $F = J$. In the Markov process literature, the state $j = 0$ is said to be \textit{absorbing}: once the process reaches this state it remains there for all time. Therefore, we refer to the event of reaching the state $j = 0$ as \textit{absorption}. Conversely, if the footprint $F$ exceeds the bound $J$, we say the process has \textit{escaped}. `Solving' the process means characterizing the probabilities of each of these outcomes---absorption or escape---in terms of the birth and death rates of the \edit{population}. 

The transition probabilities define the dynamics of the system. Consider a single, actively burning \edit{unit} out of the whole population $j(t)$. Let $\beta$ be the rate (per unit time) at which this single \edit{parcel} ignites a new firelet. This parameter might reflect the weather, fuel, and topographic conditions of a wildfire. In some short time interval $\Delta t$, the probability that the entire population $j(t)$ sees a single \edit{birth (new ignition)} is
\begin{equation}
	\pr{j \to j+1} = \beta j(t) \Delta t + \OO(\Delta t^2),
\end{equation}
which defines the aggregate birth rate $\lambda_j \equiv \beta j$. Define also the aggregate death rate $\mu_j \equiv \delta j + \gamma$. The aggregate death rate is the rate per unit time of deaths in the population, so that the probability of a single \edit{death (firelet burning out)} in the interval $\Delta t$ is 
\begin{equation}
	\pr{j \to j-1} = (\delta j(t) + \gamma) \Delta t + \OO(\Delta t^2).
\end{equation}
Here, the parameter $\delta$ is the death rate for an individual firelet per unit time; it represents the natural process of \edit{eradication, i.e., extinguishment by fuel exhaustion.} The second parameter $\gamma$ is the rate of suppression. It represents the action of \edit{eradiction} resources \edit{(firefighting)} on the fire, again per unit time. The effect of suppression on the aggregate death rate is independent of the size of the fire; the suppression affects the aggregate population, while all the individuals feel the effect of fuel exhaustion. One can already see that the suppression rate $\gamma$ affects the aggregate death rate $\delta j + \gamma$ more when the population $j(t)$ is small. 

In some short time interval $\Delta t$, the transition probabilities are
\begin{subequations}
\label{eq:transprobs}
\begin{gather}
	\pr{j\to j+1} = \lambda_j \Delta t,\\
	\pr{j \to j-1}) = \mu_j\Delta t,\\
	\pr{j \to j}= 1-(\lambda_j + \mu_j)\Delta t,
\end{gather}
\end{subequations}
where $\lambda_j = \beta j,\ \mu_j = \delta j + \gamma$. This defines the continuous-time dynamics of the population $j(t)$. Processes begin with an initial population $j(0) = N > 0$, after which stochastic transitions begin to occur. 

As a choice of units, we set the death rate $\delta = 1$. The parameters $\beta, \delta, \gamma$ are defined in terms of some unit time; setting $\delta = 1$ is effectively a choice of time units, i.e. ``minutes until \edit{death}." Equivalently one can consider this a redefinition $\beta \to \beta/\delta,\gamma \to \gamma/\delta$. One must also specify some boundary conditions for the process. When $j = 0$, \edit{the process has absorbed.} In this case $\lambda_0 = 0$, i.e. there is zero probability of \edit{another birth}. For consistency one must set $\mu_0 = 0$ for any $\gamma$ so no further deaths can occur. There is also an issue of boundary conditions as we move towards infinity in state space. While we consider the probability of the footprint reaching some finite size $J$, the analysis does not bound the population $j(t)$ in any formal way, \edit{and the parameter $J$ does not enter into the dynamics.}

\subsection{General theory of the birth-death process}
The most general birth-death process is defined by the transition probabilities above in Eqn.~\eqref{eq:transprobs}, and is studied in its generality in e.g. \cite{anderson,feller1967introduction,karlinClass}. The birth-death-suppression model is just a specific instantiation with $\lambda_j,\mu_j$ being affine functions of the population $j(t)$. Consider first the population $j(t)$. A natural statistic to introduce is the probability of the population taking a certain value at a specific time:
\begin{equation}
	p_n(t) = \pr{j(t) = n}.
\end{equation}
These population probabilities $p_n(t)$ should satisfy $0 \leq p_n(t) \leq 1$ and $\sum_n p_n(t) \leq 1$.\footnote{In a finite state space, this would be a strict equality. The issue is accumulation of probability measure at infinity.} For an initial population of definite size $j(0) = N$, one would write $p_n(0) = \delta_{n,N}$ (here, the Kronecker delta). As time progresses, these probabilities will change in accordance with the transition probabilities as
\begin{multline}
\label{eq:diffdiff}
	\frac{d}{d t} p_n(t)=\lambda_{n-1} p_{n-1}(t)+\mu_{n+1} p_{n+1}(t)\\-\left(\lambda_n+\mu_n\right) p_n(t),
\end{multline}
defining a differential-difference equation. If one can solve this equation to determine $p_n(t)$ for an arbitrary initial distribution $p_n(0)$, one can deduce any expectations about the population dynamics. To this end, construct a related object: the transition or correlation matrix $P(t)$. This matrix has elements which are the probability of transitioning from state $n$ to state $m$ in a time $t$: 
\begin{equation}
\label{eq:Pmatelems}
	P_{nm}(t) = \pr{j(t+s) = m\ |\ j(s) = n},
\end{equation}
where $n,m \geq 1$. The Markov property ensures that this definition is independent of $s$. Note the restriction $n,m \geq 1$; this is done because zero is a special (absorbing) state in the process. Despite this exclusion, the probability $p_A(t)$ of absorbing at $j = 0$ can be calculated from $p_1(t)$ via Eqn.~\eqref{eq:diffdiff}, giving
\begin{equation}
\label{eq:absorbdef}
    p_A(t) = \mu_1\int_0^t p_1(\tau)d\tau .
\end{equation}
The absorption probability $p_A(t)$ will be of central interest. 

For an initial population of definite size $j(0) = N$, the probabilities $p_n(t)$ form a row of the matrix $P(t)$:
\begin{equation}
    p_n(t) = P_{Nn}(t).
\end{equation}
One can view $p_n(t)$ for $n\geq 1$ as a row vector $\bra{p(t)}$. The initial conditions of the system $p_n(0)$ then form a vector. The transition matrix can be used to write
\begin{equation}
\label{eq:transIC}
   \bra{p(t)} = \bra{p(0)}P(t); \qquad  \frac{d}{dt}\bra{p(t)} = \bra{p(t)}\mathcal{A},
\end{equation}
also writing the differential-difference equation \eqref{eq:diffdiff} in vector notation. The (possibly infinite) matrix $\mathcal{A}$ is
\begin{equation}
\label{eq:Amat}
    \mathcal{A}=\left(\begin{array}{cccc}
-\left(\lambda_1+\mu_1\right) & \lambda_1 & 0 & 0\\
\mu_2 & -\left(\lambda_2+\mu_2\right) & \lambda_2 & 0\\
0& \ddots & \ddots & \ddots 
\end{array}\right).
\end{equation}
\edit{One should note that the probabilities $p_n(t)$, forming an infinite row vector, depend on a specific initial configuration $p_n(0)$, whereas the transition matrix $P(t)$ (forming an infinite matrix) does not.} Given the dynamical equation for the $p_n(t)$, linearity implies that the transition matrix $P(t)$ satisfies, \edit{for any initial configuration $p_n(0)$,} the first-order equation
\begin{equation}
\label{eq:mastereq}
    \frac{d}{dt}P(t) = P(t)\mathcal{A},
\end{equation}
with the initial condition $P(0) = \operatorname{Id}$, where $\operatorname{Id}$ is the identity matrix. In the Markov process literature, this is called the Kolmogorov backward equation. For a physicist, it is a first-order (Hamiltonian) dynamical equation, what statistical mechanics would refer to as a master equation.\footnote{Since $P$ is an infinite matrix, one can regard it as a (positive semidefinite) operator on a Hilbert space, and interpret $\mathcal{A}$ as a Hamiltonian: like a Hamiltonian, $\mathcal{A}$ generates time translations of the `density' matrix of our system, $P(t)$.}Solving this equation determines the dynamics of the system for any initial configuration. The analogy between Markovian evolution and traditional statistical mechanics has been made precise in recent work, e.g. \cite{korbel2021stochastic}, or in \cite{wolpert2019space}, where the generality of the master equation is discussed.

If the state space is finite, then $P$ and $\mathcal{A}$ are finite-size matrices and the master equation admits the formal solution \cite{spectral,reuterDiff}
\begin{equation}
    P(t) = \exp(\mathcal{A}t).
\end{equation}
The matrix exponential is very cumbersome to compute for state spaces of any reasonable size, numerically or otherwise. In this paper, we take the state space to be infinite, so that $P$ and $\mathcal{A}$ are infinite matrices. In this case the matrix exponential solution is not well-defined; one must take a different approach. Thankfully, in an infinite state space, certain new analytical tools are available. 

We assume throughout that $\beta$ and $\gamma$ are time-independent. Given this, the process is homogeneous and Markov. Physically, one might be interested in a situation in which the birth rate changes for some time: this would represent, for example, the effect of a high wind event increasing the spread rate of a fire. This is compatible with our assumptions so long as the birth rates are piecewise-constant. Say a process has birth rate $\beta$ for $t < \tau$ and a different birth rate $\beta'$ for $t > \tau$. The transition matrices may be stitched together, taking advantage of the Markov property to do so: 
\begin{equation}
    \bra{p(\tau + t)} = \bra{p(0)}P(\tau;\beta)P(t;\beta').
\end{equation}
With this in mind, we take the birth and suppression rates to be time-constant, mostly without loss of generality. Note however that the footprint $F(t)$, being a cumulative variable, does not have the same Markovian properties as the population $j(t)$. 

Given an expression for the transition matrix $P(t)$, and hence the probabilities $p_n(t)$, one can compute various probabilistic expectations for the system: the average population, the variance in population, or the absorption probability $p_A(t)$. Note that the absorption probability $p_A(t)$ is the probability that the population satisfies $j(T) = 0$ for any $T \leq t$; it is monotonically increasing. The actual time $T$ of absorption (c.f. extinguishment\edit{, eradication}), which we refer to as the \textit{lifetime}, is itself a variable of interest \cite{kendallGeneral}. Since the absorption probability $p_A(t)$ is the integrated probability that $T \leq t$, the lifetime $T$ is distributed as 
\begin{equation}
	d\sigma(T) \propto \frac{d}{dt}\left[p_A(T)\right]dT = \mu_1 p_1(T)dT
\end{equation}
writing $d\sigma(T)$ for the normalized probability measure on lifetimes $T$. Determining this distribution $d\sigma(T)$ allows computation of statistics like the average $\ex{T}$ or median $T_m$ lifetimes. We proceed to calculate these types of quantities in an effort to characterize the physical predictions of the Markov model. 

\subsection{The suppression-free process}
\label{sec:suppressionfreeprocess}
Without suppression, the model reduces to the pure linear birth-death process with aggregate birth-death rates
\begin{equation}
\label{eq:physicalBDrates}
    \lambda_j = \beta j; \qquad \mu_j = \delta j.
\end{equation}
The dynamics of this process are well-known; they are solved by Kendall in \cite{kendallGeneral} using a generating function approach. The solution by this method and its details are reviewed in Appendix \ref{sec:appgfunc}. In this section, we discuss the phenomenology of this suppression-free birth-death process, using the language of the wildfire interpretation to add physical relevance to the formulae. 

First, consider the evolution of the population. Let the process have some constant birth rate $\beta$ and begin with an initial population of $j(0) = N$ firelets. With a death rate $\delta = 1$, the mean population is characterized by exponential growth or decay depending on the value of the birth rate:
\begin{equation}
\label{eq:zerogamavgpop}
    \ex{j(t)} = Ne^{(\beta-1)t}.
\end{equation}
With $\beta < 1$, births happen more slowly than deaths and the population $j(t)$ reaches zero with probability 1: the fire is extinguished after some finite (but possibly long) time. This defines a phase of the process that we refer to as \textit{sub-critical}. When $\beta > 1$, the average population diverges in size exponentially and the phase is \textit{super-critical}. Finally, in the \textit{critical} $\beta = 1$ phase, the average population is constant for all time. The dynamics of the critical population consist of fluctuations in size around a constant average; indeed, when $\beta = 1$ the population variance $\Delta^2j(t)\propto t$ increases linearly in time. This type of behavior is familiar from random walks and is essentially Brownian motion in one dimension. Such a fire tends to develop a large footprint while maintaining a roughly even active size, potentially still posing a large risk. 
\begin{figure}[!h]
	\centering
	\includegraphics[width = \columnwidth]{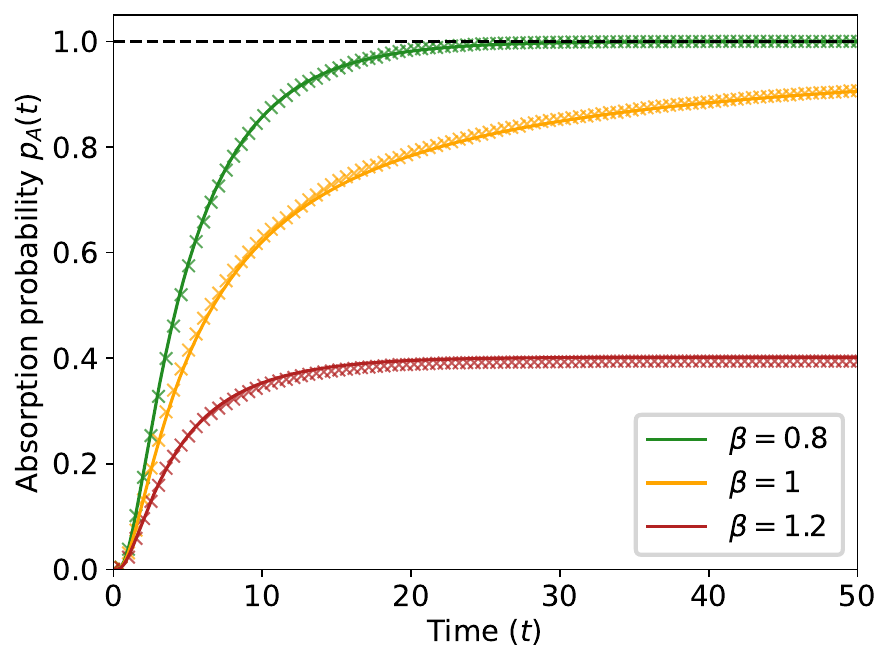}
    \caption{Absorption probability---the probability of being extinguished---versus time. Here, $j(0) = N = 5$ with curves pictured for various birth rates, \edit{alongside scatter plots of simulated data (marked with an `x') in an ensemble of $10^4$ processes}. The super- and sub-critical curves quickly approach their asymptotic limits, whereas the critical curve does so much more slowly. Both sub-critical and critical processes approach an absorption probability of one as $t \to \infty$.}
    \label{fig:zerogamabsorb}
\end{figure}

To keep a small (or even finite) footprint, the fire must be extinguished---the population must reach the state $j(t) = 0$. The likelihood of this is characterized by the absorption probability introduced earlier. Here, it is given by
\begin{gather}
    \label{eq:zerogamabsorb}
    p_A(t)=\left(\frac{1-e^{(\beta-1)t}}{1-\beta e^{(\beta-1) t}}\right)^N;\\
   \lim_{\beta \to 1}p_A(t) = \left(\frac{t}{1+t}\right)^N,
\end{gather}
and is plotted in Fig. \ref{fig:zerogamabsorb}. 

In the sub-critical phase, absorption almost always occurs after long times. Specifically, taking the asymptotic limit $t\to \infty$ one finds
\begin{equation}
\label{eq:zerogamAsymAbsorb}
    p_A(\infty) = \lim_{t\to\infty}G_0(0,t) = \begin{cases}
        \beta^{-N}, & \beta > 1;\\
        1, & \beta \leq 1.
    \end{cases}
\end{equation}
This result states that asymptotically, fires in the critical or sub-critical phases eventually extinguish. The probability of a super-critical fire extinguishing decreases exponentially in the initial size of the fire and is in general less than unity. 

% The fact that absorption happens with probability 1 in the critical case may be surprising given that the mean population does not change in time. Recall however that the critical process consists of large fluctuations around this constant mean. While the process may freely fluctuate into large values of $j(t)$, a significant downward fluctuation will result in absorption at $j = 0$, or extinguishing of the fire. Such a downward fluctuation is possible and, in the limit of infinite time, inevitable; leading all such processes to eventually absorb.

\begin{figure}[t]
    \centering
    \begin{minipage}{0.45\textwidth}
        \centering
        \includegraphics[width=\textwidth]{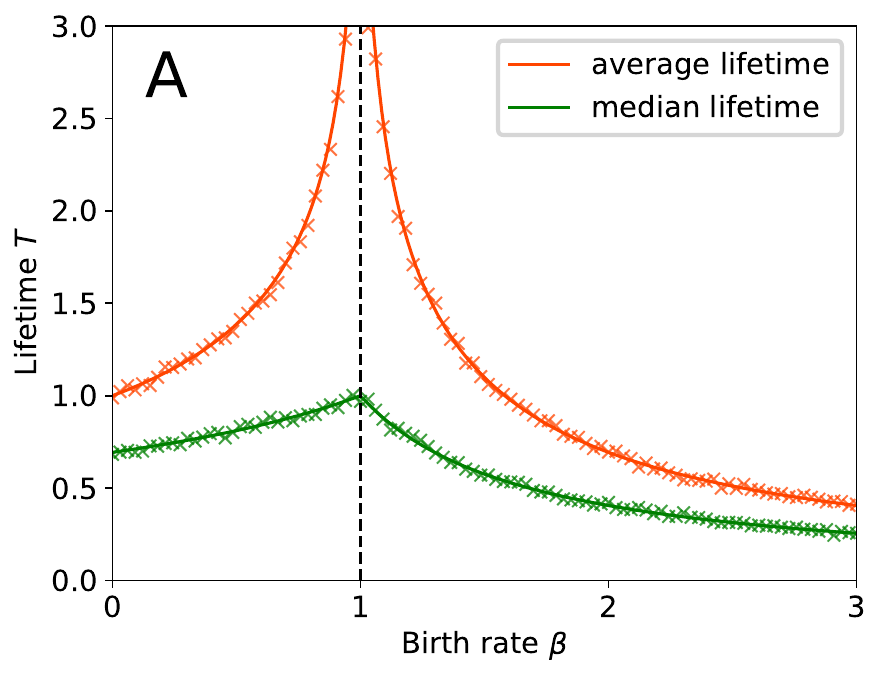} % first figure itself
        % \caption{Absorption time distribution with $N = 5$; the small fire.}
        
    \end{minipage}\hfill
    \begin{minipage}{0.45\textwidth}
        \centering
        \includegraphics[width=\textwidth]{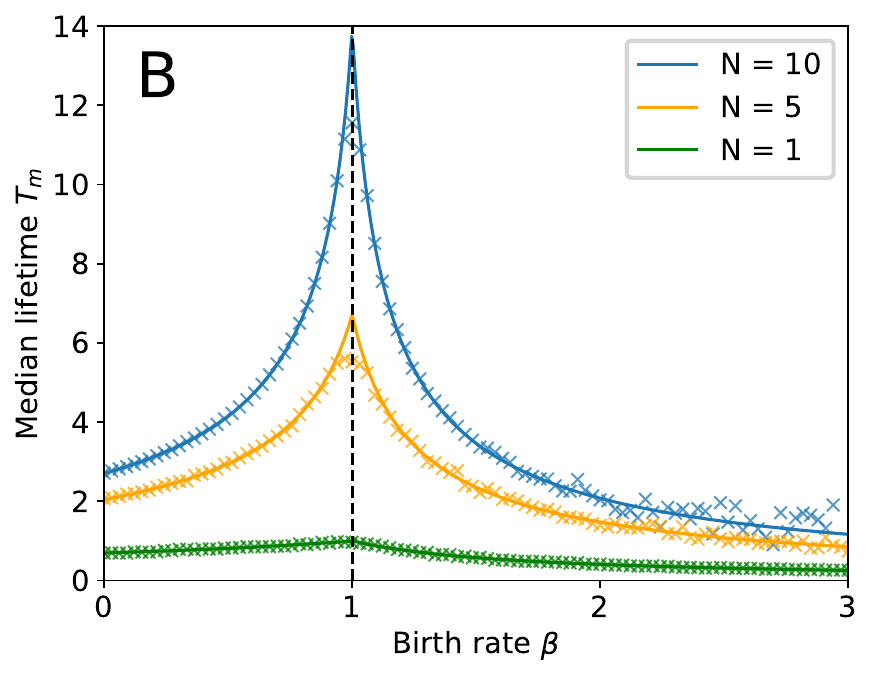} % second figure itself
        % \caption{Absorption time distribution with $N = 55$; the big fire.}
    \end{minipage}
    \caption{Lifetime statistics versus birth rate for processes which end in absorption; \edit{exact results alongside simulated data (marked with an `x') from an ensemble of $10^4$ processes. In the case $N = 5$, $10^5$ simulated processes were needed to obtain sufficiently conditioned data}. In Fig. 4A; average and median lifetimes versus birth rate $\beta$ and with $j(0) = N = 1$. At criticality, the average lifetime diverges. In Fig. 4B; the median process lifetime for various initial population sizes, on a logarithmic scale. The dependence on the initial population $N$ is strongest at criticality $\beta = 1$.}
    \label{fig:zerogamavgmedianlife}
\end{figure}

To better understand the timescales of absorption, consider the distribution of lifetimes, where the lifetime $T$ is defined as the exact time of absorption. With a sub-critical fire, it is most likely that absorption occurs early; the lifetime distributions have medians near $T_m \sim \log N$. The critical process, on the other hand, tends to take a much longer time to absorb, with the median critical lifetime going like $T_m \sim N$. Such long-lived fires would generically result in a large footprint. Note that because the distributions in question are conditioned on absorption occurring, the median lifetime in the super-critical phase also goes like $T_m \sim \log N$, with $T_m\to 0$ as $\beta \to \infty$. These properties are demonstrated by the lifetime statistics in Fig. \ref{fig:zerogamavgmedianlife}.

\edit{The distributions of absorption times for the suppression-free process share universal behavior with other absorbing Markov processes, and are in fact Gumbel-distributed \cite{hathcock2022asymptotic}. This universality class is defined by processes where the birth/death rates decrease linearly as one approaches the absorbing state. This is true for the zero-suppression process in all phases, as was noted in \cite{kessler2023extinction}.}

It may seem counter-intuitive that the super-critical median and average lifetimes are so comparable to the sub-critical ones. This is because most super-critical fires are \textit{not} absorbed at zero; those that do become absorbed likely reached $j = 0$ quickly, as a rare downward fluctuation. Hence the lifetime distributions, being predicated on absorption occurring, are quite similar away from criticality. In fact, it can be shown that super-critical processes with birth rate $\beta$, when conditioned on eventual absorption, behave like sub-critical processes with birth rate $1/\beta$ \cite{Waugh1958,linearTavare}. When absorption does not happen, the processes are formally of infinite lifetime. 

Basic statistics of these lifetime distributions are shown in Fig. \ref{fig:zerogamavgmedianlife}, where one can see that proximity to criticality leads to the longest-lived eventually absorbing processes. \edit{For high super-critical birth rates, the vast majority of processes do not absorb. The distributions of lifetimes are therefore conditioned on extremely rare events, which means they have little physical importance. This is evidenced by the degeneration of the numerical data on the right-hand side of Fig. \ref{fig:zerogamavgmedianlife}B, which was generated from an ensemble of $10^5$ simulated processes.}

\begin{figure}[!h]
	\centering
	\includegraphics[width=\columnwidth]{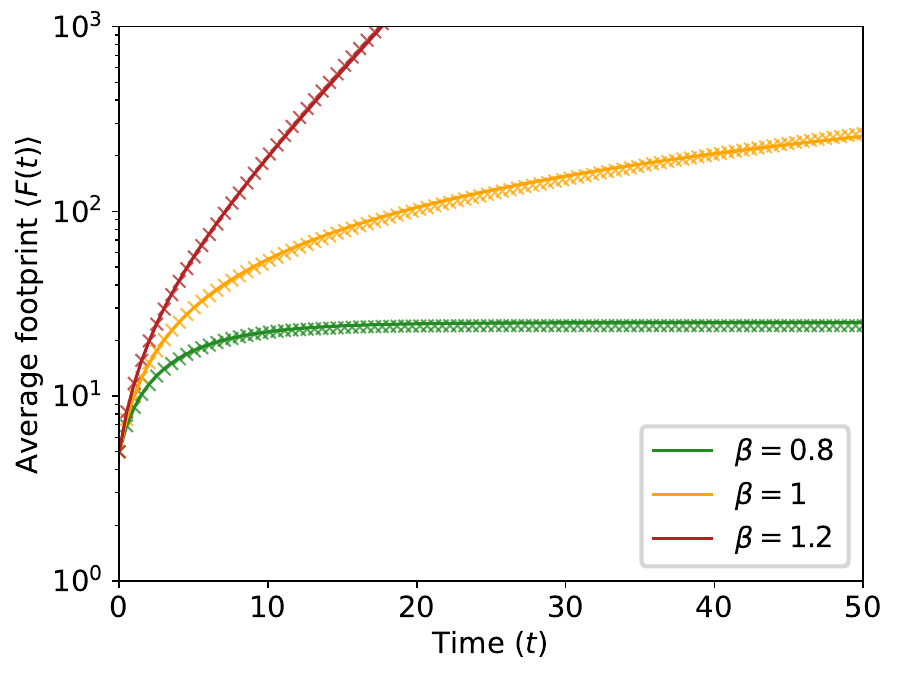} 
    \caption{Average footprint $\langle F(t)\rangle$ versus time on a logarithmic scale for various birth rates $\beta$, \edit{compared with simulated data (marked with an `x') from an ensemble of $10^3$ processes}. Each curve has $N = 5$. The sub-critical processes terminate with a finite average footprint while in the critical and super-critical phases, the average footprint continues to grow for all time, exceeding any finite bound $J$.}
    \label{fig:avgFtpt}
\end{figure}
Along with the timescale of absorption, one ideally would understand the timescale and probability of escape: when (and if) the footprint reaches some given size $J$. Exact results on the joint distribution of $j(t)$ and $F(t)$ can be obtained using the method of the generating function, though most calculations using this method are out of practical reach \cite{kendallGeneral}. The average footprint $\ex{F(t)}$, a tractable quantity, evolves as
\begin{gather}
    \label{eq:zerogamavgFagain}
   \langle F(t)\rangle=\frac{N}{\beta-1}\left(\beta e^{(\beta-1) t}-1\right);\\  \lim _{\beta \rightarrow 1}\langle F(t)\rangle=N(t+1).
\end{gather}
In the sub-critical phase there is a finite asymptotic limit whereas in the critical and super-critical phases the average footprint grows without bound, surpassing any finite size $J$. This behavior is demonstrated in Fig. \ref{fig:avgFtpt}. \edit{However, this quantity only minimally describes the distribution of footprints.}

Finally, consider the asymptotic distribution of footprints. Empirically it is observed that the cumulative distribution of wildfire footprints is a power law; the probability $P(F \geq J)$ that a given fire has footprint larger than or equal to $J$ goes like $\sim J^{-1/2}$ \cite{maxjeanWildfireHOT}. The same is true for the near critical birth-death process. In the large footprint $F,J \gg 1$ regime, one approximately has\footnote{To get this result one approximates a sum by an integral and enters the asymptotic regime $J,F \gg 1$.}
\begin{equation}
    P(F_\infty \geq J)\approx \frac{J^{-1 / 2}}{\sqrt{\pi}}
   + \frac{(1-\beta)}{2}\left(\frac{J^{-1 / 2}}{\sqrt{\pi}}-1\right),
\end{equation}
expanding to first order below the critical point $\beta = 1$. By $F_\infty$ we mean the asymptotic limit of $F(t)$, i.e. $F_\infty = \lim_{t\to\infty}F(t)$. The power law scaling is correctly reproduced, and is stable to leading order in $(1-\beta)$. 

These results characterize the probability of absorption, its timescales, and the asymptotic distribution of footprints for the linear birth-death process without suppression. Since absorption occurs with probability 1 in the sub-critical and critical phases, we have described these dynamics of these phases completely. In the following sections, we repeat this analysis for the birth-death process with suppression. 
\section{The population dynamics}
\label{sec:populationdynamics}
We now turn to the full birth-death-suppression process, where we solve the master equation
\begin{equation}
\label{eq:sec3master}
    \frac{d}{dt}P(t)=P(t) \mathcal{A},
\end{equation}
in which the matrix $\cal A$, defined in Eqn.~\eqref{eq:Amat}, has elements given by the birth and death rates $\lambda_j = \beta j,\ \mu_j = \delta j + \gamma$ (with $\delta = 1$) for each value of the population $j(t)$. For reasons of convention, we introduce a new variable $k$ to index the state space defined simply as $k = j - 1$. In terms of this new index the birth-death rates are
\begin{equation}
    \lambda_k = \beta (k+1),\quad \mu_k = k + \gamma + 1;\quad k\geq 0.
\end{equation}
The reason for this re-labeling is to be consistent with the wider literature on birth-death processes and the conventions for orthogonal polynomials. Note that now the state $j= 0$ or $k = -1$ is absorbing, and that a \edit{process} with one initial \edit{unit of the population} $N = 1$ has $j(0) = 1,\ k(0) = 0$. 

With zero suppression the method of generating function can be used to find the occupation probabilities for the population $j(t)$. In the presence of a nonzero suppression factor $\gamma \neq 0$, an inhomogeneous term is introduced to the generating function equation preventing a direct solution by this method, as demonstrated in Appendix \ref{sec:appgfunc}. Given this obstacle, to compute the transition matrix $P(t)$ one must turn to a different approach: the method of Karlin and McGregor, making use of orthogonal polynomials, which is nothing more than solving Eqn.~\eqref{eq:sec3master} by eigenvector decomposition. 
\subsubsection*{The method of Karlin and McGregor}
The (infinite) matrix $\cal A$, and the matrix $P(t)$, are almost symmetric. To be precise, \edit{there exists a set of constants $\pi_n$ such that} the matrix $P(t)$ satisfies the symmetry condition
\begin{equation}
\label{eq:symmetryCondition}
    \pi_n P_{n m}=\pi_m P_{m n}
\end{equation}
where the $\pi_n$ are defined as
\begin{equation}
	\pi_0=1, \quad \pi_n=\frac{\lambda_0 \lambda_1 \cdots \lambda_{n-1}}{\mu_1 \mu_2 \cdots \mu_n};\qquad n,m \geq 0.
\end{equation}
This symmetry relation \edit{(also referred to as the condition of detailed balance)} describes the intuitive fact that the process is time-reversed if one swaps $\lambda \leftrightarrow \mu$. Together, the backward equation \eqref{eq:sec3master} and the symmetry condition \eqref{eq:symmetryCondition} imply that the transition matrix $P(t)$ satisfies the Kolmogorov forward equation, that is
\begin{equation}
\label{eq:kolmogorovfwd}
    \frac{d}{dt}P(t) = \mathcal{A}P(t).
\end{equation}
The strategy is to solve the above by eigendecomposition of the matrix $\cal A$. Assume $\cal A$ has a complete set of eigenvectors\footnote{To be precise, one can define the symmetric operator $H$ with matrix elements $H_{nm} = \pi_n\mathcal{A}_{nm}$ to apply a version of the spectral theorem for self-adjoint operators. } $\ket{Q(x)}$ with eigenvalues $-x$ and $x \geq 0$: the eigenvalues must be non-positive so that matrix elements of $P(t)$ remain bounded as $t\to\infty$. We seek an explicit form of the eigenvectors $\ket{Q(x)}$. To make this characterization, write the abstract equation $-x\ket{Q(x)} = \mathcal{A}\ket{Q(x)} $ in its component form: 
\begin{align}
Q_0 & =1, \\
-x Q_0 & =-\left(\lambda_0+\mu_0\right) Q_0+\lambda_0 Q_1, \\
-x Q_k & =-\left(\lambda_k+\mu_k\right) Q_k+\lambda_k Q_{k+1} + \mu_k Q_{k-1}.
\end{align}
The components $Q_k(x)$ are defined by a three-term recurrence relation, where $\deg Q_k(x) = k$; they are a family of orthogonal polynomials in $x$ on the real half line $x \geq 0$. Indeed, Favard's theorem states that any $Q_k(x)$ satisfying a three-term recurrence relation like the above are orthogonal polynomials with respect to some positive spectral measure $d\sigma(x)$ \cite{favard}. The polynomials $Q_k(x)$ are therefore the object of central interest. They satisfy the normalized orthogonality relation 
\begin{equation}
\int_0^{\infty} Q_n(x) Q_m(x) d\sigma(x)=\frac{\delta_{nm}}{\pi_m}.
\end{equation}
which may be derived from their recurrence relations. From these polynomials, and their associated spectral measure $d\sigma(x)$, the matrix elements of $P(t)$ are given by the integral
\begin{equation}
\label{eq:masterPQ}
P_{nm}(t)=\pi_m \int_0^{\infty} e^{-x t} Q_n(x) Q_m(x) d\sigma(x).
\end{equation}
This is the \textit{spectral form} of the transition matrix and can easily be seen to solve the forward equation \eqref{eq:kolmogorovfwd} by virtue of the recursion satisfied by the polynomials $Q_k(x)$. 

However, determining the polynomials $Q_k(x)$ and their spectral measure may not be straightforward. For the process $\lambda_k = \beta (k+1),\ \mu_k = k + \gamma+1$, the polynomials in question can be identified---as done by Karlin and McGregor \cite{karlinear}---as a known set of polynomials of the Askey scheme, a classification of hypergeometric orthogonal polynomials \cite{askey}. To be precise, the polynomials of interest are of the Meixner and Laguerre families. Defined in Appendix \ref{sec:appPolys}, each polynomial family is associated with a different phase of the process: $\beta > 1,\ \beta = 1,\ \beta < 1$. 

Given the known expressions and identities for these polynomials one can derive explicit expressions for the transition matrix elements $P_{nm}(t)$. The following subsections make use of these results to compute statistics of the suppressed birth-death process, and to analyze the effect of suppression on the population.

\subsection{Lifetime statistics}
The starting point for determining the distribution of lifetimes and hence the probability of absorption is computing the transition matrix element $P_{n0}(t)$, a result appendicized in Eqn.~\eqref{eq:fulltranistionNoncrit}. The element $P_{n0}$ represents the probability of transitioning from a state $j = n+1$ to the $j = 1$ state, which physically represents a fire with a single active firelet.

In the right coordinates, the measure on lifetimes takes a particularly simple form. The random variable $T$, the lifetime or exact time of absorption, for processes with $j(0) = N$ is distributed as $\propto P_{N-1,0}(T)dT$. In terms of the variable $s(t) = e^{(\beta-1)t}$, the matrix element $P_{N-1,0}$ is
\begin{multline}
    P_{N-1,0}(s) = \frac{(\gamma+2)_{N-1}}{(N-1)!}s^{\gamma+1} \\ \times \left(\frac{1-\beta}{1-\beta s}\right)^{\gamma + 2}\left(\frac{1-s}{1-\beta s}\right)^{N-1}.
\end{multline}
This formula is valid for both the super- and sub-critical cases. It is convenient to define yet another variable $z(t)$ as
\begin{gather}
	\label{eq:zee}
    z(t) = \frac{1 - e^{(\beta-1)t}}{1-\beta e^{(\beta-1)t}} = \frac{1-s(t)}{1-\beta s(t)};\\ \lim_{\beta \to 1}z(t) = z_c(t) = \frac{t}{1+t}.
\end{gather}
The coordinate $z(t)$ is equivalent to the probability of absorption in the zero suppression, single initial firelet process, and so satisfies $0\leq z(t)\leq 1$ always. Additionally, for any phase of the process, $z(t)$ is monotonically increasing in time. Asymptotically,
\begin{equation}
\label{eq:zasymp}
    \lim_{t\to\infty}z(t) \equiv z_\infty = \begin{cases}
        1, & \beta \leq 1;\\
        \frac{1}{\beta}, & \beta > 1.
    \end{cases}
\end{equation}
The measure on lifetimes $d\sigma(T) \propto P_{N-1,0}(T)dT$, after a change of variables $T \to z(T)$, may be written
\begin{equation}
    d\sigma(T) \propto (1-z)^{\gamma}z^{N-1}dz(T),
\end{equation}
where the proportionality is fixed by normalization. The normalized measure is
\begin{equation}
\label{eq:lifetimemeasure}
    d\sigma(T) = \frac{1}{B(z_\infty;N,\gamma+1)}(1-z)^{\gamma}z^{N-1}dz(T).
\end{equation}
The normalization is an incomplete beta function. Defined by the integral
\begin{equation}
\label{eq:incBeta}
    B(x;a,b) = \int_0^x dt\ t^{a-1}(1-t)^{b-1},
\end{equation}
for $x = 1$ it coincides with the classical beta function of Euler:
\begin{equation}
    B(1;a,b) = B(a,b) = \frac{\Gamma(a)\Gamma(b)}{\Gamma(a+b)}.
\end{equation}
The lifetimes are therefore beta-distributed, or approximately so, with shape parameters $N$ and $\gamma + 1$. The absorption probability $p_A(t)$ is simply the cumulative distribution function, which here is given by the regularized incomplete beta function \cite{NIST:DLMF}: 
\begin{equation}
\label{eq:absorbprob}
    p_A(t)=\frac{B(z(t) ; N, \gamma+1)}{B(N, \gamma+1)}.
\end{equation}
Finally, consider the asymptotic probability of absorption. Since $\gamma > 0$ only increases the aggregate death rate, absorption should remain asymptotically certain except in the super-critical phase. Indeed, for $\beta \leq 1$, the limit is
\begin{equation}
    p_A(\infty) = \frac{B(1;N,\gamma+1)}{B(N,\gamma + 1)} = 1; \qquad (\beta \leq 1),
\end{equation}
where the beta function $B(1;a,b) = B(a,b)$ is `completed' for $z(\infty) = 1$. In the super-critical case the situation is more complicated as the beta function remains incomplete; the exact expression is Eqn.~\eqref{eq:absorbprob} with $z= 1/\beta$. In the special case $N = 1$, the expression simplifies to 
\begin{equation}
\label{eq:supercritasymabsorb}
    p_A(\infty) = 1 - \left(\frac{\beta - 1}{\beta}\right)^{\gamma+1};\qquad (\beta > 1).
\end{equation}
Note that this expression, the probability of the process ending at $j = 0$, is the complementary probability to the process ending `at infinity.' The measure at infinity and its interpretation will be discussed further in Section \ref{sec:predfootprint}. 
\subsection{The effect of suppression}
\begin{figure}[!h]
    \centering
    \begin{minipage}{0.45\textwidth}
        \centering        \includegraphics[width=\textwidth]{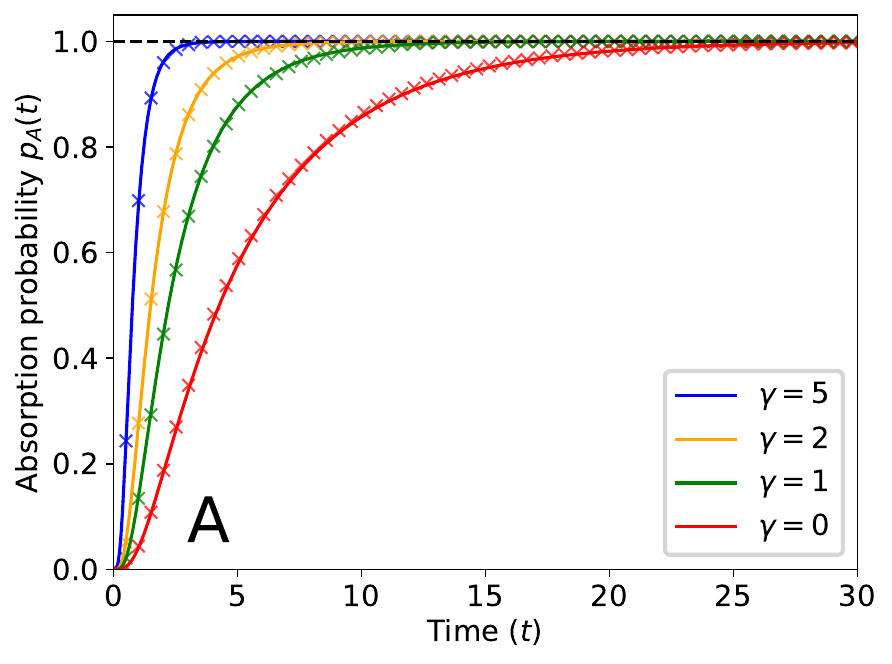} % first figure itself
        % \caption{Absorption time distribution with $N = 5$; the small fire.}
        
    \end{minipage}\hfill
    \begin{minipage}{0.45\textwidth}
        \centering
        \includegraphics[width=\textwidth]{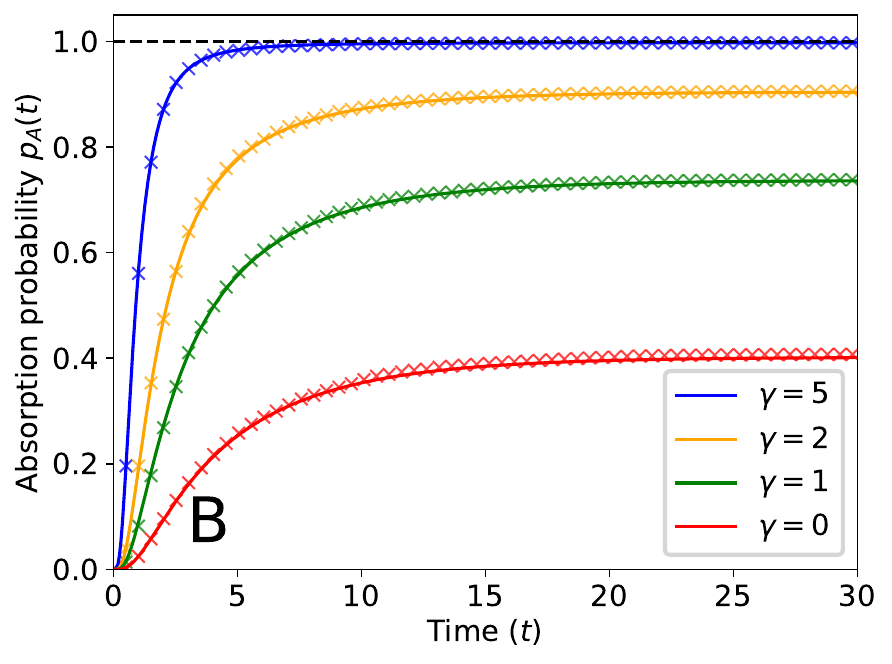} % second figure itself
        % \caption{Absorption time distribution with $N = 55$; the big fire.}
    \end{minipage}
    \caption{Absorption probabilities versus time at nonzero suppression $\gamma$ \edit{alongside simulated data (marked with an `x') from an ensemble of $10^4$ processes}. Shown is the effect of suppression on the probability of absorption in a sub-critical process ($A$, with $\beta = 0.8$) and a super-critical process ($B$, with $\beta = 1.2$). Both began with an initial population $N = 5$. In $B$, with a super-critical birth rate, suppression increases limiting value of the absorption probability and hence the overall share of absorbing processes.}
    \label{fig:absorbProb}
\end{figure}

Having computed the distributions of lifetimes and probability of absorption, one can now ask: what is the effect of suppression $\gamma > 0$ on the population dynamics? Without suppression, the average population $\ex{j(t)}$ is characterized by exponential growth or decline depending on the phase of the process. With \edit{$N = 1$}, the average population with suppression is 
\begin{gather}
    \ex{j(t)} = e^{(\beta-1)t}[1-z(t)]^\gamma;\\ \lim_{\beta \to 1}\ex{j(t)} = \left(\frac{1}{1+t}\right)^\gamma.
\end{gather}
With $\gamma \geq 0$ and $z(t) < 1$, the average population in the presence of suppression is always less than it would be with $\gamma = 0$, and the average population goes to zero in the limit of infinite supppression $\gamma \to \infty$ for any fixed value of $\beta$. Indeed, the presence of suppression greatly increases the chance that small fires absorb at zero, decreasing the time at which the absorption probabilities become reasonably close to their asymptotic values. This is demonstrated in Fig. \ref{fig:absorbProb}, where one can see that suppression also significantly increases the asymptotic probability of absorption for super-critical processes.

\begin{figure}[!h]
	\centering
	\includegraphics[width=\columnwidth]{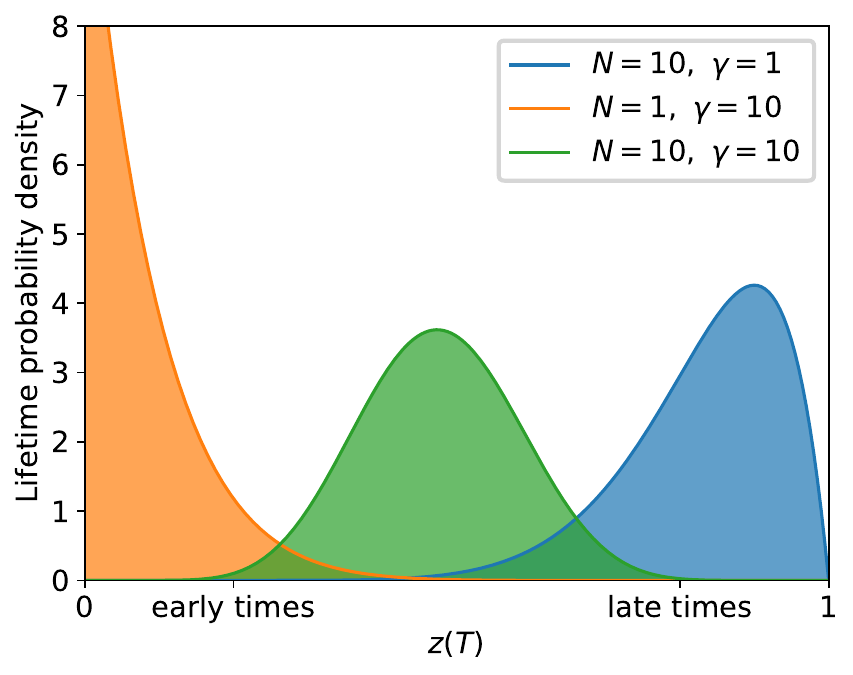}
	\caption{The exact distribution of lifetimes for various values of the initial population $N$ and suppression rate $\gamma$. Lifetimes are beta-distributed with respect to the coordinate $z(T)$ defined in Eqn.~\eqref{eq:zee}; here the measures are pictured with $\beta = 1$. The normal limit of the beta distribution, where $N\gg1$ and $\gamma \gg 1$, is apparent in green. }
    \label{fig:beta}
\end{figure}

To see the interplay between the initial size $N$ of the fire and the suppression $\gamma$, consider the normalized distribution of lifetimes for $\beta \leq 1$. Here, $z(T)$ runs from 0 to 1 with the beta distribution
\begin{equation}
    \label{eq:subcritlifetime}
    d\sigma(T) = \frac{1}{B(N,\gamma+1)}(1-z)^\gamma z^{N-1}dz(T),
\end{equation}
where the lifetime $T$ may be found from $z(T)$ via
\begin{equation}
    T = \frac{1}{\beta -1}\log\left[\frac{1-z(T)}{1-\beta z(T)}\right].
\end{equation}
When $N$ becomes large compared to $\gamma$, the mass of the distribution \eqref{eq:subcritlifetime} becomes concentrated around $z \lesssim 1$. This means that lifetimes in this regime are likely to be longer, consistent with the physical expectation that larger fires burn out more slowly. In the strict large size limit $N \gg 1$, suppression is neglible: the aggregate death rate $\mu_j = j + \gamma$ is dominated by the first term proportional to $N$. The limit as $\gamma \to 0$ is not singular and commutes with the limit $N \gg 1$. On the other hand, when $\gamma$ is large compared to $N$, the distribution becomes peaked near $z \gtrsim 0$, telling us that highly suppressed \edit{processes absorb} quickly. None of this is physically surprising, and is demonstrated by the orange and blue distributions of Fig. \ref{fig:beta}.

However, there is another limit: simultaneously taking $N\gg1$ and $\gamma \gg 1$. Recalling that $\gamma$ is defined in units of the death rate of a single firelet, this limit corresponds to a large initial population coupled with suppression which \edit{eradicates the population much more quickly than natural deaths would occur} (a physically reasonable regime \edit{for fire suppression}). In this limit, the distribution on lifetimes converges to a normal distribution with mean $\ex{z}=  N/(N+\gamma)$ and variance which scales like $\sim 1/N$. The measure \eqref{eq:subcritlifetime} is shown in Fig. \ref{fig:beta} for different values of the parameters $N,\gamma$, demonstrating the different regimes.  

\begin{figure}[h!]
	\centering  
    \includegraphics[width=\columnwidth]{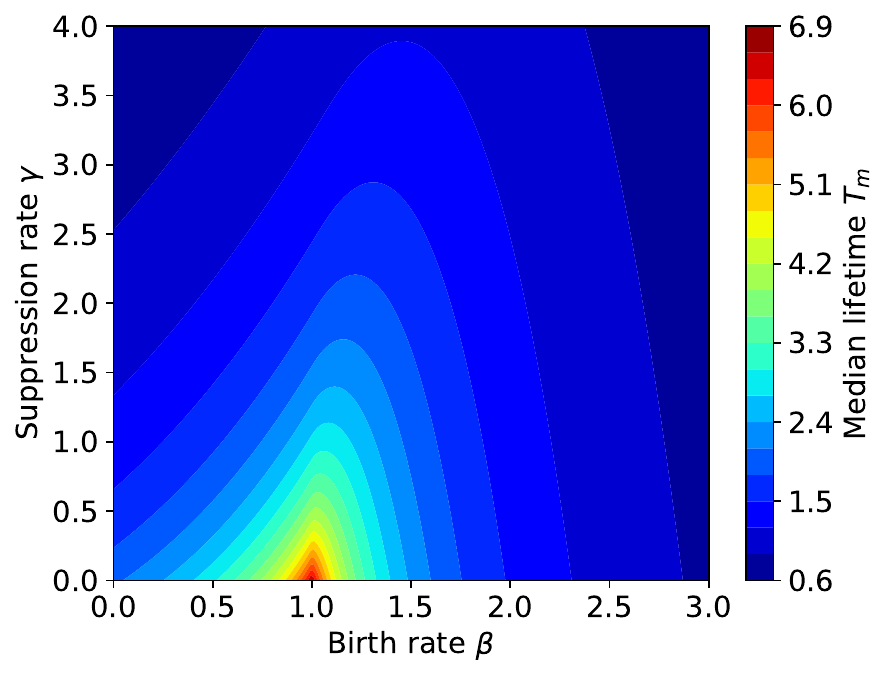} % first figure itself
    \caption{Contour plot of the median lifetime in the $(\beta, \gamma)$ plane, pictured for a fire with $N = 5$ and conditioned on eventual absorption. The longest-lived finite processes are near un-suppressed criticality.}
    \label{fig:medianContour}
\end{figure}

The median and mean coincide in the simultaneous limit $N,\gamma \gg 1$ and one has the median lifetime
\begin{gather}
    T_m \approx \frac{1}{1-\beta}\log\left[1 + \frac{N}{\gamma}(1-\beta)\right];\\ \lim_{\beta \to 1}T_m \approx \frac{N}{\gamma}.
\end{gather}
As in the zero suppression case, the median lifetimes display a logarithmic dependence on the initial population $N$ away from criticality which at criticality is enhanced to a stronger, linear dependence. However, the suppression rate $\gamma$ now acts as a multiplicative dressing which reduces the effective initial population. Therefore, to obtain a median lifetime which is $\OO(1)$, the suppression rate, as multiples of the natural death rate, must be of the same order as the initial size. The median lifetime may be represented exactly as an inverse regularized beta function; a contour plot of the exact median lifetime $T_m$ in the $\beta, \gamma$ plane is shown in Fig. \ref{fig:medianContour}.

The statistics on lifetimes describe only those processes which eventually absorb. However, the time until absorption may be infinite: at criticality, it is on average infinite. Any process with an infinite lifetime is clearly unrealistic---it does not consider any physical bound on the population $j(t)$ or the footprint $F(t)$. In a real modeling scenario, systems are not of infinite size; there will be a finite \edit{effective} size $J$ to any system. The presence of this \edit{threshold} prevents sojourns of infinite time. In the following section, we develop the theory of the footprint $F(t)$ and address these issues of infinitude.

\section{Predicting the footprint}
\label{sec:predfootprint}

Recall that the footprint $F(t)$ is defined to share all the positive transitions of $j(t)$. It is therefore cumulative, monotonically increasing ad infinitum unless absorption at $j = 0$ occurs. The notion of a cumulative population for the birth-death process was introduced by Kendall in \cite{kendall1949stochastic,kendallGeneral}\edit{, where its interpretation was that of measuring the severity of an epidemic.} In the wildfire interpretation, the cumulative population (or footprint) counts the total burned area and serves as a measure of the overall severity of the fire event. 

In Appendix \ref{sec:appgfunc}, generating function methods are used to obtain predictions about the asymptotic distribution of footprints in the zero suppression process, \edit{ recreating the results of Kendall.} As before, for the population dynamics, this approach is unsuitable for the suppressed process with $\gamma > 0$. The generating function equation depends on an unknown function. 

The assumption is that the goal of suppression is to extinguish a fire while keeping the footprint $F(t)$ less than some \edit{threshold} $J$. To make this precise, define the \textit{escape probability}, or $P_J(N)$, by
\begin{equation}
    \label{eq:burnProb}
   P_J(N)=\{\text {probability of } F(\infty) \geq J \ \big| \ j(0)=N\}.
\end{equation}
This is the probability that the footprint at some point exceeds \edit{the threshold} $J$ of burnable substrate. Minimizing the escape probability is a natural metric for successful suppression. 

A process which escapes, having $F(\infty) \geq J$, may or may not eventually absorb, while a process which does not escape (with $F(\infty) < J$) must end in absorption. The definition of the escape probability does not reference time (or time of absorption) but rather the asymptotic outcome of a given process. Computing $P_J(N)$ is therefore a matter of counting \textit{transitions} of the Markov process rather than time. However, being agnostic to the time of absorption, this quantity alone does not give any estimate of when the footprint $F(t)$ may reach some value. With no time variable, the birth-death process becomes a random walk on the state space. We turn now to analyzing this random walk.
\subsection{The random walk polynomials}
For any continuous-time birth-death process with rates $(\lambda_k,\mu_k)$ with $k \geq 0$ (recall $k = j- 1$ where $j$ is the number of firelets) there exists an embedded Markov chain: the \textit{jump chain}. It is a random walk on the state space where the transition probabilities are
\begin{gather}
    p_k \equiv \Pr\{k \to k+1\} =  \frac{\lambda_k}{\lambda_k + \mu_k};\\ q_k \equiv \Pr\{k \to k -1\} = \frac{\mu_k}{\lambda_k + \mu_k}.
\end{gather}
The continuous-time process solved in the previous section is just this jump chain coupled with holding times in each state which are exponentially distributed $\Delta t \propto \exp((\lambda_k + \mu_k)^{-1})$. This is how one simulates a continuous-time process on a computer. 

Let $k(n)$ be the state of the process after $n$ transitions: in the present context, $j= k + 1$ still represents the actively burning population, but we are no longer keeping track of time $t$. The state depends instead on the transition count $n$. Consider the aggregate birth and death rates given by
\begin{equation}
\label{eq:shiftedAgRates}
    \lambda_k = \beta(k+1);\qquad \mu_k = k + \gamma + 1.
\end{equation}
Recall that here, the $k = -1$ state is absorbing, with $p_{-1} = q_{-1} \equiv 0$. The jump variable absorbs after some number of transitions $n_T$, i.e. $k(n_T) = -1$, after which there are no further transitions. The goal is to characterize the distribution of transition counts $n_T$ at absorption: this will also determine the distribution of footprints at absorption. The problem is analogous to determining the distribution of lifetimes $T$ in the continuous-time process, except now one seeks a distribution over the positive integers.

In the embedded jump chain, the object of interest is the $n$-step transition matrix $S(n)$ with elements
\begin{equation}
    S_{\ell,\ell'}(n) = \Pr\{k(m+n) = \ell'\ |\ k(m) = \ell\},\quad \ell,\ell' \geq 0,
\end{equation}
with the analogy $S(n) \sim P(t)$. It obeys a semi-group condition
\begin{equation}
    \label{eq:altmaster}
    S(n+1) = S(1) S(n)
\end{equation}
instead of a master equation. 

\edit{In the absence of suppression, the random walk associated to the birth-death process is a simple random walk: the probabilities $p_k,\ q_k$ are independent of the state $k$. The transition matrix $S(n)$ can therefore be determined either by the method of generating function or standard binomial arguments: it becomes equivalent to the ``gambler's ruin" problem, see e.g. \cite{ross1995stochastic}. However, with a non-zero suppression factor, the probabilities $p_k,\ q_k$ are state-dependent and the random walk is not simple. As with the continuous-time process, the presence of a suppression term introduces an unknown function to the generating function equation and renders a direct solution infeasible; these more standard approaches will not work.}

The master equation governing the continuous-time process was solved by eigenspace decomposition.  This amounted to defining and determining a set of orthogonal polynomials, an approach introduced by Karlin and McGregor \cite{karlinDiff}. In the later work \cite{randomwalksKarlin}, the same authors show that the orthogonal polynomial approach can also solve the random walk problem. 

The procedure is essentially the same as in the continuous-time process: the transition probabilities define a set of polynomials $W_k(x)$ via the relations
\begin{align}
W_0(x) & =1, \\
x W_0(x) & =p_0 W_1(x), \\
x W_k(x) & =q_k W_{k-1}(x)+p_k W_{k+1}(x).
\end{align}
The structure of these recurrence relations and the relation $p_k + q_k = 1$ implies that the $W_k(x)$ are orthogonal on the interval $[-1,1]$ with respect to some even measure $d\sigma(x)$:
\begin{equation}
    \int_{-1}^1 W_\ell(x) W_{\ell'}(x) d\sigma(x)=h_\ell\delta_{\ell\ell'}.
\end{equation}
Given the polynomials $W_k(x)$ and their spectral measure $d\sigma(x)$, the $n$-step transition matrix elements have the spectral form
\begin{equation}
    S_{\ell,\ell'}(n) = \frac{1}{h_{\ell'}}\int_{-1}^1 x^n W_\ell(x)W_{\ell'}(x) d\sigma(x).
\end{equation}
The immediate focus is to compute this object, followed by relating properties of this matrix $S(n)$ to the distribution of footprints $F$. 

Previously, when applying the method of orthogonal polynomials to the continuous-time process, the polynomial families of interest were known and classified: they were Meixner polynomials away from criticality and Laguerre polynomials at criticality. Both these fall into the Askey scheme of hypergeometric orthogonal polynomials \cite{askey}. The identification was made by examining the recurrence relation given by the birth and death rates and comparing it to known recurrence relations for previously classified polynomials. In the present case, the recurrence is
\begin{multline}
    \label{eq:ftptRecurrence}
    xW_n(x) = \frac{\beta(n+1)}{(\beta+1)(n+1) + \gamma}W_{n+1}(x) \\+ \frac{n+\gamma +1}{(\beta+1)(n+1) + \gamma}W_{n-1}(x).
\end{multline}
The recurrence has two parameters $\beta, \gamma$, and a particular (rational) dependence on $n$ in its coefficients. An examination of all the polynomials in the Askey scheme reveals that this recurrence is not so classified. However, the $W_n(x)$ are equivalent to a previously studied set of polynomials outside the Askey scheme\footnote{We are indebted to W. Van Assche for his correspondence, pointing out this equivalence.}: the Pollazcek polynomials \cite{szego,chihara2011introduction}. The Pollazcek polynomials $P_n^\lambda(x;a,b)$ have been studied in numerous physical \cite{koornwinder1989meixner,Bank} and mathematical contexts \cite{van1990pollaczek,rui1996asymptotic,li2001asymptotics,yermolayeva1999spectral,araaya2004meixner}. The exact correspondence is
\begin{equation}
	W_n\left(\frac{2 \sqrt{\beta}}{\beta+1} x\right)=P_n^\lambda(x ; a, b) \beta^{-n / 2},
\end{equation}
where the parameters $\lambda,\ a,\ b$ are related to the birth-death-suppression parameters as
\begin{equation}
	b=0, \quad \lambda=1+\frac{\gamma}{2}, \quad \lambda+a=1+\frac{\gamma}{\beta+1}.
\end{equation}
Expressions for the Pollazcek polynomials and their measure of orthogonality have been found by various methods, but the parametrizations used are quite different from the present set. In the interest of examining the structure of these polynomials from the perspective of the birth-death-suppression process, in Appendix \ref{sec:firewalk} we explicitly construct the polynomials $W_n(x)$ and their measure analytically. \edit{The reader is directed to the appendices for background, while in the present section we will continue to build calculational tools from those results.}

The analysis required to construct the polynomials is inspired by the closely related Section 4 of Askey and Ismail's work \cite{ismail} which does the same for the birth-death-immigration (BDI) process. This closely related process has aggregate birth and death rates $\lambda_j = \beta j + m,\ \mu_j = \delta j$, where $m$ is interpreted as an immigration rate: the $j$-independent rate at which new units are added to the population. The suppression rate is roughly the opposite of the immigration rate, where the suppression $\gamma$ is the $j$-independent rate at which units are removed from the population. As mentioned in the introduction, the ergodic continuous-time BDI process is dual to the continuous-time birth-death-suppression process.\footnote{There exists a duality mapping processes with $\mu_0 > 0$ to those with $\mu_0 = 0$, see Sec. 8.2 in \cite{anderson}.} As a result, the phenomenology of the orthogonal polynomials for the BDI process is very similar to that of the birth-death-suppression process analyzed here. However, the duality of the two continuous-time processes does not naturally descend to a duality between the embedded jump chains of the two processes.

\subsection{The footprint}
The Pollazcek polynomials $W_n(x)$ characterize the jump chain dynamics of the birth-death-suppression process. \edit{Their explicit description and complete orthogonality relation \eqref{eq:firewalkcompleteortho} is derived and given in Appendix \ref{sec:firewalk}. 

The phenomenology of the polynomials changes between the phases of the process. In the sub-critical and critical phase $\beta \leq 1$, they have a continuous measure of orthogonality supported only on the closed interval $[-I_\beta,I_\beta]$ where $I_\beta = 2\sqrt{\beta}/(1+\beta)$. At criticality $\beta = 1$, they are equivalent to the Gegenbauer polynomials $C_n^\lambda(x)$ as
\begin{equation}
    W_n^c(x ; \gamma)=C_n^{1+\gamma / 2}(x),
\end{equation}
and are supported on the whole interval $[-1,1]$. The Gegenbauer polynomials fall in the Askey scheme; they are a special case of the Jacobi polynomials. 
Their measure of orthogonality is given by
\begin{equation}
    \label{eq:gegenMeasure}
    d\sigma(x)=\frac{\Gamma\left(\frac{\gamma}{2}+2\right)}{\sqrt{\pi} \Gamma\left(\frac{\gamma+3}{2}\right)}\left(1-x^2\right)^{(\gamma+1) / 2} d x.
\end{equation}
However, in the super-critical phase $\beta < 1$, the continuous support retracts back to $|x| \leq I_\beta$. In its place a discrete measure develops at infinitely many points $x_k$ with discrete measure weights $\Delta_k$. In the zero suppression limit $\gamma \to 0$, the weights of the discrete measure also disappear. }

An explicit expression for $W_n(x)$ is given in Eqn.~\eqref{eq:firewalkexpr}. The continuous measure $w(x)dx$ is given in Eqn.~\eqref{eq:firewalkMeasure} and the discrete measure weights $\Delta_k$ in Eqn.~\eqref{eq:TmeasWeights}. The elements of the $n$-step transition matrix $S(n)$ therefore have the spectral form
\begin{multline}
    S_{\ell,\ell'}(n)=\frac{1}{h_{\ell'}} \int_{-I_\beta}^{I_\beta} x^n W_{\ell}(x) W_{\ell'}(x) w(x)dx \\+ \frac{1}{h_{\ell'}} \sum_{k=0}^{\infty} (x_k)^n\Delta_k W_\ell(x_k) W_{\ell'}(x_k) \\
    + \frac{1}{h_{\ell'}} \sum_{k=0}^{\infty} (-x_k)^n\Delta_k W_\ell(-x_k) W_{\ell'}(-x_k).
\end{multline}
The constants $h_n$, which are analogous to the (reciprocal of the) constants $\pi_n$ are given by
\begin{equation}
    h_n = \frac{(\gamma+2)_n}{\beta^{n}n!}\cdot\frac{1+\beta + \gamma}{\gamma + (\beta+1)(n+1)};\quad h_0 = 1.
\end{equation}
With all this machinery one can now explicitly (numerically) compute arbitrary matrix elements of $S(n)$. In turn these matrix elements describe aspects of the footprint dynamics. 

To see this connection, consider first the analog of the lifetime distribution: the distribution of transition counts $n_T$ at absorption. In particular, let $R^n_N$ be the probability that, given $j(0) = N$, the process absorbs with exactly $n$ transitions \cite{karlinear,randomwalksKarlin}. This quantity, analogous to the absorption probability, is given by 
\begin{align}
    R^n_N &= q_0S_{N-1,0}(n-1) \\&= \frac{\gamma+1}{\beta+\gamma+1} \int_{-1}^1 x^{n-1} W_{N-1}(x) d\sigma(x).
\end{align}
The $R^n_N$ are directly related to the footprint $F$ at absorption. Take a process which begins with an initial population $N$ and absorbs after $n_T$ transitions. From the perspective of the jump chain, the entire process is just a sequence of births $j \to j+1$ and deaths $j \to j - 1$, in some particular order. If the total number of births is $n_B$ and the total number of deaths is $n_D$, the process, ending in absorption, must have $n_D - n_B = N$, in addition to the basic statement $n_B + n_D = n_T$. The footprint $F$ at absorption, by definition, is the number $n_B$ of births plus the initial size $N$. One therefore has the direct constraint
\begin{equation}
    n_T = 2F - N,
\end{equation}
relating the number of transitions $n_T$ of an absorbing process to the initial population $N$ and the footprint $F$ at absorption. Therefore, the probability that a process with $j(0) = N$ has footprint equal to $F$ at absorption is equivalent to the probability that the process absorbs in exactly $2F - N$ transitions, or just $R^{2F-N}_N$. 

\begin{figure}[!h]
    \centering
    \includegraphics[scale = 0.6]{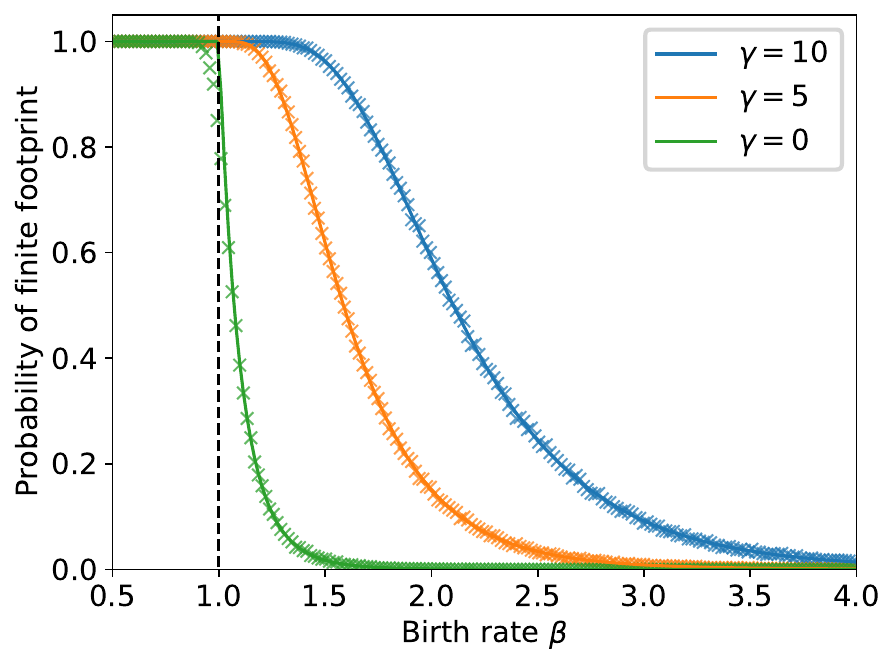}
    \caption{The probability of a finite footprint versus birth rate. Equivalently, the probability of eventual absorption, pictured for a fire with $N = 10$ and three values of suppression $\gamma$; \edit{simulated results (marked with  an `x') from an ensemble of $10^4$ processes}. The critical birth rate $\beta = 1$ is demarcated as a dashed line. The difference between the curves and $1$ is the probability measure `at infinity.' Suppression extends the flat portion of the curves on the left where almost all processes have finite footprint, which is true of all processes with $\beta < 1$.}
    \label{fig:measureInfinity}
\end{figure}

While all sub-critical processes eventually absorb, a large number of super-critical processes continue population growth without bound. To quantify this behavior, one can use the $R^n_N$ to compute the probability of absorbing with a finite number of transitions, or equivalently a finite footprint. This is simply the sum over $n$ of all the $R_N^n$, a quantity which is not necessarily equal to 1. It is of course equivalent to the asymptotic probability of absorption, as any process with a bounded footprint must eventually absorb. We therefore have in all phases of the process the equality 
\begin{align}
    \Pr\{F < \infty\ |\ j(0) = N\} &= \Pr\{j(\infty) = 0\ |\ j(0) = N\} \nonumber \\ q_0\int_{-1}^1 \frac{W_{N-1}(x)}{1-x} d\sigma(x) &= \frac{B(z_\infty;N,\gamma+1)}{B(N,\gamma+1)},
\end{align}
where $z_\infty = \operatorname{min}(1,1/\beta)$. This quantity is plotted in Fig. \ref{fig:measureInfinity} versus birth rate for a few values of the suppression parameter; it is clear that once the super-critical regime is entered, a significant amount of measure accumulates quickly at infinity.

The escape probability of Eqn.~\eqref{eq:burnProb}, that $F \geq J$ at absorption, parametrizes \edit{by $J$ a portion of the complement} to a finite process. This includes the measure at infinity. The probability $P_J(N)$ can thus be expressed in terms of a spectral integral as
\begin{multline}
    \label{eq:exactBurnProb}
    P_J(N) = 1 - \sum_{n = 0}^{2J - N} R^n_N =\\ 1-q_0 \int_{-1}^1 \left(\frac{1-x^{2J-N}}{1-x}\right)W_{N-1}(x)d\sigma(x).
\end{multline}
As a demonstration of this machinery, we briefly revisit the asymptotic distribution of footprints. With zero suppression this was given by the power law $P_J \sim J^{-1/2}$ near criticality. Here, consider the $N = 1$ escape probability \eqref{eq:exactBurnProb} at criticality $\beta = 1$. Using the Gegenbauer measure of Eqn.~\eqref{eq:gegenMeasure}, by direct integration one finds
\begin{multline}
    P_J(1) = \frac{\Gamma(J-1/2)\Gamma(1+\gamma/2)}{\sqrt{\pi}\Gamma(J + \gamma/2)}\\ \sim \frac{\Gamma(1+\frac\gamma2)}{\sqrt{\pi J^{1+\gamma}}} + \OO(J^{-3/2}),
\end{multline}
where the correct power law, with a modification due to the suppression factor, emerges in the $J \gg 1$ limit. For small initial sizes, nonzero suppression strongly affects the absorption probability and hence the distribution of footprints at large $J$. If $N$ is large compared to $\gamma$, then this effect will be attenuated. However, finding simple analytic results in this regime is not straightforward.

In the critical case one can explicitly compute the average number of transitions at absorption $\ex{n_T}$, which is by linearity related to the average footprint at absorption $\ex{F} = (N + \ex{n_T})/2$. For a single initial firelet $N = 1$ at critical conditions $\beta= 1$, the average number of transitions is given by the integral
\begin{align}
    \ex{n_T} &= \sum_{n=1}^\infty n R^n_1 \nonumber \\&= \frac{\gamma+1}{\gamma + 2}\int_{-1}^1\frac{\Gamma\left(\frac{\gamma}{2}+2\right)}{\sqrt{\pi} \Gamma\left(\frac{\gamma+3}{2}\right)}\cdot\frac{\left(1-x^2\right)^{(\gamma+1) / 2}}{(1-x)^2} d x \nonumber\\&= \frac{\gamma + 1}{\gamma -1};\quad \gamma > 1.
\end{align}
Interestingly, the average transition count (and hence average footprint) diverges for $\gamma \leq 1$. We showed earlier that the unsuppressed critical process had an infinite average lifetime. This reiterates that result (at zero suppression) but takes into account that for $\gamma > 1$ one has $\lambda_1/\mu_1  \leq 1/2$, which causes sufficiently many processes to absorb early that the transition and hence footprint distribution has finite mean. 

The transition count and thus footprint at absorption is not the only quantity over which the matrix $S(n)$ affords traction. The matrix element $S_{ij}(n)$ describes the probability of transitioning from state $i$ to state $j$ after $n$ transitions. This is equivalent to the probability of transitioning from state $i$ to state $j$ after increasing the footprint by $n_B = (n+j-i)/2$, i.e. after $n_B$ births $j \to j+1$. By combining these dynamics with the continuous-time transition matrix, it could be possible to make statements about the time-dependent distribution of footprints rather than only its asymptotic behavior. This is beyond the scope of the current work.

Finally, in a modeling scenario one may need to compute the escape probability for an ensemble of processes, possibly with varying footprints. Without perfect knowledge of the time-dependent footprint distribution, this must be done approximately. If a process begins with a definite population $N$ and threshold $J$,  after some finite time $t_0$ it evolves to give probabilities $p_n(t_0)$ for each possible population state $j(t_0) = n$.\footnote{Strictly speaking one should have $J\gg N \gg 1$ and $t_0$ small so that the probability of absorption is small and almost no processes have already saturated the bound $F \leq J$.} By the Markov property, one may regard this as an ensemble of new processes with different initial populations $N_i$, each with an ensemble probability $p_{N_i}(t_0)$. However, the footprint variable is cumulative. Therefore, the ensemble process with initial population $N_i$ should already have footprint $F_i \geq \max(N,N_i)$. This is an underestimate of the true footprint distribution. The escape probability at time $t_0$ is thus (under)estimated by
\begin{multline}
    P_J(t_0;N) \geq \sum_{n=1}^N P_{J-N}(n)p_n(t_0) \\+ \sum_{n = N+1}^J P_{J-n}(n)p_n(t_0),
\end{multline}
an approximation that we will make use of in the following section.

Having determined the relevant polynomials and their measures, computing essentially any probabilistic prediction of the continuous-time or discrete birth-death-suppression process is now a matter of arithmetic, or rather computing spectral integrals. This concludes the theoretical development of the solution of the process. \edit{In the final sections we show some sample applications and describe some future directions.}

\begin{figure*}[t]
\centering
\begin{minipage}{\textwidth}
   \includegraphics[width=\textwidth]{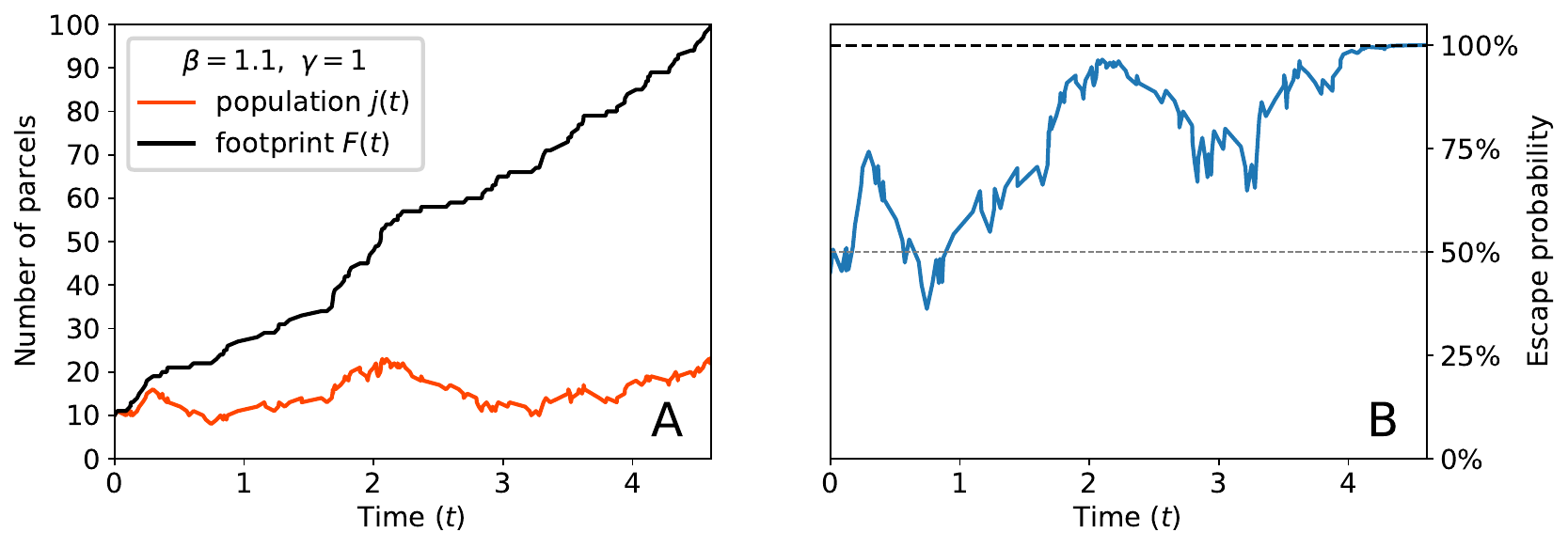}
   % \caption{testing}
\end{minipage}
\begin{minipage}{\textwidth}
   \includegraphics[width=\textwidth]{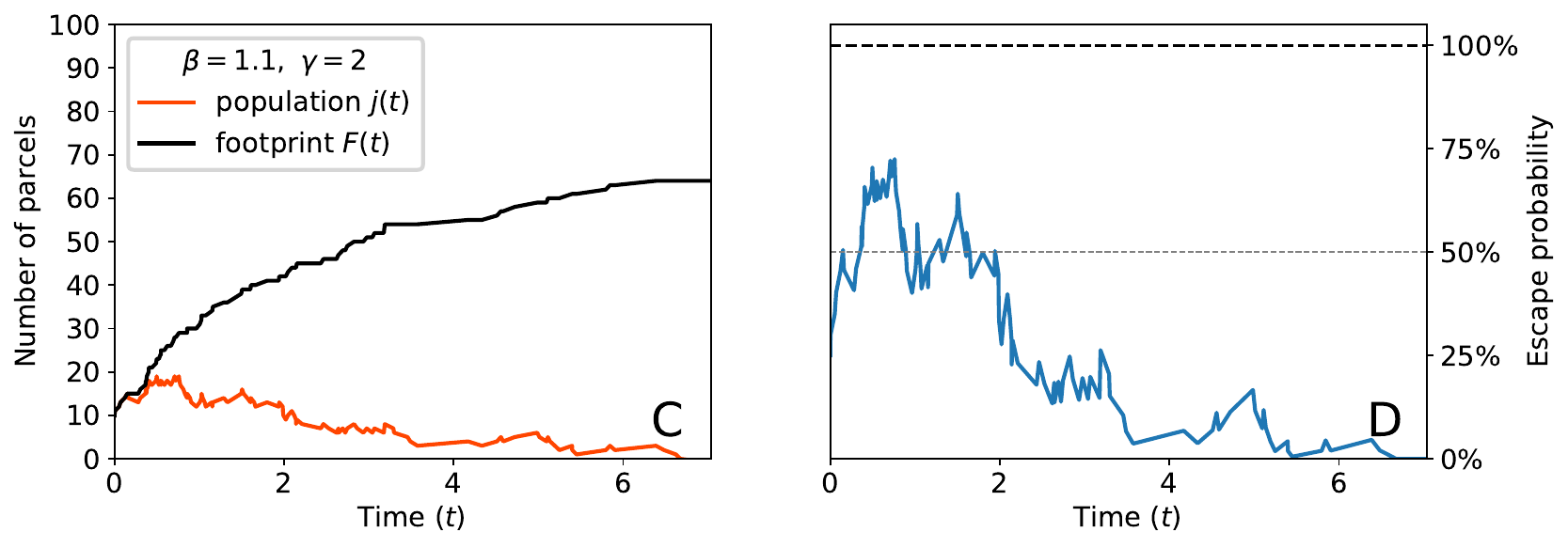}
   % \caption{Second image}
   \caption{Dynamic risk estimation over the lifetime of a fire event. Pictured are simulated birth-death-suppression processes (in A, C) and the escape probability $P_J(N)$ recomputed after each transition, serving as a real-time risk metric (in B, D). Both processes have $\beta = 1.1,\ N = 10,\ J = 100$. The process which escapes (A) has $\gamma = 1$, while the process which absorbs (C) has $\gamma = 2$.}
   \label{fig:escapePlot}
   % \label{fig:image2}
\end{minipage}
\end{figure*}

\edit{With numerical traction over the continuous-time and discrete-time processes, one can begin to ask questions about bounded, finite-lifetime birth-death-suppression processes. Developing the theory of the cumulative population or `footprint' is crucial to being able to bound these processes to some finite threshold. These tools allow one to predict the expected time until extinguishment, the distribution of footprints, and the effects of differing levels of suppression all in a dynamic framework. Having developed the theory of the population $j(t)$ and the footprint $F(t)$, we turn to applying these concepts with the wildfire interpretation at front of mind. Within the scope of the probabilistic predictions developed here the model exhibits clear trade-offs and optimal policy choices---tensions and strategies that are also reflected in decision processes related to actual fire events.}
\edit{\section{Dynamic risk and optimization}}

\label{sec:scenarios}
 
\subsubsection*{Dynamic risk estimation}
As a first application of the technology developed in this work, we show how the  `escape' probability of 
Eqn.~\eqref{eq:exactBurnProb} can be dynamically computed alongside the stochastic trajectory of a birth-death process. In Fig. \ref{fig:escapePlot}, plots B, D show the escape probability recomputed after each transition of the stochastic trajectories in A, C, which show simulated birth-death-suppression processes. The escape probability $P_J(N)$ quantifies the chance that the fire expands to have footprint above some bound $J$. \edit{In the wildfire interpretation, $J$ represents the amount of burnable substrate or effective distance to built infrastructure. In an epidemic interpretation, the bound $J$ may represent the severity at which a certain containment protocol must be enacted, for example.}

To be precise, the right-hand plots in Fig. \ref{fig:escapePlot} do the following: one first computes the escape probability $P_J(N)$ for a process of initial size $N$ with a cumulative threshold $J$. After each transition, one updates $J \to J - j(t) + F(t)$ and $N \to j(t)$ and recomputes the escape probability. \edit{This takes advantage of the Markov property of the population dynamics and accounts for the previous growth of the footprint.} As the outcome of the fire (being extinguished or escaping) becomes almost certain at later times, the predicted risk curves approach either zero or one. One can see that small changes in the actively burning population can have a large effect on the escape probability overall. 

This type of calculation represents a real-time risk metric which takes into account the burnable substrate $J$, the cumulative footprint $F(t)$, and the size of the actively burning population $j(t)$. It is updated immediately as conditions change and new information is available. This is a simple example of using the birth-death-suppression model for real-time risk forecasting. By modulating the suppression rate $\gamma$ used to compute the escape probability, one can dynamically compare the effects of policy choices by observing the effect on the escape probability. 

In future work, we aim to describe in greater detail the trade-offs and dynamic optimization problems associated with a changing suppression rate over the course of a process. This elevates the dynamics from a homogeneous Markov chain to a Markov decision process over which reinforcement learning could identify optimal policies. Of course, such learning requires a cost (or reward) function. Real suppression is not free: there is a cost, and the cost could even be the removal of suppression from another, simultaneous event. The presence of a cost or reward will create optimal strategies. The nature of the cost function and hence of the optimal strategies is an important consideration in the application of this work. 
\begin{figure}[!h]
    \centering
    \begin{minipage}{0.45\textwidth}
        \centering
        \includegraphics[scale = 0.9,width=\textwidth]{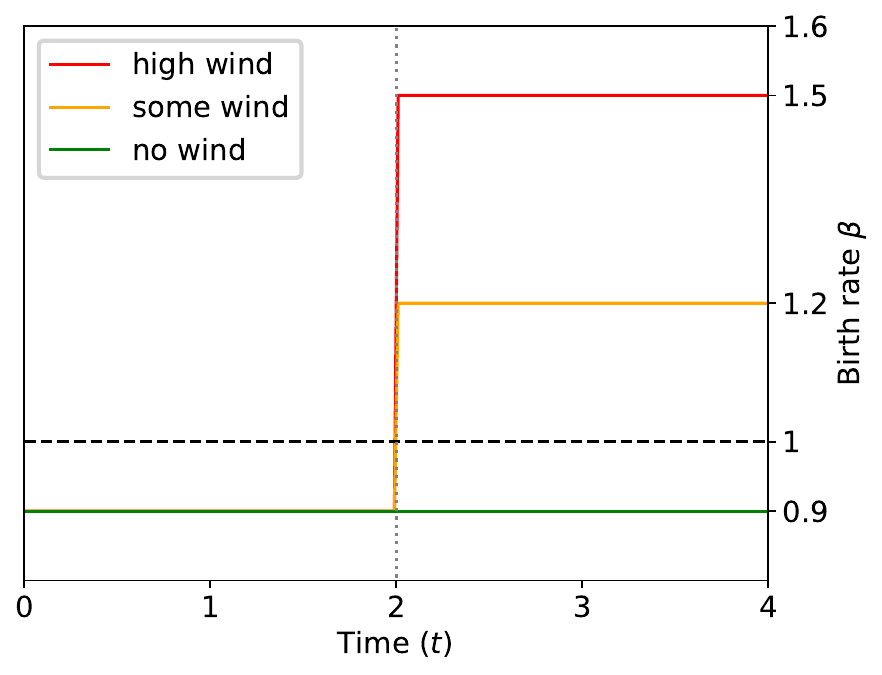} % second figure itself
        \caption{The onset of dangerous conditions modeled as a step-function birth rate. We begin with $\beta = 0.9$ until $t = 2$, at which point the birth rate increases to $\beta' > 1$, reflecting e.g. a wind event increasing the spread rate of a fire. Two different increases are shown. }
        \label{fig:steps}
    \end{minipage}
    \begin{minipage}{0.45\textwidth}
        \centering        
        \includegraphics[width=\textwidth]{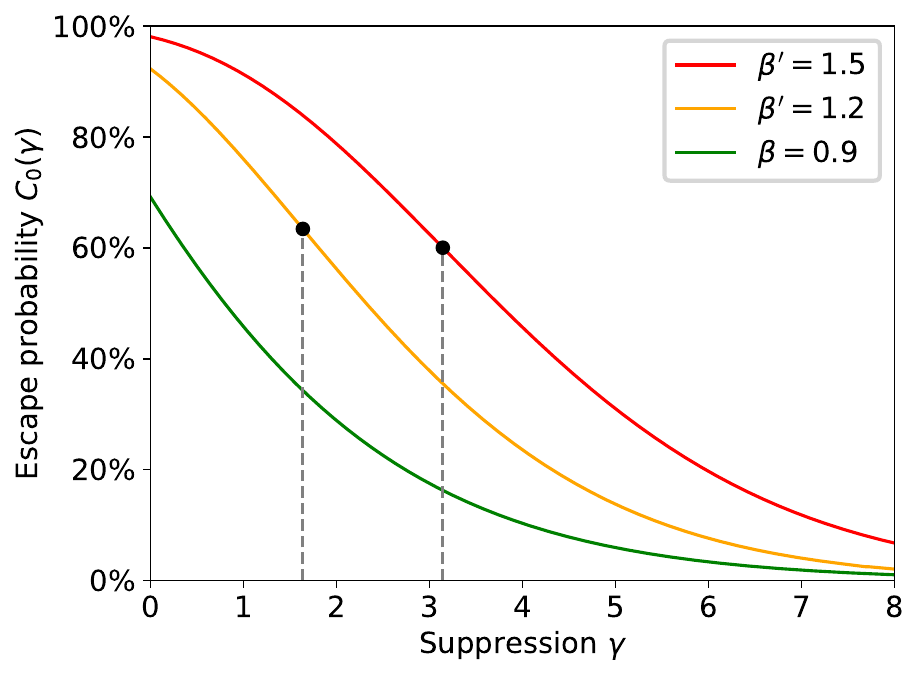} % first figure itself
        \caption{Escape probability versus suppression rate. Here we have an initial population $N = 20$ and threshold $J = 100$. The green curve has a constant birth rate $\beta = 0.9$ while the orange, red curves transition to a higher birth rate $\beta' > 1$ at time $t = 2$ as in Fig. \ref{fig:steps}. The point of diminishing returns is shown.}
        \label{fig:convex}
    \end{minipage}
\end{figure}

\subsubsection*{Optimal suppression and adverse conditions}
As a crude example of policy optimization, consider a policy which consists of choosing a constant suppression rate $\gamma$ for the entire lifetime of a fire. A primitive measure of the severity of a fire event is given by the escape probability $P_J(N)$:
\begin{equation}
	C_0(\gamma) = P_J(N;\gamma),
\end{equation}
which of course depends on the applied suppression $\gamma$. To begin, let the process be in a sub-critical phase with a constant birth rate; for definiteness we take $\beta= 0.9$ with $N = 20$ initial firelets in a burnable area of size $J = 100$. Physically, this represents a modestly sized ignition in mild fire conditions, where it will eventually self-extinguish. Of course, the greater the suppression, the lower the escape probability. In this case, applying more suppression always yields diminishing returns; the `cost' function $C_0(\gamma)$ is convex. This is the green curve in Fig. \ref{fig:convex}. 
\begin{figure}[b]
    \centering
    \includegraphics[scale = 0.6]{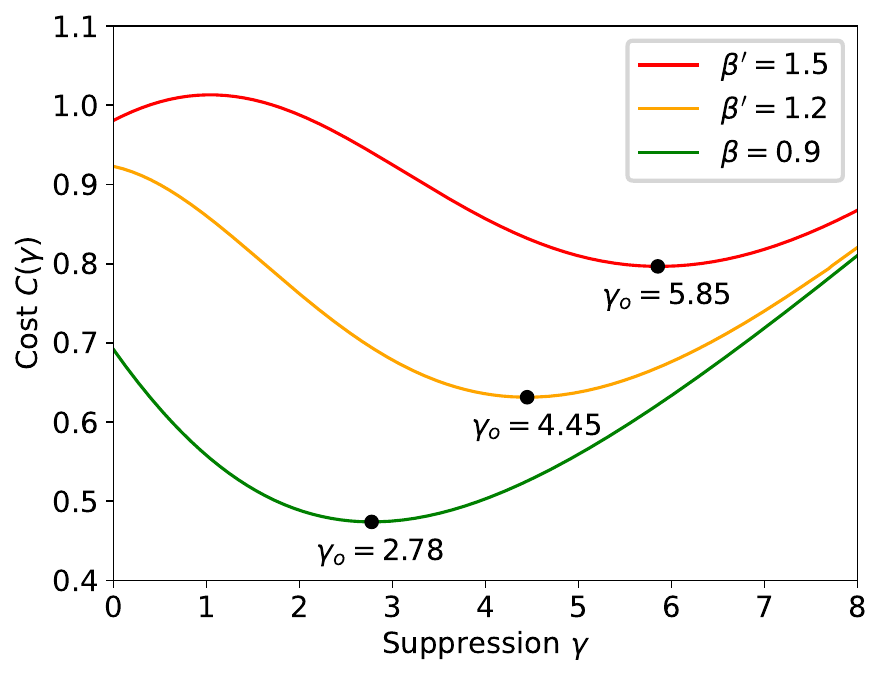}
    
    \caption{Optimal suppression policies coincide with minima of the cost function. The cost is computed as in Eqn.~\eqref{eq:linearcost} with $\gamma_\star = 10$ using the escape probability curves of Fig. \ref{fig:convex}. The optimal (constant) suppression strategy corresponds to the minimum $\gamma_o$ in the cost versus suppression curve. The red curve also displays a local maximum. }
    \label{fig:optimal}
\end{figure}

The situation becomes more interesting if one considers an onset of dangerous fire conditions, i.e. if at some time $t$ the birth rate increases into a super-critical phase $\beta \to \beta' > 1$, as depicted in Fig. \ref{fig:steps}. This models the impact of a wind event on the spread of the fire. After this increase, one recomputes the escape probability, now with the higher birth rate $\beta'$. This is done in ensemble, using the expected distribution of active populations, as described at the end of Section \ref{sec:predfootprint}. 

Recomputing the escape probability in the face of a wind event yields new cost curves which are pictured in Fig. \ref{fig:convex} alongside the `no wind' curve. With changing conditions, the escape probability exhibits an inflection point, also marked in Fig. \ref{fig:convex}. The inflection point is the point of diminishing returns, which is now nonzero. This reflects the intuitive idea that early, pre-wind event suppression may pay off disproportionately in the face of dangerous conditions. By decreasing the active size of the fire early on, one decreases the impact that an increased birth rate has on the overall severity of the fire event. 

With no cost to suppression, the obvious (and unrealistic) optimal strategy is to apply as much suppression as possible. A slightly more realistic cost function is obtained by adding a (linear) cost profile to suppression. If the maximum available suppression is $\gamma_\star$, a nicely scaled cost function is 
\begin{equation}
    \label{eq:linearcost}
    C(\gamma) = P_J(N;\gamma)  + \gamma/\gamma_\star.
\end{equation}
The addition of a cost to suppression immediately creates local minima and hence optimal (constant) suppression strategies, shown in Fig. \ref{fig:optimal}. Beyond the local minima, the red curve (corresponding to an onset of high wind conditions) actually displays a local maximum: the cost of mild suppression $\gamma \approx 1,2$ is greater than the cost of zero suppression. This happens because the effect of such low suppression is minimal on the escape probability, but does incur a cost. The cost is lower if nothing is done. Realistically, the choice of whether to apply suppression to a given event will include the choice to not apply it elsewhere. By identifying local maxima in cost functions, one can determine when removing suppression from one event in favor of another may be the best policy choice. 

The cost function and constant suppression strategies outlined here are extremely coarse, and essentially the simplest possible choices: constant suppression, a linear cost profile, etc. Lots of generalizations exist. The suppression rate may become time dependent and react to conditions. A wind event as modeled in Fig. \ref{fig:steps} could be forecasted for only a finite time, changing the eventual outlook for the fire. The cost function can be modified with a non-linear profile with respect to suppression or made risk-averse by considering the variance in the population and penalizing high-variance scenarios.

This is all at the level of analyzing strategies for a single, isolated fire event. One may furthermore extend the analysis to a set of multiple birth-death processes, modeling the management of resources across a multi-fire event. This allows resource allocation between events with dynamic risk predictions, including the time-delay associated to transferring resources between events. A multi-event model of this type would provide a quantitative and dynamically updating decision support tool to agencies which manage a set of fires of varying sizes, conditions, and risk levels. 
\pagebreak
\section{Conclusion}
In this work we review and extend the theoretical results of Karlin, MacGregor \cite{karlinear}, and Kendall \cite{kendallGeneral} on the birth-death-suppression Markov process. This process is closely related to many similar linear birth-death processes which were solved by various authors in relatively quick succession many years ago. For historical or accidental reasons, the birth-death-suppression process considered here was less studied in the present parametrization and went without the present wildfire interpretation.  

We derive explicit expressions for the transition matrices for both the continuous-time and discrete-time birth-death-suppression processes. To obtain the latter, we applied a resolvent-based complex-analytic approach. This introduced a set of polynomials equivalent to the Pollazcek orthogonal polynomials; we describe their computation, asymptotics and measure of orthogonality in terms of the birth and suppression parameters. These polynomials are used to express physical predictions about the process as spectral integrals which can be evaluated numerically and in some cases exactly. In particular, the polynomials allow us to determine the asymptotic distribution of footprints (the cumulative population) as functions of the global parameters of the process. This had previously only been done in the zero suppression process \cite{kendallGeneral}. 

The analysis is in general closely related to and inspired by work on the birth-death-immigration (BDI) process, the process studied by Askey and Ismail \cite{ismail}. Whereas the suppression rate appears as an additive increase in the aggregate death rate, in the BDI process the immigration rate appears as an additive increase to the aggregate birth rate. One could interpret an immigration rate as characterizing `spotting' or new ignition in the wildfire context, and could theoretically solve the process with both `immigration' and suppression. 

The BDI process has also been extended to capture the dynamics of immigration between multiple population centers \cite{anderson,Tavar1987TheBP}. This same type of generalization could be made to the birth-death-suppression process, with the fire interpretation of describing the management of multiple simultaneous fire events. Already in previous work this type of resource allocation optimization was carried out for a two-fire event \cite{petrovic}; the $n$-fold generalization should be just a matter of bookkeeping. Another important aspect of the tools developed here is their Markovian property, which makes them dynamic. Not only can one describe the probabilities of certain outcomes of the process parametrically, but these probabilities can be continuously updated as new information about a given real-life stochastic process is obtained. The simplicity of the model means that these updates and re-calculations can occur immediately and with minimal computational cost. 

There exist a wide array of questions related to resource allocation, dynamic policy optimization, and the analysis of trade-offs that can now be quantitatively addressed. We have focused on the wildfire interpretation, but we stress that the model is general enough to describe a range of phenomena. The original motivation for the introduction of a cumulative population or `footprint,' as we have referred to it, was in the context of mathematical biology \cite{kendall1949stochastic}. Indeed, one can draw an analogy between the wildfire interpretation and, for example, the susceptible-infectious-recovered (SIR) model of disease dynamics. The firelets $j(t)$ are the `infectious' states, the burnable substrate $J - F(t)$ is the `susceptible' population, and the footprint $F$ represents the `recovered' group. One could apply the same formulas used in this work in the context of wildfire to ask questions about SIR dynamics: what is the probability of infecting all susceptible individuals in a finite group? What is the optimal level of treatment (suppression) to maximize the probability of eradication (absorption at $j = 0$) with some cost profile to the treatment? There exist a wide range of applications of which this type of stochastic model affords quantitative analysis.

The focus of the work has been to develop the theory of the birth-death-suppression process, but the motivation and guiding context for the authors remains wildfire modeling. As a model of fire, our process is temporal in nature: its main credence comes from its universality as a model of any continuously fluctuating population, not from its accurate representation of the spatial dynamics of fire. There of course exist far more accurate and widely used high-fidelity physics-based fire models \cite{finney2006overview,finney1998farsite,linn2020quic,richards1995general,pastor2003mathematical,Linn2002StudyingWB,peterson2009using,papadopoulos2011comparative} \edit{which far outperform the present model in the simulation of physical fire.} Some of the analysis herein could reasonably be augmented by such high-fidelity models, and they could be applied to gain a quantitative understanding of the relationship between the Markov process parameters and the measurable data of real fires, fuels, topography, and weather conditions. The practical reality of firefighting is far removed from the decision analysis presented herein. However, it is our hope that the dynamic risk forecasting and quantitative control we have developed here lead to tools which support streamlined, quantitatively informed decisions that help us live in the modern climactic reality. 

\acknowledgments
We thank W. Van Assche for his correspondence and expertise on orthogonal polynomials. This work was supported in part by the Office of Secretary of Defense Strategic Environmental Research and Development Program (SERDP) Project \#RC21-1233.

\newpage
\appendix 

\section{The zero suppression process}
\label{sec:appgfunc}
The zero suppression process is the linear birth-death process with aggregate birth and death rates
\begin{equation}
\label{eq:simplelinear}
    \lambda_j = \beta j; \quad \mu_j = \delta j .
\end{equation}
This process has been solved by a variety of methods \cite{anderson}. For the solution as the limit of a finite system, see \cite{spectral}. Here, we focus on the generating function approach of \cite{kendallGeneral}, and show how results cited in Section \ref{sec:suppressionfreeprocess} may be derived such that the work is self-contained. 

The present focus is to determine the population probabilities $p_n(t) \equiv \pr{j(t) = n}$ given definite initial configurations $p_n(0) = \delta_{n,N}$ (equivalently $j(0) = N$). The dynamics are governed by 
\begin{multline}
    \label{eq:appdiffdiff}
    \frac{d}{d t} p_n(t)=\lambda_{n-1} p_{n-1}(t)+\mu_{n+1} p_{n+1}(t)\\-\left(\lambda_n+\mu_n\right) p_n(t).
\end{multline}
This differential-difference equation may be solved by employing a generating function; define
\begin{equation}
    G(t,z) = \sum_{n=0}^\infty p_{n}(t)z^n
\end{equation}
in terms of the population probabilities, where $p_0(t) \equiv p_A(t)$ is the absorption probability. By ansatz this generating function is analytic as $z\to 0$. For an initial population $j(0) = N$, the generating function satisfies the boundary condition $G(0,z) = z^N$. The function $G(t,z)$ allows computation of many quantities of interest:
\begin{align}
\left\langle j^k(t)\right\rangle&=\lim _{z \rightarrow 1}\left(z \frac{\partial}{\partial z}\right)^k G(t,z), \\
p_n(t)&=\frac{1}{n !} \lim_{z\to0}\frac{\partial^n}{\partial z^n}G(t, z).
\end{align}
Equation \eqref{eq:appdiffdiff} defines relations between the series coefficients of $G(t,z)$. For the moment, keep $\gamma \neq 0$ and use the birth-death rates $\lambda_n = \beta n,\ \mu_n = n + \gamma$. Taking the time derivative of $G(t,z)$ and applying the recurrence relation descends to a partial differential equation for the generating function:
\begin{multline}
    \frac{\partial G}{\partial t} = (\beta z-1)(z-1)\frac{\partial G}{\partial z} \\+ \gamma\left(\frac{1-z}{z}\right)(G(t,z) - p_A(t)).
    \label{eq:genfuncdiffeq}
\end{multline}
This is a linear, non-homogeneous first order equation; in theory, it may be solved by the method of characteristics. The term proportional to $\gamma$ appears to display a singularity as $z \to 0$. This non-analytic divergence is removed by the subtraction of the absorption probability $p_A(t)$ from $G(t,z)$. However, without an explicit form of the absorption probability $p_A(t)$, one cannot solve Eqn.~\eqref{eq:genfuncdiffeq} explicitly. In the presence of nonzero suppression different techniques are needed. However, setting $\gamma = 0$, the simple linear process of \eqref{eq:simplelinear} gives the homogeneous PDE 
\begin{equation}
    \frac{\partial G}{\partial t}=(\beta z-1)(z-1) \frac{\partial G}{\partial z}.
\end{equation}
For an initial population $j(0) = N$, and hence $G(0,z) = z^N$, this equation is solved by
\begin{equation}
    \label{eq:genfunc0}
    G(t,z) = \left[\frac{(z-1) e^{(\beta -1) t}-\beta z+1}{\beta(z-1) e^{(\beta - 1) t}-\beta z+1}\right]^N.
\end{equation}
Taking the limit as $\beta \to 1$ gives the critical generating function
\begin{equation}
    \label{eq:critgenfunc0}
    G^c(t,z) = \left[\frac{z + t(1-z)}{1 + t(1-z)}\right]^N.
\end{equation}
\subsection{Population statistics}
The average population is given by 
\begin{equation}
\label{eq:zerogamAvgPop}
    \langle j(t)\rangle = \lim_{z\to1}\partial_zG(z,t) = Ne^{(\beta-1)t},
\end{equation}
which defines the super-critical $\beta > 1$ and sub-critical $\beta < 1$ phases of exponential growth and extinction respectively. In the critical case $\beta = 1$, the mean population is constant in time. The population variance is given by 
\begin{align}
\label{eq:zerogamPopVar}
    \Delta j^2(t) &= \lim_{z\to1}\left(\partial_z^2G(z,t) + \partial_z G(z,t) - [\partial_zG(z,t)]^2\right) \nonumber\\&= \frac{\beta + 1}{\beta -1}Ne^{(\beta-1)t}(e^{(\beta-1)t}-1).
\end{align}
For $\beta > 1$, the variance grows exponentially at late times, and is monotonically increasing. The critical variance is linear in time $\Delta j_c^2 = 2Nt$ and similarly increases without bound. However, the sub-critical variance peaks at some finite time.

The time-dependent probability of absorption $p_A(t)$ is trivial to calculate by just plugging in $z = 0$ to both generating functions:
\begin{gather}
\label{eq:zerogamAbsorb}
    p_A(t) = \left(\frac{1-e^{(\beta-1)  t}}{1-\beta  e^{(\beta-1)  t}}\right)^N, \\ \lim_{\beta \to 1}p_A(t) = \left(\frac{t}{t+1}\right)^N.
\end{gather}
By taking the limit $t\to\infty$, one obtains the asymptotic absorption probability:
\begin{equation}
\label{eq:zerogamAsymAbsorb}
    p_A(\infty) = \lim_{t\to\infty}G_0(0,t) = \begin{cases}
        \beta^{-N}, & \beta > 1\\
        1, & \beta \leq 1,
    \end{cases}
\end{equation}
which states that absorption (extinguishing) is all but certain except in the super-critical case. Finally, consider the distribution of lifetimes until absorption $\propto p_1(T)dT$. With an initial population to $j(0) = N$, the normalized distribution of lifetimes $T$ is 
\begin{multline}
\label{eq:zerogamlifetimeMeasure}
    d\sigma(T) = N\max(\beta,1)^N(\beta -1)^2e^{(\beta-1)T}\\\times\frac{(1-e^{(\beta-1)T})^{N-1}}{(1-\beta e^{(\beta-1)T})^{N+1}}
    dT
\end{multline}
for $\beta \neq 1$, reducing in the critical case to
\begin{equation}
    d\sigma(T) = N \frac{T^{N-1}}{(1+T)^{N+1}}dT.
\end{equation}
First, let $N = 1$ to analyze the small fire limit. It is convenient to make the change of variables $s = e^{(\beta - 1)T}$; note that the limit $\lim_{T\to\infty}s$ depends on the value of $\beta$. In particular, as $T \to \infty$ one has $s \to 0$ in a sub-critical phase ($\beta < 1$) and $s \to \infty$ in a super-critical phase ($\beta > 1$). The single initial population measure for lifetimes becomes 
\begin{equation}
    d\sigma(T) = \max(1,\beta)|\beta-1|\frac{1}{(1-\beta s)^2}ds.
\end{equation}
From this one can write down expressions for the mean and median lifetimes of the fire. Note that the median lifetime $T_m$ is the time such that $p_A(\infty) = 2p_A(T_m)$: the time after which half of all processes in ensemble will have absorbed.  

The average lifetime is infinite in the critical case, and otherwise is given by
\begin{equation}
\label{eq:zerogamAvgLife}
    \ex{T} = \begin{cases}
        -\frac{1}{\beta}\log(1-\beta), & \beta < 1;\\
        -\log(1-1/\beta), & \beta > 1.
    \end{cases}
\end{equation}
Recall that this represents the average lifetime of fires which end in absorption; it does not count the fires which formally diverge in population size. One can also explicitly compute the median lifetime for a fire with $j(0) = 1$:
\begin{equation}
\label{eq:zerogamMedianLife}
    T_m = \begin{cases}
        \frac{1}{1-\beta}\log(2-\beta), & \beta < 1;\\
        \frac{1}{\beta -1}\log(2-1/\beta), & \beta > 1,
    \end{cases}
\end{equation}
where, in the above, the limit as $\beta \to 1$ exists and is equal to $1$. In fact, one can solve $p_A(\infty) = 2p_A(T_m)$ exactly to obtain the median lifetime. In the limit of large initial size $N \gg 1$, the leading order contributions are
\begin{equation*}
    \label{eq:zerogamlargeNmedian}
    T_m \approx_{N\gg1} \begin{cases}
        \frac{1}{1-\beta}\log N - \frac{1}{1-\beta}\log\left(\frac{1-\beta}{\log2}\right), & \beta < 1;\\
        \frac{N}{\log2} - 1/2, & \beta = 1;\\
        \frac{1}{\beta-1}\log N + \frac{1}{\beta -1}\log\left(\frac{1-1/\beta}{\log2}\right), & \beta > 1,
    \end{cases}
\end{equation*}
discarding terms of order $\OO(N^{-1})$. In the large population limit the difference between sub- and super-critical phases is washed out, and the lifetimes increase only logarithmically in the size of the initial population. The next-to-leading order terms differ between the sub- and super-critical phases and become relevant near criticality, where the dependence on the initial population is much stronger. 

This completes discussion of the dynamics of the suppression-free population. In Section \ref{sec:suppressionfreeprocess}, these results are summarized along with some visualizations. Here, we move on to constructing the suppression-free dynamics of the cumulative population or footprint $F(t)$. 

\subsection{Footprint statistics}
The generating function approach may also be employed to determine at least the low moments of the footprint distribution. Recall that the footprint $F(t)$ shares all the births of the population $j(t)$ but none of the deaths. It is natural to define the joint probability matrix $\cal P$ with elements $\mathcal{P}_{j,F}(t)$ which are the probability of having population $j$ and footprint $F$ at time $t$. These matrix elements satisfy a differential-difference relation as before:
\begin{multline}
    \label{eq:jointProbdiffdiff}
    \frac{d}{dt}\mathcal{P}_{j, F}(t)=\lambda_{j-1} \mathcal{P}_{j-1, F-1}(t)+\mu_{j+1} \mathcal{P}_{j+1, F}(t)\\-\left(\lambda_j+\mu_j\right) \mathcal{P}_{j, F}(t).
\end{multline}
To solve this by generating function, define a function of three variables
\begin{equation}
    \label{eq:jointGenFunc}
    \Psi(t, z, w)=\sum_{j, F \geq 0} \mathcal{P}_{j, F}(t) z^j w^F.
\end{equation}
Restricting to the case $\gamma = 0$, this joint generating function solves the differential equation
\begin{equation}
\label{eq:jointgenfunceq}
    \frac{\partial \Psi}{\partial t}=\left(\beta z^2 w-z(\beta+1)+1\right) \frac{\partial \Psi}{\partial z},
\end{equation}
where, for an initial population $j(0) = F(0) = N$, there is the initial condition $\Psi(0,z,w) = (zw)^N$. This equation has an exact solution, but it is difficult to determine closed form expressions for its expansion coefficients \cite{kendallGeneral}. Instead, one can extract a few important predictions about the footprint by solving for the cumulant generating function $K(t,u,v)$:
\begin{equation}
    e^{K(t,u,v)} = \Psi(t,e^u,e^v).
\end{equation}
Substituting this into Eqn.~\eqref{eq:jointgenfunceq} shows that $K$ satisfies the equation
\begin{equation}
    \label{eq:jointcumgenfunceq}
    \frac{\partial K}{\partial t}=\left[\beta\left(e^{u+v}-1\right)-\left(1-e^{-u}\right)\right] \frac{\partial K}{\partial u}
\end{equation}
along with the initial condition $K(0,u,v) = N(u+v)$. The cumulant generating function, by definition, has series coefficients given by the various cumulants of the joint distribution of $j,F$. To quadratic order these are
\begin{multline}
\label{eq:cumulants}
    K  = u \ex{j(t)} + v\ex{F(t)} + \frac{1}{2}u^2 \Delta j^2(t) \\+ uv \operatorname{Cov}(j,F) + \frac{1}{2}v^2\Delta F^2(t)  + \cdots
\end{multline}
 The strategy is then to solve Eqn.~\eqref{eq:jointcumgenfunceq} order-by-order to determine the low cumulants listed above. One finds
\begin{gather}
    \ex{F(t)} = \frac{N}{\beta -1}(\beta e^{(\beta-1)t}-1), \\ \lim_{\beta\to1}\ex{F(t)} = N(t+1).
\end{gather}
Only in the sub-critical case does the asymptotic average footprint have a finite value. One may obtain explicit forms for the other cumulants, but they are not especially enlightening.

Finally, consider the asymptotic distribution of footprints. To find this, one should solve Eqn.~\eqref{eq:jointgenfunceq} and then take the limit $t \to \infty$; the solution is known and the limit may be taken explicitly. Setting $z = 1$, $N = 1$ and entering the asymptotic time regime, one finds
\begin{align}
\label{eq:jointgenfuncasymform}
    \Psi(\infty,1,w) &= \sum_{F=1}^\infty \pr{F(\infty) = F}w^F \nonumber\\&= \frac{1}{2\beta}(1+ \beta - \sqrt{(\beta+1)^2-4\beta w}).
\end{align}
For an absorbing process with $\beta \leq 1$, one can explicitly evaluate the series coefficients of the above. In the limit of large footprint $F \gg 1$, they are approximately given by
\begin{equation}
    \pr{F(\infty) = F} \sim \left(\frac{4\beta}{(1+\beta)^2}\right)^FF^{-3/2}.
\end{equation}
The cumulative distribution of footprints, which we call the `escape probability' $P(F\geq J)$, is then given by the integral of the above. The quantity $P(F\geq J)$ expresses the probability that the footprint of a given process will end up larger than some value $J$. This is the distribution which is empirically expected to follow a power law $\sim J^{-\alpha}$ with exponent close to $\alpha \approx 1/2$. 

In the eventually absorbing regime $\beta \leq 1$, write $\beta = 1-a$ and consider small, positive $a$, expanding away from criticality into the sub-critical regime. The escape probability can then be expanded to leading order in $a$ as
\begin{equation}
    \label{eq:zerogamburnprob}
    P(F(\infty)\geq J) \approx \frac{J^{-1/2}}{\sqrt{\pi}} + \frac{a}{2}\left(\frac{J^{-1/2}}{\sqrt{\pi}}-1\right),
\end{equation}
which reproduces the desired power law scaling exactly in the critical $\beta = 1$ phase, and approximately for $\beta \lesssim 1$. This characterizes the asymptotic distribution of footprints, but not the timescales associated with saturation. Since the solution to Eqn.~\eqref{eq:jointgenfunceq} is known, one could in principle compute the exact coefficients $P_{j,F}(t)$. This is beyond the scope of this work, and would likely be unreasonably difficult.
\section{Explicit transition matrices}
\label{sec:appPolys}
In this section we review the method of Karlin and MacGregor for solving the population dynamics in the birth-death-suppression process \cite{karlinear,karlinClass}, starting by defining some of the families of orthogonal polynomials associated to each phase of the process. These polynomials fall into the Askey scheme classification of \cite{askey}. The first are the Meixner polynomials $\phi_n(x)$, which are defined in terms of the Gauss hypergeometric function: 
\begin{equation}
    \label{eq:meixner}
    \phi_n(x) \equiv \phi_n(x;b,g) = {}_2F_1(-n,-x;b;1-1/g).
\end{equation}
Here, $b > 0$ and $0 < g < 1$. Recall that
\begin{gather}
    { }_2 F_1(a, b ; c ; z)=\sum_{n=0}^{\infty} \frac{(a)_n(b)_n}{(c)_n} \frac{z^n}{n !};\\ (a)_n = \frac{\Gamma(a+n)}{\Gamma(a)},
\end{gather}
so that when $a$ is a negative integer, the Pochhammer symbol $(a)_n$ vanishes at some order and the function ${}_2F_1$ is a polynomial in $z$. These polynomials are orthogonal with respect to a discrete measure with atoms at $k = 0,1,2,\ldots$ and inner product
\begin{equation}
    \label{eq:meixnerinner}
    \sum_{k=0}^\infty \phi_n(k)\phi_m(k)\rho_k = \delta_{nm}\frac{n!}{(b)_n g^n},
\end{equation}
where the spectral measure $\rho_k$ is 
\begin{equation}
    \label{eq:meixnerspecmeas}
    \rho_k = (1-g)^b\frac{(b)_k}{k!}g^k.
\end{equation}
Most importantly, the polynomials $\phi_n(x)$ satisfy the following recurrence relation: 
\begin{multline}
    \label{eq:meixnerrec}
    -x \frac{(1-g)}{g} \phi_n(x)=\frac{n}{g} \phi_{n-1}\\-\left(n+\frac{n}{g}+b\right) \phi_n+(n+b) \phi_{n+1}.
\end{multline}
The other set of polynomials needed are the associated Laguerre polynomials, defined as 
\begin{align}
    \label{eq:laguerredef}
    L_n^\alpha(x)&=\frac{e^x x^{-\alpha}}{n !} \frac{d^n}{d x^n}\left(e^{-x} x^{n+\alpha}\right) \\&= \sum_{m=0}^n(-1)^m \frac{(n+\alpha) !}{(n-m) !(\alpha+m) ! m !} x^m.
\end{align}
Their spectral measure is continuous, leading to the orthogonality relation
\begin{equation}
    \label{eq:laguerreortho}
    \int_0^{\infty} e^{-x} x^\alpha L_n^\alpha(x) L_m^\alpha(x) d x=\frac{(n+\alpha) !}{n !} \delta_{m n}.
\end{equation}
These functions satisfy the three-term recurrence relation
\begin{multline}
    \label{eq:laguerrerec}
    -xL^\alpha_n(x) = (n+\alpha)L_{n-1}^\alpha(x) + (n+1)L^\alpha_n(x)
    \\- (2n + \alpha + 1)L_n^\alpha(x).
\end{multline}
It so happens that in general, the measures of orthogonality develop discrete parts when the process moves away from criticality. 
\subsection{Polynomials for the suppression process}
Here, we identify the set of orthogonal polynomials for each phase of the process and present their properly normalized measures of orthogonality, which should satisfy
\begin{equation}
    \int_0^\infty  Q_n(x)Q_m(x)d\sigma(x) = \frac{\delta_{nm}}{\pi_n}.
\end{equation}
With $\lambda_n = \beta (n+1)$ and $\mu_n = n + \gamma+1$, the constants $\pi_n$ are given by
\begin{equation}
    \pi_0 = 1;\quad \pi_n = \frac{\beta^{n}\cdot n!}{(\gamma+2)_{n}},
\end{equation}
where the index $n$ runs over $0,1,2,\ldots$, with $n = j-1$ where $j$ is the number of firelets. The polynomials $Q_n(x)$ satisfy the three-term recurrence relation
\begin{multline}
    -xQ_{n}(x) = (n+\gamma + 1)Q_{n-1}(x) + \beta (n+1)Q_{n+1}(x)\\ - (n + \beta n  + \beta + \gamma + 1)Q_{n}(x).
\end{multline}
\subsubsection*{Critical phase}
In the critical phase $\beta = 1$, the recursion reduces to
\begin{multline}
    -xQ_{n}(x) = (n+\gamma + 1)Q_{n-1}(x)  +(n+1)Q_{n}(x)\\- (2n + \gamma + 2)Q_{n}(x).
\end{multline}
Comparing to the Laguerre recurrence in Eqn.~\eqref{eq:laguerrerec}, one can easily identify $Q_n(x) = L^\alpha_n(x)$, where $\alpha = \gamma + 1$. Thus, for the critical process, the desired family of orthogonal polynomials is
\begin{equation}
    \label{eq:critpolys}
    Q_n(x) = L_{n}^{\gamma + 1}(x).
\end{equation}
The normalized measure of orthogonality is
\begin{equation}
    d\sigma(x) = \frac{1}{\Gamma(\gamma+2)}e^{-x}x^{\gamma+1}dx,
\end{equation}
which is continuous over the real half-line. 
\subsubsection*{Sub-critical phase}
Now consider the case $\beta < 1$. Define a modified form of the Meixner polynomials
\begin{equation}
    \xi_n(x) = \frac{(b)_n}{n!}\phi_n\left(\frac{x}{1-\beta} - b + 1;b,\beta\right),
\end{equation}
which satisfy a recurrence relation slightly modified from \eqref{eq:meixnerrec}, working out as
\begin{multline}
    -x\xi_n(x) = (n + b - 1)\xi_{n-1}(x) + \beta(n+1)\xi_{n+1}(x)\\- (\beta n + \beta + n + b - 1)\xi_n(x).
\end{multline}
This matches the recursion for the $Q_n(x)$ provided $b = \gamma + 2$. So, in the $\beta < 1$ case, the polynomials are
\begin{equation}
    Q_n(x) = \frac{(\gamma + 2)_{n}}{n!}\phi_{n}\left(\frac{x}{1-\beta} - \gamma - 1;\gamma + 2,\beta\right).
\end{equation}
Their normalized measure of orthogonality is discrete with support at $\rho_k \equiv(1-\beta)(k + \gamma + 1)$ for $k \in \mathbb{Z}_+$; it may be written
\begin{equation}
    d\sigma(x) = (1-\beta)^{\gamma+2}\sum_{k=0}^\infty\frac{(\gamma+2)_k}{k!}\beta^k\delta\left(x-\rho_k\right)dx.
\end{equation}
\subsubsection*{Super-critical case}
For $\beta > 1$, define an altered set of Meixner polynomials by
\begin{equation}
    T_n(x) = \frac{(b)_n}{n!\beta^n}\phi_n\left(\frac{x}{\beta -1} - 1;b,\frac{1}{\beta}\right),
\end{equation}
which satisfy the same recursion as the $\xi_n(x)$, again identifying $b = \gamma + 2$. The difference is the arguments of the $\phi_n$. The conclusion is that, for the super-critical $\beta > 1$ case, one has the orthogonal polynomials
\begin{equation}
    Q_n(x) = \frac{(\gamma + 2)_{n}}{n!\beta^{n}}\phi_{n}\left(\frac{x}{\beta -1} - 1;\gamma + 2,\frac{1}{\beta}\right).
\end{equation}
The normalized measure of orthogonality has atoms at $r_k \equiv (\beta -1)(k+1)$ for $k \in \mathbb{Z}_+$, and is given by
\begin{equation}
    d\sigma(x) = \left(\frac{\beta -1}{\beta}\right)^{\gamma+2}\sum_{k=0}^\infty \frac{(\gamma+2)_k}{\beta^kk!}\delta(x-r_k)dx.
\end{equation}
\subsection{Explicit transition matrices}
With explicit forms of orthogonal polynomials for the process $(\lambda_n,\mu_n) = (\beta (n+1),n + \gamma+1)$, all that is left is to apply the spectral formula for the transition matrix:
\begin{equation}
    P_{nm}(t)=\pi_m \int_0^{\infty} e^{-x t} Q_n(x) Q_m(x) d\sigma(x).
\end{equation}
We leave out the computation and simply state the results of \cite{karlinear}, correcting various errata and writing in our parametrization. All of the expressions given here are valid only for $0\leq k \leq \ell$; by using the transposition formula $P_{k\ell} = (\pi_\ell/\pi_k)P_{\ell k}$ one can easily generate all other elements. While these expressions are cumbersome, they can be easily evaluated numerically. 
\subsubsection*{Non-critical cases}
Recall the auxiliary variable $s(t) = \exp((\beta-1)t)$. When $\beta < 1$, we have $s(\infty) = 0$; when $\beta > 1$, instead one has $s(\infty) = \infty$. Define also the auxiliary function
\begin{equation}
    X(s) = \frac{(\beta s - 1)(s - \beta)}{\beta(s - 1)^2},
\end{equation}
which has the limits $X(0) = X(\infty) = 1$. This becomes the argument of the following hypergeometric function:
\begin{multline}
\label{eq:auxiliaryS}
    F_{k\ell}(t) = \frac{\Gamma(k + \ell + \gamma + 2)}{\Gamma(\ell+1)\Gamma(k+\gamma + 2)}\\\times{}_2F_1\left(-k,-\ell;-1-k-\ell-\gamma;X[s(t)]\right).
\end{multline}
Numerically, this expression is poorly scaled and troublesome: one should take logarithms for stability. The full transition matrix is the expression
\begin{multline}
\label{eq:fulltranistionNoncrit}
    P_{k\ell}(t) = \frac{\pi_\ell}{\pi_k}\beta^{k}s^{\gamma+1}\\ \times \left(\frac{1-\beta}{1-\beta s}\right)^{\gamma+2}\left(\frac{1-s}{1-\beta s}\right)^{k+\ell}F_{k\ell}(t),
\end{multline}
which is valid for both cases of $\beta \neq 1$ and in the above form for $k \leq \ell$. 
\subsubsection*{Critical case}
To find the critical transition matrix elements, one may simply take the limit as $\beta \to 1$ of the above expressions. One finds
\begin{equation}
    \lim_{\beta \to 1} X(t)  = 1 - \frac{1}{t^2};\qquad \lim_{\beta \to 1} \frac{1-s}{1-\beta s} = \frac{t}{1+t}. 
\end{equation}
Defining the auxiliary function
\begin{multline}
    F^c_{k \ell}(t)=\frac{\Gamma(k+\ell+\gamma+2)}{\Gamma(\ell+1) \Gamma(k+\gamma+2)}\\\times{ }_2 F_1(-k, -\ell ; -1-k-\ell-\gamma ; 1-1/t^2),
\end{multline}
the critical transition matrix elements are
\begin{equation}
    P^c_{k\ell}(t) =  \frac{\pi_\ell}{\pi_k}\left(\frac{1}{1+t}\right)^{\gamma+2}\left(\frac{t}{1+t}\right)^{k+\ell}F^c_{k\ell}(t).
\end{equation}
Python code which implements these functions for easy evaluation is available on request. 
\subsection{Low population moments}
The generating function $G(t,z)$ for the occupation probabilities $p_n(t)$ satisfies the inhomogeneous equation
\begin{multline}
    \frac{\partial G}{\partial t}=(\beta z-1)(z-1) \frac{\partial G}{\partial z}\\+\gamma\left(\frac{1-z}{z}\right)\left(G(t, z)-p_A(t)\right).
\end{multline}
The cumulant generating function, defined as $K(t,u) = \log G(t,e^u)$, therefore satisfies the related equation
\begin{multline}
    \frac{\partial K}{\partial t} = (\beta e^u - (\beta + 1) + e^{-u})\frac{\partial K}{\partial u} \\+ \gamma(e^{-u}-1)(1 - e^{-K}p_A(t)).
\end{multline}
Expanding the function $K(t,u)$ to quadratic order
\begin{equation}
    K(t,u) = \ex{j(t)}u + \frac{1}{2}\Delta j(t)^2 u^2 + \OO(u^3),
\end{equation}
and solving order-by-order yields coupled equations for the average population $\ex{j(t)}$ and the variance $\Delta j(t)^2$. The first of these, governing the average population $\ex{j(t)}$, is 
\begin{equation}
    \frac{\partial}{\partial t}\ex{j(t)} = (\beta -1)\ex{j(t)} + \gamma(p_A(t) -1),
\end{equation}
which is solved by 
\begin{multline}
    \ex{j(t)} = Ne^{(\beta-1)t} \\+ \gamma \int_0^t d\tau\ e^{(\beta-1)(t-\tau)}(p_A(\tau)-1).
\end{multline}
In the simple case $N =1$, we have
\begin{equation}
    p_A(t)=1-(1-z(t))^{\gamma+1};
\quad z(t) = \frac{1-e^{(\beta-1)t}}{1-\beta e^{(\beta-1)t}},
\end{equation}
which leads to the average population
\begin{equation}
    \langle j(t)\rangle=e^{(\beta-1) t}[1-z(t)]^\gamma.
\end{equation}
\edit{\section{The random walk polynomials}}
\label{sec:firewalk}
\edit{In this appendix we construct analytically the random walk polynomials for the birth-death-suppression process. These polynomials are equivalent under reparametrization to the Pollaczek polynomials, as discussed in Section \ref{sec:predfootprint}. 

We use the method of Stieltjes inversion to construct the measure of orthogonality given expressions for the polynomials $W_n(x)$. 

\subsubsection{The critical polynomials}
To begin, consider the critical phase: when $\beta = 1$, the recurrence relation Eqn.~\eqref{eq:ftptRecurrence} simplifies to
\begin{multline}
    x W^c_n(x)=\frac{n+1}{2(n+1)+\gamma} W^c_{n+1}(x)\\+\frac{n+\gamma+1}{2(n+1)+\gamma} W^c_{n-1}(x).
\end{multline}
The critical polynomials $W^c_n(x)$ are equivalent to a known set of orthogonal polynomials: the Gegenbauer or ultraspherical polynomials $C^\lambda_n(x)$. The exact correspondence is:
\begin{equation}
    W^c_n(x;\gamma) = C^{1+\gamma/2}_n(x).
\end{equation}
A generalization of the spherical harmonics, the Gegenbauer polynomials fall in the Askey scheme and many of their properties are well-known (for reference, see 1.8.1 in \cite{askey}). They are a special case of the Jacobi polynomials. Their generating function is 
\begin{equation}
    \frac{1}{\left(1-2 x z+z^2\right)^{1+\gamma / 2}} = \sum_{n=0}^\infty W^c_n(x;\gamma)z^n.
     \label{eq:gegenbauerGenerating}
\end{equation}
% recalling the basic definitions of the (Gauss) hypergeometric and Pochhammer functions:
% \begin{equation}
%     { }_2 F_1(a, b ; c ; z)=\sum_{n=0}^{\infty} \frac{(a)_n(b)_n}{(c)_n} \frac{z^n}{n !} ; \quad(a)_n=\frac{\Gamma(a+n)}{\Gamma(a)}.
% \end{equation}
These critical random walk polynomials are orthogonal with respect to the normalized, continuous measure 
\begin{equation}
    \label{eq:gegenMeasure}
    d\sigma(x)=\frac{\Gamma\left(\frac{\gamma}{2}+2\right)}{\sqrt{\pi} \Gamma\left(\frac{\gamma+3}{2}\right)}\left(1-x^2\right)^{(\gamma+1) / 2} d x.
\end{equation}
supported on the interval $[-1,1]$. One could proceed to immediately calculate with these expressions. Instead, their introduction here will be used to verify the non-critical results by showing that they agree in the critical limit. 

\subsubsection{The non-critical polynomials}
Away from criticality the $W_n(x)$ do not admit a simple hypergeometric formula. For convenience, we now refer to the general, non-critical polynomials $W_n(x;\beta,\gamma) \equiv W_n(x)$ as the \textit{firewalk} polynomials, recalling of course that they are equivalent to the Pollazcek family. One can find an expression for these firewalk polynomials $W_n(x;\beta, \gamma)$ by the method of generating function: define
\begin{equation}
    \Phi(z,x;\beta, \gamma) = \sum_{n=0}^\infty W_n(x;\beta,\gamma)z^n.
\end{equation}
The recurrence relation \eqref{eq:ftptRecurrence} satisfied by the $W_n$ descends to a differential equation for the generating function:
\begin{equation}
    \label{eq:ftptGenFunc}
    \frac{\partial \Phi}{\partial z} = \frac{x(\beta+\gamma+1)-z(\gamma+2)}{z^2+\beta-x z(\beta+1) } \Phi,
\end{equation}
along with the boundary condition $\Phi(0,x;\beta,\gamma) = W_0(x) = 1$. First, consider the singular points of the equation above. These occur at the roots of the quadratic $z^2-x z(\beta+1)+\beta$. The discriminant $D(x)$ of this quadratic is 
\begin{equation}
   D(x) = \sqrt{x^2(\beta+1)^2-4 \beta},
\end{equation}
where for consistency we take the positive root always. The discriminant vanishes at $x = I_\beta$, where
\begin{equation}
    I_\beta = \frac{2\sqrt{\beta}}{1+\beta};\quad I_\beta \leq 1.
\end{equation}
Specifically, $D(x)$ is real for $|x| \geq I_\beta$ and pure imaginary otherwise. The bound $I_\beta =1$ only at criticality when $\beta = 1$. To solve Eqn.~\eqref{eq:ftptGenFunc}, one computes the partial fraction decomposition
\begin{equation}
    \label{eq:partialfraction}
    \frac{x(\beta+\gamma+1)-z(\gamma+2)}{z^2-x z(\beta+1)+\beta}=\frac{A}{z-u}+\frac{B}{z-v}.
\end{equation}
The roots $u,v$ of the quadratic $z^2-x z(\beta+1)+\beta$ are
\begin{equation}
    u=\frac{x(\beta+1)}{2}+\frac{D(x)}{2}; \quad v=\frac{x(\beta+1)}{2}-\frac{D(x)}{2},
\end{equation}
noting the transformation rule $u(-x) = -v(x)$. The relative magnitude of these roots depends on $x$: if $x \in [-I_\beta,I_\beta]$ the discriminant $D$ is pure imaginary and $|u| = |v|$. If $x > I_\beta$ then $|v| < |u|$; by symmetry for $x < -I_\beta$ one has $|u| < |v|$. The dominant root (that of smallest magnitude) is relevant when constructing the measure of orthogonality.  The coefficients $A,B$ of the partial fraction decomposition \eqref{eq:partialfraction} are given by
\begin{gather}
    A=-\frac{\gamma+2}{2}-\frac{x \gamma(\beta-1)}{2 D(x)}; \\ B=-\frac{\gamma+2}{2}+\frac{x \gamma(\beta-1)}{2 D(x)},
\end{gather}
where under $x \mapsto -x$ the coefficients $A \leftrightarrow B$ are interchanged.
% Some convenient algebraic relations between these quantities are
% \begin{equation}
%     A+B=-(\gamma+2), \quad u-v=D(x), \quad u v=\beta.
% \end{equation}
The equation to be solved is thus rewritten as
\begin{equation}
% \label{eq:ftptGfuncSol}
    \frac{\partial \Phi}{\partial z}=\left(\frac{A}{z-u}+\frac{B}{z-v}\right) \Phi,
\end{equation}
which, along with the boundary condition $\Phi(0,x) = 1$, determines
\begin{equation}
    \Phi(z,x;\beta, \gamma) = (1-z / u)^A(1-z / v)^B.
    \label{eq:firewalkGenerating}
\end{equation}
A first check is that this function reproduces the generating function \eqref{eq:gegenbauerGenerating} in the limit $\beta \to 1$. In the critical limit, the parameters reduce to
\begin{gather}
    \lim_{\beta \to 1}A = \lim_{\beta\to1}B = - (1 + \gamma/2);\\ \lim_{\beta \to 1}u = x + i\sqrt{1-x^2},\\ \lim_{\beta \to 1}v = x - i\sqrt{1-x^2},
\end{gather}
which allows one to verify that, indeed, as $\beta \to 1$ the generating function $\Phi$ simplifies to
\begin{equation}
    \lim_{\beta\to1}\Phi(z, x ; \beta, \gamma) = \frac{1}{\left(1-2 x z+z^2\right)^{1+\gamma / 2}},
\end{equation}
agreeing with the Gegenbauer generating function in Eqn.~\eqref{eq:gegenbauerGenerating}. The firewalk or Pollazcek polynomials $W_n(x;\beta,\gamma)$ can therefore be considered a deformation of the Gegenbauer polynomials parametrized by the birth rate $\beta$. 

To find a closed form, one must determine the $n$-th series coefficient of \eqref{eq:firewalkGenerating}. An expression is
\begin{equation}
    W_n(x)=(-B)_n \frac{v^{-n}}{n !}{ }_2 F_1(-n,-A ;-n+B+1 ; v / u).
\end{equation}
By the use of some hypergeometric identities \cite{szego,NIST:DLMF}, this is equivalent to the form
\begin{multline}
\label{eq:firewalkexpr}
W_n(x)=\frac{(\gamma+2)_n}{n !} u^{-n}\\ \times {}_2F_1(-n,-B, \gamma+2 ;-u D(x) / \beta).
\end{multline}
While this expression does not obviously give polynomials in $x$, direct calculation shows
\begin{gather}
    W_1(x) = \frac{x(\beta+\gamma+1)}{\beta};\\W_2(x)=\frac{x^2(\beta+\gamma+1)(2 \beta+\gamma+2)-\beta(\gamma+2)}{2 \beta^2};
\end{gather}
noting that the $W_n$ are odd for odd $n$ and even for even $n$, consistent with their measure of orthogonality being even. But determining this expression for the firewalk polynomials $W_n(x)$ is only the first step. In order to evaluate matrix elements of the jump chain transition matrix $S(n)$, we also must determine the correct measure of orthogonality. 

\subsection{The measure by Stieltjes inversion}
By Favard's theorem there exists a real, even, positive measure of orthogonality $d\sigma(x)$ on $[-1,1]$ associated to the firewalk polynomials $W_n(x)$. From such a measure, the Stieltjes transformation defines a function $\chi(z)$, the \textit{resolvent} via
\begin{equation}
\label{eq:resolventDefinition}
    \chi(z) = \int_{-1}^1 \frac{d\sigma(x)}{z-x}.
\end{equation}
This function is analytic in the complex $z$ plane away from the interval $[-1,1]$, where certain singularities arise exactly where the measure of orthogonality is supported. Generically, the measure may be a mix of continuous and discrete parts:
\begin{gather}
    \label{eq:measureFacts}
    d\sigma(x) = w(x)dx + \sum_{k} \Delta_k\delta(x-x_k) dx ,\\ \chi(z) = \int \frac{w(x)}{z-x}dx + \sum_{k}\frac{\Delta_k}{z-x_k}.
\end{gather}
At points $x_k$ where the measure has some discrete atomic weight $\Delta_k$ the resolvent $\chi(z)$ therefore has a simple pole with residue $\Delta_k$. On the other hand, where the measure has continuous support the function $\chi(z)$ has a branch cut. The transformation from measure to resolvent can thus be inverted by finding the discontinuity in $\chi(z)$ across the cut, or the residue of $\chi(z)$ at its poles. In \cite{ismail}, the authors carry out this calculus for the birth, death, and immigration process. For a more thorough review of the analytic methods used here, see also \cite{analyticcombs,szego,Bank,chihara2011introduction}.
\subsubsection{Computing the resolvent}
The starting point for computing the resolvent $\chi(z)$ is to compute the firewalk polynomials `of the second kind,' $W^*_n(x)$. They satisfy the same recurrence relations \eqref{eq:ftptRecurrence} but with the altered initial conditions
\begin{equation}
    W_0^*(x)=0 ; \quad W_1^*(x) = \frac{1}{p_0}=\frac{\beta+\gamma+1}{\beta}.
\end{equation}
Any set of orthogonal polynomials has an partner set `of the second kind' defined in this manner. A key theorem due to Markov \cite{szego} is that the resolvent $\chi(z)$ may be written in terms of the limit
\begin{equation}
    \chi(z) = \lim_{n\to\infty}\frac{W^*_n(z)}{W_n(z)},
\end{equation}
which allows computation of $\chi(z)$ without knowledge of the measure. Computing these asymptotics for each set of polynomials $W_n,W_n^*$ involves analyzing their respective generating functions and applying the method of Darboux. This lemma for estimating the asymptotic series coefficients of a complex function is reviewed in Appendix \ref{sec:appDarboux}. 

As the $W_n^*(x)$ satisfy the same recurrence as the $W_n(x)$, their generating function $\Phi^*$ satisfies a very similar differential equation. The difference is in the initial conditions for the recurrence which is reflected by the introduction of an inhomogeneous term. In particular, the $W_n^*(x)$ are generated by $\Phi^*(z)$ which solves
\begin{multline}
    \frac{\partial \Phi^*}{\partial z}=\frac{x(\beta+\gamma+1)-z(\gamma+2)}{z^2-x z(\beta+1)+\beta} \Phi^*\\+\frac{\beta+\gamma+1}{z^2-x z(\beta+1)+\beta}.
\end{multline}
Using the same partial fraction decomposition as before one can write
\begin{equation}
    \frac{\partial \Phi^*}{\partial z}=\left[\frac{A}{z-u}+\frac{B}{z-v}\right] \Phi^*+\frac{\beta+\gamma+1}{(z-u)(z-v)}.
\end{equation}
The solution is $\Phi^* = f(z)\Phi$ where $\Phi$ solves the homogeneous equation. With the boundary condition $\Phi^*(0) = 0$ one finds
\begin{multline}
    \Phi^*(x, z)=(\beta+\gamma+1)(u-z)^A(v-z)^B\\ \times \int_0^z d t\ (u-t)^{-A-1}(v-t)^{-B-1}.
\end{multline}
With the generating functions $\Phi,\Phi^*$ the asymptotics of both the $W_n(x),\ W_n^*(x)$ can now be determined by an application of Darboux's method. This is done in Appendix \ref{sec:appDarboux}. For $\re x>0$ and off the cut $[-I_\beta, I_\beta]$, the results are
\begin{gather}
    \lim _{n \rightarrow \infty} W_n(x) \sim(1-v / u)^A \frac{n^{-B-1}}{\Gamma(-B)} v^{-n},\\
    \lim _{n \rightarrow \infty} W_n^*(x) \sim(\beta+\gamma+1) \frac{D^A}{\Gamma(-B)} v^{B-n} n^{-B-1} \\ \qquad\qquad \qquad\times\int_0^v d t\ (u-t)^{-A-1}(v-t)^{-B-1}.
\end{gather}
The resolvent $\chi(z)$ is constructed by taking the ratio of these, recalling that $u,v,A,B,D$ should all be considered as functions of a complex variable $z$:
\begin{multline}
\label{eq:ftptResolvent}
    \chi(z)=\lim _{n \rightarrow \infty} \frac{W_n^*(z)}{W_n(z)}=(\beta+\gamma+1) u^A v^B \\ \times \int_0^v d t\ (u-t)^{-A-1}(v-t)^{-B-1};\quad \re z \geq 0.
\end{multline}
This expression, valid for $\operatorname{Re} z > 0, \operatorname{Im} z \neq 0$, is analytic for $z$ away from the cut $[-1,1]$ on the real line. It is precisely at this singular locus that the resolvent gives the data of the measure. 
\subsubsection{Stieltjes inversion}
With the resolvent computed, one can now proceed to compute the elements of the measure. It so happens that there are both continuous and discrete parts, sourced by the branch cuts and poles of the resolvent $\chi(z)$ respectively. Formally, the Stieltjes transformation $d\sigma(x) \mapsto \chi(z)$ may be inverted by 
\begin{equation}
    \sigma(x_2)-\sigma(x_1)=\lim _{\varepsilon \rightarrow 0^{+}}  \int_{x_1}^{x_2}dx\ \frac{1}{\pi}\operatorname{Im}\left[\chi(x+i \varepsilon)\right]. 
\end{equation}
This inversion formula is physically analogous to the optical theorem, where $\chi(z)$ represents the scattering amplitude, and the measure $d\sigma(x)$ is analogous to the cross-section. First consider the continuous part of the measure which is supported on the real interval $[-I_\beta, I_\beta]$. Taking the limit $x_1 \to x_2$ in the above yields
\begin{equation}
    w(x) \equiv \frac{d\sigma(x)}{d x}= \lim _{\varepsilon \rightarrow 0^{+}}\frac{1}{\pi}\operatorname{Im}[\chi(x+i\varepsilon)];\quad |x| \leq I_\beta,
\end{equation}
where $x$ lies on the branch cut. As one approaches the cut $\varepsilon\to 0$, we have $u \to \Bar{v}$ and $A \to \Bar{B}$. This means that the resolvent prefactor $u^Av^B$ as well as the integrand are real as the cut is approached---they are symmetric in $u\leftrightarrow v,\ A\leftrightarrow B$. The integral bounds, however, induce a nonzero imaginary part;
\begin{multline}
2i \im \int_0^v dt\ f(t) = \int_0^v dt\ f(t) - \overline{\int_0^v dt\ f(t)} \\
= \int_0^v dt\ f(t) + \int_{\Bar{v}}^0 dt\ f(t) = \int_u^v dt\ f(t),
\end{multline}
where the overbar represents complex conjugation. As a result, the continuous part of the measure of orthogonality is
\begin{multline}
    w(x)=\frac{\beta+\gamma+1}{2 \pi i} u^A v^B \\ \times\int_v^u d t\ (u-t)^{-A-1}(v-t)^{-B-1},
\end{multline}
for $x$ in the interval $[-I_\beta, I_\beta]$. 
\begin{figure*}[t]
    \centering
    \includegraphics[width = 2\columnwidth]{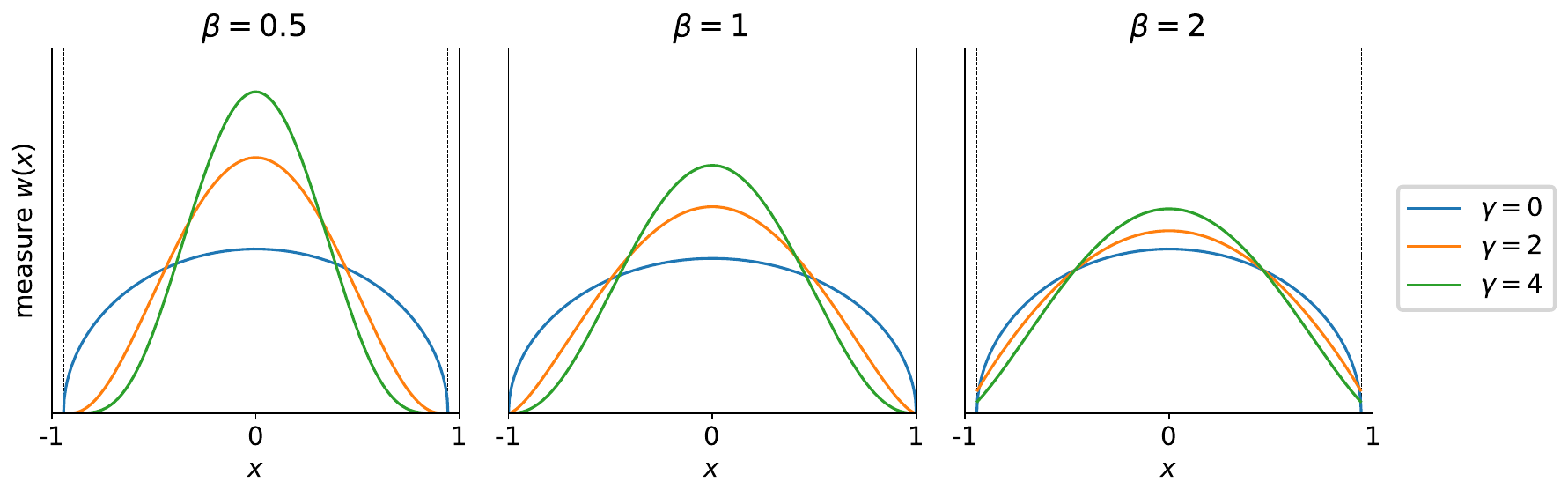}
    \caption{The continuous measure of orthogonality $w(x) = d\sigma/dx$ of the firewalk polynomials $W_n(x)$ for various birth rates $\beta$ and suppression rates $\gamma$. The dashed lines at $x = \pm I_\beta$ show that the measure is not supported on the entire interval $x \in [-1,1]$. Larger suppression values can be seen to sharpen the peak of the distribution near zero.}
    \label{fig:measureCollage}
\end{figure*}
The definite integral, after some rescaling, is an Euler beta function. In particular, one finds the expression
\begin{multline}
    \label{eq:firewalkMeasure}
    w(x) =\frac{\beta+\gamma+1}{2 \pi i} u^A v^B \\\times \frac{(u-v)^{-A}}{(v-u)^{B+1}} \frac{\Gamma(-A) \Gamma(-B)}{\Gamma(\gamma+2)},
\end{multline}
noting that the formula is singular for $A,B \in \ZZ_+$. While obtuse, this measure, supported only on $[-I_\beta,I_\beta]$ defines a smooth weight function plotted in Fig. \ref{fig:measureCollage} for various values of the parameters. In the limit $\beta \to 1$ it should reduce to the Gegenbauer polynomial measure \eqref{eq:gegenMeasure}. One may check that in the critical limit $\beta \to 1$, 
\begin{gather}
u^A v^B \rightarrow 1 ; \\ \frac{(u-v)^{-A}}{(v-u)^{B+1}} \rightarrow 2^{\gamma+1} i\left(1-x^2\right)^{(\gamma+1) / 2}; \\
\frac{\Gamma(-A) \Gamma(-B)}{\Gamma(\gamma+2)} \rightarrow \frac{\sqrt{\pi} 2^{-(\gamma+1)} \Gamma(\gamma / 2+1)}{\Gamma(\gamma / 2+3 / 2)},
\end{gather}
after applying some gamma function identities. Therefore one finds
\begin{align}
    \lim _{\beta \rightarrow 1} w(x)&=\frac{\gamma+2}{2 \pi i} \cdot i\left(1-x^2\right)^{(\gamma+1) / 2} \cdot \frac{\sqrt{\pi} \Gamma(\gamma / 2+1)}{\Gamma\left(\frac{\gamma+3}{2}\right)}\nonumber\\&=\frac{\Gamma\left(\frac{\gamma}{2}+2\right)}{\sqrt{\pi} \Gamma\left(\frac{\gamma+3}{2}\right)}\left(1-x^2\right)^{(\gamma+1) / 2},
\end{align}
which agrees perfectly with 
the known Gegenbauer measure. However, the work is not yet done. The continuous measure just constructed is not supported for $I_\beta < |x| < 1$. The full measure of orthogonality for the firewalk polynomials does have support in this region, but only at a discrete set of points. These points occur are precisely the poles of the resolvent. 

\subsubsection{The discrete measure}
The continuous measure found in Eqn.~\eqref{eq:firewalkMeasure} is even, as one can explicitly check by considering the transformation of $u,v,A,B$ under $x \mapsto -x$. It is undefined at the points where $A,B$ are non-negative integers, which suggests that at these loci the full measure has discrete weight and thus discontinuous derivative. From the resolvent formula
\begin{multline}
	\label{eq:resolventInt}
    \chi(z) = (\beta+\gamma+1) u^A v^B \\\times\int_0^v d t\ (u-t)^{-A-1}(v-t)^{-B-1} ; \quad \re z \geq 0,
\end{multline}
one can see that a singularity arises in the integrand as $t\to v$. This occurs in particular when $B = k$ is a non-negative integer, and only where this form is valid in the right half-plane. Transforming $z \mapsto -z$, one finds
\begin{multline}
    \chi(z) = -(\beta+\gamma+1) u^A v^B\\\times \int_0^u d t\ (u-t)^{-A-1}(v-t)^{-B-1} ; \quad \re z \leq 0
\end{multline}
so that in the left half-plane, a singularity arises as $t\to u$ and when $A = k \in \ZZ_+$. These singularities therefore occur at $\pm x_k$ satisfying
\begin{equation}
    k =- \frac{\gamma + 2}{2} + \frac{|x_k|\gamma(\beta-1)}{2D(x_k)};\quad k\in \ZZ_+.
\end{equation}
Since $\gamma \geq 0$ always, this equation only has solutions when $\beta > 1$. Therefore the discrete measure develops only for $\beta > 1,\ \gamma > 0$, noting that $A = B= -1$ uniformly in the case of zero suppression. Since the measure is even, we restrict to considering $x > 0,\ \beta > 1$, and the singularities associated with $B = k \in \ZZ_+$. These occur at the values $\pm x_k$ satisfying
\begin{equation}
    x_k ^2= \frac{\beta(\gamma+2(k+1))^2}{(\gamma+(k+1)(\beta+1))(\beta \gamma+(k+1)(\beta+1))}.
\end{equation}
For all $k$ one has $I_\beta < x_k < 1$, with $x_k \to I_\beta$ as $k \to \infty$. In fact, the $x_k$ approach $I_\beta$ quite quickly as $k$ increases. 

One now should compute the residue of the resolvent at the points $x_k$. By examining the resolvent expression in Eqn.~\eqref{eq:resolventInt}, it is clear that with $B = k$ the relevant singularity arises when $t \to v$ in the integral. Let $s = v - t$; then one has
\begin{multline}
    \label{eq:resolventMidway}
    \chi(z)=\frac{\beta+\gamma+1}{\beta}\left(\frac{u}{u-v}\right)^{A+1} v^{B+1} \\ \times\int_0^v d s\left(1+\frac{s}{u-v}\right)^{-A-1} s^{-B-1},
\end{multline}
with the goal being to isolate the singularity and extract its residue. From here, one can expand the first part of the integrand as an infinite sum, writing
\begin{multline}
    \label{eq:resolventMidway2}
    \int_0^v d s\left(1+\frac{s}{u-v}\right)^{-A-1} s^{-B-1} \\= \sum_{n=0}^{\infty} \frac{(A+1)_n}{n !}\left(\frac{1}{v-u}\right)^n\left[\int_0^v d s\ s^{n-B-1}\right].
\end{multline}
Here, one should regard the definite integral as its analytic continuation to all $B \notin \ZZ_+$ in the sense of Hadamard \cite{ismail,Bank}. The quantities $u,v,A,B$ all take finite, real values at the $|x_k| \geq I_\beta$. The measure weight is the residue of the resolvent as $z\to x_k$; by definition this is the coefficient of $(z-x_k)^{-1}$ in the expansion of the function around $z = x_k$. Near this point, one has
\begin{gather}
    \lim_{z\to x_k}B = k + \alpha(z-x_k) + \cdots;\\ \alpha = \frac{dB}{dz}\bigg|_{z=x_k} = -2 \beta \gamma(\beta-1) D(x_k)^{-3}.
\end{gather}
Formally evaluating the integral near $B = k$ yields the expansion
\begin{multline}
    \int ds\ s^{n-B-1} = \frac{1}{n-B}s^{n-B} = \frac{1}{n-k -\alpha(z-x_k)}\\\times[s^{n-k} - \alpha s^{n-k}\log(s)(z-x_k) + \cdots].
\end{multline}
For any $n \neq k$, this function is formally analytic as $z \to x_k$. A nonzero residue at $x_k$ is therefore generated only by the term with $n = k$. Setting $n = k$ in Eqns. \eqref{eq:resolventMidway}, \eqref{eq:resolventMidway2} and evaluating the definite integral one finds that the residue $\Delta_k$ at $x_k$ is
\begin{multline}
    \Delta_k=\frac{\beta+\gamma+1}{\beta}\left(\frac{u}{u-v}\right)^{A+1}\\\times \frac{v^{k+1}}{(v-u)^k} \frac{(A+1)_k}{k !}\left.\left(-\frac{d x}{d B}\right)\right|_{x=x_k},
\end{multline}
which by even-ness is also the weight of the measure at $-x_k$. Writing $u_k\equiv u(x_k)$ and evaluating everything at $x_k$ one finds that the weight $\Delta_k$ of the discrete measure at $\pm x_k$ is
\begin{multline}
\label{eq:TmeasWeights}
\Delta_k=\frac{\beta+\gamma+1}{\beta^{\gamma+3}}\left(u_k\right)^{-k}\left(v_k\right)^{k+\gamma+2} \\\times\frac{(\gamma+2)_k}{k !} \cdot \frac{D_k^{\gamma+4}}{2 \gamma(\beta-1)};\quad \beta > 1,
\end{multline}
with the understanding that $\Delta_k \equiv 0$ for $\beta \leq 1$. Along with the continuous measure in Eqn.~\eqref{eq:firewalkMeasure}, these discrete weights fully characterize the orthogonality relation for the $W_n(x)$:
\begin{multline}
    \int_{-I_\beta}^{I_\beta} W_n(x) W_m(x) w(x) dx
    +\sum_{k=0}^{\infty} \Delta_k W_n( x_k) W_m( x_k)\\+\sum_{k=0}^{\infty}\Delta_k W_n(-x_k)W_m(-x_k)=h_n \delta_{n, m}.
    \label{eq:firewalkcompleteortho}
\end{multline}
In principle, one can now calculate arbitrary matrix elements of the $n$-step transition matrix $S(n)$. However, these expressions rarely, if ever, have a reasonable analytical form. They are better suited to numerical analysis.}
\section{The method of Darboux}
\label{sec:appDarboux}
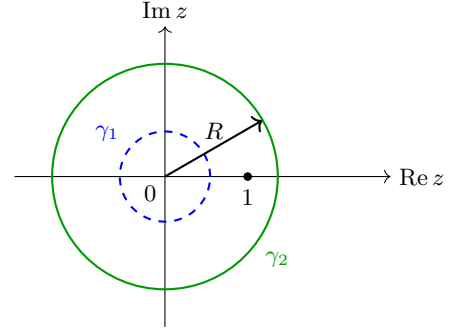
\begin{figure}[!h]
    \centering
    \begin{tikzpicture}
    % Draw axes
    \draw[->] (-2, 0) -- (3, 0) node[anchor=west] {$\re z$};
    \draw[->] (0, -2) -- (0, 2) node[anchor=south] {$\im z$};

    % Singularity at z=1
    \node[draw, circle, fill, inner sep=1pt, label=below:{$1$}] at (1.1, 0) {};

    % Label 0 at the origin
    \node[anchor=north east] at (0, 0) {0};

    % Contour gamma_1
    \draw[blue, dashed, thick, postaction={decorate}, decoration={
        markings,
        mark=at position 0.4 with {\node[blue, anchor=south east] {$\gamma_1$};}
    }] (0, 0) circle (0.6);
    % Arrow R
    \draw[->, thick, black] (0, 0) -- (30:1.5) node[midway, anchor = south] {$R$};
    % Contour gamma_2
    \draw[green!60!black, thick, postaction={decorate}, decoration={
        markings,
        mark=at position 0.9 with {\node[green!60!black, anchor=north west] {$\gamma_2$};}
    }] (0,0) circle (1.5);
    \end{tikzpicture}
    \caption{Contours for Darboux's method. By choosing a comparison function $g(z)$ to subtract off the singularity at $z = 1$, the function $f(z) - g(z)$ can be extended to the contour $\gamma_2$, where large $n$ asymptotics may be taken. }
    \label{fig:contours}
\end{figure}

In this Appendix we review the method of Darboux for estimating the asymptotic coefficients of a meromorphic function $f(z)$. For a thorough reference, see \cite{analyticcombs,flajolet2006hybrid,wong2001asymptotic}. Here, we develop only the tools necessary to compute the measure in Sec. \ref{sec:predfootprint}, for completeness and pedagogical purposes. Begin with a meromorphic function $f(z)$ analytic at the origin: the goal is to estimate its asymptotic series coefficients. For simplicity, let the dominant singularity (that of smallest magnitude) of $f(z)$ be at $z = 1$, so that the series
\begin{equation}
\label{eq:fseries}
    f(z) = \sum_{n=0}^\infty c_n z^n
\end{equation}
converges inside the unit disk. Of course, any meromorphic function may be holomorphically mapped to this form. It turns out that the asymptotic behavior of the $c_n$ depends strongly on the nature of the singularities of the function $f(z)$. To see why this is the case, recall that the constants $c_n \equiv [z^n]f(z)$ may be computed by the following contour integral
\begin{equation}
    \label{eq:cauchy}
    c_n = [z^n]f(z) = \frac{1}{2\pi i}\oint_{\gamma_1} f(z)\frac{dz}{z^{n+1}},
\end{equation}
where the contour $\gamma_1$ encircles the origin, as shown in Fig. \ref{fig:contours}. 
Define another function $g(z)$ which is analytic for $|z| < 1$ and such that $f(z) - g(z)$ is continuous on $|z| = 1$. This defines a sort of `minimal subtraction' of the singularity at $z = 1$. Now, by construction, the function $h(z) = f(z) - g(z)$ is analytic in the disk $|z| \leq R$ for some $R > 1$. The series coefficients of $f(z) - g(z)$ are therefore
\begin{equation*}
    f_n - g_n = [z^n]h(z) = \oint_{\gamma_1}h(z)\frac{dz}{z^{n+1}} = \oint_{\gamma_2}h(z)\frac{dz}{z^{n+1}},
\end{equation*}
where the contour of integration can be radially extended to $\gamma_2$, located at a radius $R > 1$. By parametrizing the contour in polar coordinates, one bounds the difference $f_n - g_n$ as
\begin{align}
    |f_n - g_n|  &= \bigg|\oint_{\gamma_2} h(z) \frac{d z}{z^{n+1}}\bigg| \\&= R^{-n}\bigg|\int_0^{2\pi}h(z(\phi))e^{-in\phi}d\phi\bigg| \\&\leq R^{-n}\int_0^{2\pi} |h(z)| d\phi,
\end{align}
where, in the final step, the integral does not depend on $n$. An asymptotic statement can now be made: as $n \to \infty$, since $R > 1$, we have
\begin{equation}
    |f_n - g_n| \lesssim \OO(R^{-n}) \implies f_n \sim g_n.
\end{equation}
This lemma is often referred to as Darboux's method \cite{ismail}. Informally, it states that a comparison function $g(z)$ which shares the same dominant singularity as $f(z)$ has asymptotically identical series coefficients. To make use of this, one can compute the asymptotic series coefficients for a singularity of a standard form, then use this theorem to transfer the results to a more complex function. Consider the basic singularity
\begin{equation}
    \label{eq:basicSingularity}
    (1-z/z_0)^\alpha = \sum_{n=0}^\infty (-1)^n\binom{\alpha}{n}\left(\frac{z}{z_0}\right)^n.
\end{equation}
Here, the explicit form of the series coefficients is known, so one can easily extract the asymptotics by applying the Stirling approximation:
\begin{align}
    (-1)^n\binom{\alpha}{n} &= \frac{(-\alpha)_n}{n !} \\&\sim \frac{n^{-\alpha -1}}{\Gamma(-\alpha)}\left[ 1 + \frac{\alpha(\alpha+1)}{2n} +\cdots \right].  
\end{align}
This gives the general result
\begin{equation}
\label{eq:generalSeriesResult}
    [z^n](1-z/z_0)^\alpha\ \sim\ z_0^{-n}\frac{n^{-\alpha-1}}{\Gamma(-\alpha)}; \qquad \alpha \notin \ZZ_+.
\end{equation}
For a derivation of this formula using only explicit contour integration, see \cite{wong2001asymptotic}. 
Note that this vanishes when the exponent $\alpha$ is a non-negative integer. The reason is straightforward: the left hand side of \eqref{eq:basicSingularity} is a polynomial in $z$ of degree $\alpha$, so its asymptotic series coefficients are identically zero \cite{analyticcombs}. In this special case, the leading term in the limit $n \to \infty$ is simply the term of highest degree. 
\subsubsection*{Firewalk polynomial asymptotics}
Having established the asymptotic approximation for the standard singularity, we can turn to the generating function \eqref{eq:firewalkGenerating} for the firewalk polynomials $W_n(x)$: 
\begin{equation}
    \label{eq:appfiregenerating}
    \Phi(z,x) = (1-z/u)^A(1-z/v)^B = \sum_{n=0}^\infty W_n(x) z^n.
\end{equation}
This function has two singularities which may be coincident in norm depending on the value of $x$; one has
\begin{gather}
    u=\frac{x(\beta+1)}{2}+\frac{D(x)}{2} ; \quad v=\frac{x(\beta+1)}{2}-\frac{D(x)}{2};\\D(x) = \sqrt{x^2(\beta+1)^2-4\beta}.
\end{gather}
Recall that for $x$ real and $|x| \leq I_\beta$, the roots $u,v$ are coincident in norm with $u = \Bar{v}$. For positive $x > I_\beta$, one has $|v| < |u|$, whereas for negative $x < -I_\beta$ the inequality is reversed. Take $x \notin \RR$ and $\operatorname{Re}x>0$; this means that $v$ is the dominant singularity of the generating function \eqref{eq:appfiregenerating}. One should construct a comparison function $g(z)$ which shares the singularity at $z = v$. To do so, expand the generating function $\Phi(z,x)$ around the point $z = v$, taking only the leading term as $z \to v$: 
\begin{equation}
    g(z) = (1-v/u)^A(1-z/v)^B.
\end{equation}
One can now simply apply the result \eqref{eq:generalSeriesResult} to determine the asymptotic coefficients of $g(z)$; Darboux's lemma ensures that these are also the asymptotics of the $W_n(x)$. Therefore, one finds
\begin{equation}
    \lim _{n \rightarrow \infty} W_n(x) \sim(1-v / u)^A \frac{n^{-B-1}}{\Gamma(-B)} v^{-n},
\end{equation}
with the caveat that this formula vanishes for $B \in \ZZ_+$. For the firewalk polynomials of the second kind $W^*_n(x)$, one has the generating function
\begin{multline}
    \Phi^*(x, z)=(\beta+\gamma+1)(u-z)^A(v-z)^B \\\times\int_0^z d t\ (u-t)^{-A-1}(v-t)^{-B-1}.
\end{multline}
By the same arguments as above, for positive $x$, the root $v$ is the dominant singularity of this function. The procedure is therefore essentially the same. To define a comparison function, expand $\Phi^*$ around the point $z = v$:
\begin{multline}
    g(z) = (\beta + \gamma + 1)D^Av^B(1-z/v)^B\\ \times\int_0^v dt\ (u-t)^{-A-1}(v-t)^{-B-1},
\end{multline}
to which one can apply the result for the standard singularity. This gives the asymptotic result
\begin{multline}
    \lim _{n \rightarrow \infty}W_n^*(x) \sim(\beta+\gamma+1) \frac{D^A}{\Gamma(-B)} v^{B-n} n^{-B-1} \\\times\int_0^v d t\ (u-t)^{-A-1}(v-t)^{-B-1}.
\end{multline}
This completes the asymptotics needed to compute the resolvent for the firewalk polynomial measure.

\bibliographystyle{apsrev4-1}
\bibliography{updated_refs}% Produces the bibliography via BibTeX.`

\begin{thebibliography}{70}%
\makeatletter
\providecommand \@ifxundefined [1]{%
 \@ifx{#1\undefined}
}%
\providecommand \@ifnum [1]{%
 \ifnum #1\expandafter \@firstoftwo
 \else \expandafter \@secondoftwo
 \fi
}%
\providecommand \@ifx [1]{%
 \ifx #1\expandafter \@firstoftwo
 \else \expandafter \@secondoftwo
 \fi
}%
\providecommand \natexlab [1]{#1}%
\providecommand \enquote  [1]{``#1''}%
\providecommand \bibnamefont  [1]{#1}%
\providecommand \bibfnamefont [1]{#1}%
\providecommand \citenamefont [1]{#1}%
\providecommand \href@noop [0]{\@secondoftwo}%
\providecommand \href [0]{\begingroup \@sanitize@url \@href}%
\providecommand \@href[1]{\@@startlink{#1}\@@href}%
\providecommand \@@href[1]{\endgroup#1\@@endlink}%
\providecommand \@sanitize@url [0]{\catcode `\\12\catcode `\$12\catcode
  `\&12\catcode `\#12\catcode `\^12\catcode `\_12\catcode `\%12\relax}%
\providecommand \@@startlink[1]{}%
\providecommand \@@endlink[0]{}%
\providecommand \url  [0]{\begingroup\@sanitize@url \@url }%
\providecommand \@url [1]{\endgroup\@href {#1}{\urlprefix }}%
\providecommand \urlprefix  [0]{URL }%
\providecommand \Eprint [0]{\href }%
\providecommand \doibase [0]{http://dx.doi.org/}%
\providecommand \selectlanguage [0]{\@gobble}%
\providecommand \bibinfo  [0]{\@secondoftwo}%
\providecommand \bibfield  [0]{\@secondoftwo}%
\providecommand \translation [1]{[#1]}%
\providecommand \BibitemOpen [0]{}%
\providecommand \bibitemStop [0]{}%
\providecommand \bibitemNoStop [0]{.\EOS\space}%
\providecommand \EOS [0]{\spacefactor3000\relax}%
\providecommand \BibitemShut  [1]{\csname bibitem#1\endcsname}%
\let\auto@bib@innerbib\@empty
%</preamble>
\bibitem [{\citenamefont {Anderson}(1991)}]{anderson}%
  \BibitemOpen
  \bibfield  {author} {\bibinfo {author} {\bibfnamefont {W.~J.}\ \bibnamefont
  {Anderson}},\ }\href@noop {} {\emph {\bibinfo {title} {Continuous-Time Markov
  Chains}}}\ (\bibinfo  {publisher} {Springer-Verlag},\ \bibinfo {year}
  {1991})\BibitemShut {NoStop}%
\bibitem [{\citenamefont {Feller}(1967)}]{feller1967introduction}%
  \BibitemOpen
  \bibfield  {author} {\bibinfo {author} {\bibfnamefont {W.}~\bibnamefont
  {Feller}},\ }\href@noop {} {\emph {\bibinfo {title} {An introduction to
  probability theory and its applications}}},\ Vol.~\bibinfo {volume} {1}\
  (\bibinfo  {publisher} {John Wiley and Sons, Inc.},\ \bibinfo {year}
  {1967})\BibitemShut {NoStop}%
\bibitem [{\citenamefont {Kendall}(1949)}]{kendall1949stochastic}%
  \BibitemOpen
  \bibfield  {author} {\bibinfo {author} {\bibfnamefont {D.~G.}\ \bibnamefont
  {Kendall}},\ }\href@noop {} {\bibfield  {journal} {\bibinfo  {journal}
  {Journal of the Royal Statistical Society. Series B (Methodological)}\
  }\textbf {\bibinfo {volume} {11}},\ \bibinfo {pages} {230} (\bibinfo {year}
  {1949})}\BibitemShut {NoStop}%
\bibitem [{\citenamefont {Crawford}\ and\ \citenamefont
  {Suchard}(2012)}]{Crawford2012}%
  \BibitemOpen
  \bibfield  {author} {\bibinfo {author} {\bibfnamefont {F.~W.}\ \bibnamefont
  {Crawford}}\ and\ \bibinfo {author} {\bibfnamefont {M.~A.}\ \bibnamefont
  {Suchard}},\ }\href {\doibase 10.1007/s00285-011-0471-z} {\bibfield
  {journal} {\bibinfo  {journal} {Journal of Mathematical Biology}\ }\textbf
  {\bibinfo {volume} {65}},\ \bibinfo {pages} {553} (\bibinfo {year}
  {2012})}\BibitemShut {NoStop}%
\bibitem [{\citenamefont {Petrovic}\ \emph {et~al.}(2012)\citenamefont
  {Petrovic}, \citenamefont {Alderson},\ and\ \citenamefont
  {Carlson}}]{petrovic}%
  \BibitemOpen
  \bibfield  {author} {\bibinfo {author} {\bibfnamefont {N.}~\bibnamefont
  {Petrovic}}, \bibinfo {author} {\bibfnamefont {D.~L.}\ \bibnamefont
  {Alderson}}, \ and\ \bibinfo {author} {\bibfnamefont {J.~M.}\ \bibnamefont
  {Carlson}},\ }\href@noop {} {\bibfield  {journal} {\bibinfo  {journal} {PloS
  one}\ }\textbf {\bibinfo {volume} {7}},\ \bibinfo {pages} {e33285} (\bibinfo
  {year} {2012})}\BibitemShut {NoStop}%
\bibitem [{\citenamefont {Catchpole}\ \emph {et~al.}(1989)\citenamefont
  {Catchpole}, \citenamefont {Hatton},\ and\ \citenamefont
  {Catchpole}}]{CATCHPOLE1989101}%
  \BibitemOpen
  \bibfield  {author} {\bibinfo {author} {\bibfnamefont {E.}~\bibnamefont
  {Catchpole}}, \bibinfo {author} {\bibfnamefont {T.}~\bibnamefont {Hatton}}, \
  and\ \bibinfo {author} {\bibfnamefont {W.}~\bibnamefont {Catchpole}},\ }\href
  {\doibase https://doi.org/10.1016/0304-3800(89)90062-8} {\bibfield  {journal}
  {\bibinfo  {journal} {Ecological Modelling}\ }\textbf {\bibinfo {volume}
  {48}},\ \bibinfo {pages} {101} (\bibinfo {year} {1989})}\BibitemShut
  {NoStop}%
\bibitem [{\citenamefont {Altamimi}\ \emph {et~al.}(2022)\citenamefont
  {Altamimi}, \citenamefont {Lagoa}, \citenamefont {Borges}, \citenamefont
  {McDill}, \citenamefont {Andriotis},\ and\ \citenamefont
  {Papakonstantinou}}]{largeScaleMitigation}%
  \BibitemOpen
  \bibfield  {author} {\bibinfo {author} {\bibfnamefont {A.}~\bibnamefont
  {Altamimi}}, \bibinfo {author} {\bibfnamefont {C.}~\bibnamefont {Lagoa}},
  \bibinfo {author} {\bibfnamefont {J.~G.}\ \bibnamefont {Borges}}, \bibinfo
  {author} {\bibfnamefont {M.~E.}\ \bibnamefont {McDill}}, \bibinfo {author}
  {\bibfnamefont {C.~P.}\ \bibnamefont {Andriotis}}, \ and\ \bibinfo {author}
  {\bibfnamefont {K.~G.}\ \bibnamefont {Papakonstantinou}},\ }\href
  {https://www.frontiersin.org/articles/10.3389/ffgc.2022.734330} {\bibfield
  {journal} {\bibinfo  {journal} {Frontiers in Forests and Global Change}\
  }\textbf {\bibinfo {volume} {5}} (\bibinfo {year} {2022})}\BibitemShut
  {NoStop}%
\bibitem [{\citenamefont {Somanath}\ \emph {et~al.}(2014)\citenamefont
  {Somanath}, \citenamefont {Karaman},\ and\ \citenamefont
  {Youcef-Toumi}}]{somanath2014controlling}%
  \BibitemOpen
  \bibfield  {author} {\bibinfo {author} {\bibfnamefont {A.}~\bibnamefont
  {Somanath}}, \bibinfo {author} {\bibfnamefont {S.}~\bibnamefont {Karaman}}, \
  and\ \bibinfo {author} {\bibfnamefont {K.}~\bibnamefont {Youcef-Toumi}},\
  }in\ \href@noop {} {\emph {\bibinfo {booktitle} {53rd IEEE Conference on
  Decision and Control}}}\ (\bibinfo {organization} {IEEE},\ \bibinfo {year}
  {2014})\ pp.\ \bibinfo {pages} {1432--1437}\BibitemShut {NoStop}%
\bibitem [{\citenamefont {Thompson}(2013)}]{thompson2013modeling}%
  \BibitemOpen
  \bibfield  {author} {\bibinfo {author} {\bibfnamefont {M.~P.}\ \bibnamefont
  {Thompson}},\ }\href@noop {} {\bibfield  {journal} {\bibinfo  {journal} {PloS
  one}\ }\textbf {\bibinfo {volume} {8}},\ \bibinfo {pages} {e63297} (\bibinfo
  {year} {2013})}\BibitemShut {NoStop}%
\bibitem [{\citenamefont {Abdelmalak}\ and\ \citenamefont
  {Benidris}(2021)}]{abdelmalak2021markov}%
  \BibitemOpen
  \bibfield  {author} {\bibinfo {author} {\bibfnamefont {M.}~\bibnamefont
  {Abdelmalak}}\ and\ \bibinfo {author} {\bibfnamefont {M.}~\bibnamefont
  {Benidris}},\ }in\ \href@noop {} {\emph {\bibinfo {booktitle} {2021 IEEE
  Industry Applications Society Annual Meeting (IAS)}}}\ (\bibinfo
  {organization} {IEEE},\ \bibinfo {year} {2021})\ pp.\ \bibinfo {pages}
  {1--6}\BibitemShut {NoStop}%
\bibitem [{\citenamefont {Meng}\ \emph {et~al.}(2015)\citenamefont {Meng},
  \citenamefont {Deng},\ and\ \citenamefont {Shi}}]{meng2015mapping}%
  \BibitemOpen
  \bibfield  {author} {\bibinfo {author} {\bibfnamefont {Y.}~\bibnamefont
  {Meng}}, \bibinfo {author} {\bibfnamefont {Y.}~\bibnamefont {Deng}}, \ and\
  \bibinfo {author} {\bibfnamefont {P.}~\bibnamefont {Shi}},\ }\href@noop {}
  {\bibfield  {journal} {\bibinfo  {journal} {World atlas of natural disaster
  risk}\ } (\bibinfo {year} {2015})}\BibitemShut {NoStop}%
\bibitem [{\citenamefont {D{\'\i}az-Avalos}\ and\ \citenamefont
  {Juan}(2022)}]{diaz2022modeling}%
  \BibitemOpen
  \bibfield  {author} {\bibinfo {author} {\bibfnamefont {C.}~\bibnamefont
  {D{\'\i}az-Avalos}}\ and\ \bibinfo {author} {\bibfnamefont {P.}~\bibnamefont
  {Juan}},\ }\href@noop {} {\bibfield  {journal} {\bibinfo  {journal}
  {Environmetrics}\ }\textbf {\bibinfo {volume} {33}},\ \bibinfo {pages}
  {e2774} (\bibinfo {year} {2022})}\BibitemShut {NoStop}%
\bibitem [{\citenamefont {Diao}\ \emph {et~al.}(2020)\citenamefont {Diao},
  \citenamefont {Singla}, \citenamefont {Mukhopadhyay}, \citenamefont {Eldawy},
  \citenamefont {Shachter},\ and\ \citenamefont
  {Kochenderfer}}]{diao2020uncertainty}%
  \BibitemOpen
  \bibfield  {author} {\bibinfo {author} {\bibfnamefont {T.}~\bibnamefont
  {Diao}}, \bibinfo {author} {\bibfnamefont {S.}~\bibnamefont {Singla}},
  \bibinfo {author} {\bibfnamefont {A.}~\bibnamefont {Mukhopadhyay}}, \bibinfo
  {author} {\bibfnamefont {A.}~\bibnamefont {Eldawy}}, \bibinfo {author}
  {\bibfnamefont {R.}~\bibnamefont {Shachter}}, \ and\ \bibinfo {author}
  {\bibfnamefont {M.}~\bibnamefont {Kochenderfer}},\ }\href@noop {} {\enquote
  {\bibinfo {title} {Uncertainty aware wildfire management},}\ } (\bibinfo
  {year} {2020}),\ \Eprint {http://arxiv.org/abs/2010.07915} {arXiv:2010.07915
  [cs.AI]} \BibitemShut {NoStop}%
\bibitem [{\citenamefont {Griffith}\ \emph {et~al.}(2017)\citenamefont
  {Griffith}, \citenamefont {Kochenderfer}, \citenamefont {Moss}, \citenamefont
  {Mi{\v{s}}ic}, \citenamefont {Gupta},\ and\ \citenamefont
  {Bertsimas}}]{griffith2017automated}%
  \BibitemOpen
  \bibfield  {author} {\bibinfo {author} {\bibfnamefont {J.~D.}\ \bibnamefont
  {Griffith}}, \bibinfo {author} {\bibfnamefont {M.~J.}\ \bibnamefont
  {Kochenderfer}}, \bibinfo {author} {\bibfnamefont {R.~J.}\ \bibnamefont
  {Moss}}, \bibinfo {author} {\bibfnamefont {V.~V.}\ \bibnamefont
  {Mi{\v{s}}ic}}, \bibinfo {author} {\bibfnamefont {V.}~\bibnamefont {Gupta}},
  \ and\ \bibinfo {author} {\bibfnamefont {D.}~\bibnamefont {Bertsimas}},\
  }\href@noop {} {\bibfield  {journal} {\bibinfo  {journal} {Lincoln Laboratory
  Journal}\ }\textbf {\bibinfo {volume} {22}},\ \bibinfo {pages} {38} (\bibinfo
  {year} {2017})}\BibitemShut {NoStop}%
\bibitem [{\citenamefont {Soderlund}\ and\ \citenamefont
  {Kumar}(2018)}]{soderlund2018markovian}%
  \BibitemOpen
  \bibfield  {author} {\bibinfo {author} {\bibfnamefont {A.~A.}\ \bibnamefont
  {Soderlund}}\ and\ \bibinfo {author} {\bibfnamefont {M.}~\bibnamefont
  {Kumar}},\ }in\ \href@noop {} {\emph {\bibinfo {booktitle} {2018 IEEE
  Conference on Decision and Control (CDC)}}}\ (\bibinfo {organization}
  {IEEE},\ \bibinfo {year} {2018})\ pp.\ \bibinfo {pages}
  {5592--5597}\BibitemShut {NoStop}%
\bibitem [{\citenamefont {Kendall}(1956)}]{kendall1956deterministic}%
  \BibitemOpen
  \bibfield  {author} {\bibinfo {author} {\bibfnamefont {D.~G.}\ \bibnamefont
  {Kendall}},\ }in\ \href@noop {} {\emph {\bibinfo {booktitle} {Proceedings of
  the third Berkeley symposium on mathematical statistics and probability}}},\
  Vol.~\bibinfo {volume} {4}\ (\bibinfo {organization} {University of
  California Press Berkeley},\ \bibinfo {year} {1956})\ pp.\ \bibinfo {pages}
  {149--165}\BibitemShut {NoStop}%
\bibitem [{\citenamefont {Moritz}\ \emph {et~al.}(2005)\citenamefont {Moritz},
  \citenamefont {Morais}, \citenamefont {Summerell}, \citenamefont {Carlson},\
  and\ \citenamefont {Doyle}}]{maxjeanWildfireHOT}%
  \BibitemOpen
  \bibfield  {author} {\bibinfo {author} {\bibfnamefont {M.~A.}\ \bibnamefont
  {Moritz}}, \bibinfo {author} {\bibfnamefont {M.~E.}\ \bibnamefont {Morais}},
  \bibinfo {author} {\bibfnamefont {L.~A.}\ \bibnamefont {Summerell}}, \bibinfo
  {author} {\bibfnamefont {J.~M.}\ \bibnamefont {Carlson}}, \ and\ \bibinfo
  {author} {\bibfnamefont {J.}~\bibnamefont {Doyle}},\ }\href {\doibase
  10.1073/pnas.0508985102} {\bibfield  {journal} {\bibinfo  {journal}
  {Proceedings of the National Academy of Sciences}\ }\textbf {\bibinfo
  {volume} {102}},\ \bibinfo {pages} {17912} (\bibinfo {year}
  {2005})}\BibitemShut {NoStop}%
\bibitem [{\citenamefont {Pisarenko}\ and\ \citenamefont
  {Rodkin}(2010)}]{heavytails}%
  \BibitemOpen
  \bibfield  {author} {\bibinfo {author} {\bibfnamefont {V.}~\bibnamefont
  {Pisarenko}}\ and\ \bibinfo {author} {\bibfnamefont {M.}~\bibnamefont
  {Rodkin}},\ }\href {\doibase 10.1007/978-90-481-9171-0} {\emph {\bibinfo
  {title} {Heavy-Tailed Distributions in Disaster Analysis}}},\ Vol.~\bibinfo
  {volume} {30}\ (\bibinfo {year} {2010})\BibitemShut {NoStop}%
\bibitem [{\citenamefont {Hergarten}(2004)}]{aspectsriskassesment}%
  \BibitemOpen
  \bibfield  {author} {\bibinfo {author} {\bibfnamefont {S.}~\bibnamefont
  {Hergarten}},\ }\href {\doibase 10.5194/nhess-4-309-2004} {\bibfield
  {journal} {\bibinfo  {journal} {Natural Hazards and Earth System Science}\
  }\textbf {\bibinfo {volume} {4}} (\bibinfo {year} {2004}),\
  10.5194/nhess-4-309-2004}\BibitemShut {NoStop}%
\bibitem [{\citenamefont {Mitzenmacher}(2003)}]{generative}%
  \BibitemOpen
  \bibfield  {author} {\bibinfo {author} {\bibfnamefont {M.}~\bibnamefont
  {Mitzenmacher}},\ }\href {\doibase im/1089229510} {\bibfield  {journal}
  {\bibinfo  {journal} {Internet Mathematics}\ }\textbf {\bibinfo {volume}
  {1}},\ \bibinfo {pages} {226} (\bibinfo {year} {2003})}\BibitemShut {NoStop}%
\bibitem [{\citenamefont {Willinger}\ \emph {et~al.}(2004)\citenamefont
  {Willinger}, \citenamefont {Alderson}, \citenamefont {Doyle},\ and\
  \citenamefont {Li}}]{morenormal}%
  \BibitemOpen
  \bibfield  {author} {\bibinfo {author} {\bibfnamefont {W.}~\bibnamefont
  {Willinger}}, \bibinfo {author} {\bibfnamefont {D.}~\bibnamefont {Alderson}},
  \bibinfo {author} {\bibfnamefont {J.}~\bibnamefont {Doyle}}, \ and\ \bibinfo
  {author} {\bibfnamefont {L.}~\bibnamefont {Li}},\ }in\ \href {\doibase
  10.1109/WSC.2004.1371310} {\emph {\bibinfo {booktitle} {Proceedings of the
  2004 Winter Simulation Conference, 2004.}}},\ Vol.~\bibinfo {volume} {1}\
  (\bibinfo {year} {2004})\ p.\ \bibinfo {pages} {141}\BibitemShut {NoStop}%
\bibitem [{\citenamefont {Bak}(2013)}]{bak2013nature}%
  \BibitemOpen
  \bibfield  {author} {\bibinfo {author} {\bibfnamefont {P.}~\bibnamefont
  {Bak}},\ }\href@noop {} {\emph {\bibinfo {title} {How nature works: the
  science of self-organized criticality}}}\ (\bibinfo  {publisher} {Springer
  Science \& Business Media},\ \bibinfo {year} {2013})\BibitemShut {NoStop}%
\bibitem [{\citenamefont {Stumpf}\ and\ \citenamefont
  {Porter}(2012)}]{stumpf2012critical}%
  \BibitemOpen
  \bibfield  {author} {\bibinfo {author} {\bibfnamefont {M.~P.}\ \bibnamefont
  {Stumpf}}\ and\ \bibinfo {author} {\bibfnamefont {M.~A.}\ \bibnamefont
  {Porter}},\ }\href@noop {} {\bibfield  {journal} {\bibinfo  {journal}
  {Science}\ }\textbf {\bibinfo {volume} {335}},\ \bibinfo {pages} {665}
  (\bibinfo {year} {2012})}\BibitemShut {NoStop}%
\bibitem [{\citenamefont {Carlson}\ and\ \citenamefont {Doyle}(1999)}]{hot2}%
  \BibitemOpen
  \bibfield  {author} {\bibinfo {author} {\bibfnamefont {J.~M.}\ \bibnamefont
  {Carlson}}\ and\ \bibinfo {author} {\bibfnamefont {J.}~\bibnamefont
  {Doyle}},\ }\href {\doibase 10.1103/PhysRevE.60.1412} {\bibfield  {journal}
  {\bibinfo  {journal} {Phys. Rev. E}\ }\textbf {\bibinfo {volume} {60}},\
  \bibinfo {pages} {1412} (\bibinfo {year} {1999})}\BibitemShut {NoStop}%
\bibitem [{\citenamefont {Doyle}\ and\ \citenamefont {Carlson}(2000)}]{hot1}%
  \BibitemOpen
  \bibfield  {author} {\bibinfo {author} {\bibfnamefont {J.}~\bibnamefont
  {Doyle}}\ and\ \bibinfo {author} {\bibfnamefont {J.~M.}\ \bibnamefont
  {Carlson}},\ }\href {\doibase 10.1103/PhysRevLett.84.5656} {\bibfield
  {journal} {\bibinfo  {journal} {Phys. Rev. Lett.}\ }\textbf {\bibinfo
  {volume} {84}},\ \bibinfo {pages} {5656} (\bibinfo {year}
  {2000})}\BibitemShut {NoStop}%
\bibitem [{\citenamefont {Kendall}(1948)}]{kendallGeneral}%
  \BibitemOpen
  \bibfield  {author} {\bibinfo {author} {\bibfnamefont {D.~G.}\ \bibnamefont
  {Kendall}},\ }\href {http://www.jstor.org/stable/2236051} {\bibfield
  {journal} {\bibinfo  {journal} {The Annals of Mathematical Statistics}\
  }\textbf {\bibinfo {volume} {19}},\ \bibinfo {pages} {1} (\bibinfo {year}
  {1948})}\BibitemShut {NoStop}%
\bibitem [{\citenamefont {Karlin}\ and\ \citenamefont
  {McGregor}(1957{\natexlab{a}})}]{karlinClass}%
  \BibitemOpen
  \bibfield  {author} {\bibinfo {author} {\bibfnamefont {S.}~\bibnamefont
  {Karlin}}\ and\ \bibinfo {author} {\bibfnamefont {J.}~\bibnamefont
  {McGregor}},\ }\href {http://www.jstor.org/stable/1993021} {\bibfield
  {journal} {\bibinfo  {journal} {Transactions of the American Mathematical
  Society}\ }\textbf {\bibinfo {volume} {86}},\ \bibinfo {pages} {366}
  (\bibinfo {year} {1957}{\natexlab{a}})}\BibitemShut {NoStop}%
\bibitem [{\citenamefont {Karlin}\ and\ \citenamefont
  {McGregor}(1957{\natexlab{b}})}]{karlinDiff}%
  \BibitemOpen
  \bibfield  {author} {\bibinfo {author} {\bibfnamefont {S.}~\bibnamefont
  {Karlin}}\ and\ \bibinfo {author} {\bibfnamefont {J.~L.}\ \bibnamefont
  {McGregor}},\ }\href {http://www.jstor.org/stable/1992942} {\bibfield
  {journal} {\bibinfo  {journal} {Transactions of the American Mathematical
  Society}\ }\textbf {\bibinfo {volume} {85}},\ \bibinfo {pages} {489}
  (\bibinfo {year} {1957}{\natexlab{b}})}\BibitemShut {NoStop}%
\bibitem [{\citenamefont {Karlin}\ and\ \citenamefont
  {McGregor}(1958)}]{karlinear}%
  \BibitemOpen
  \bibfield  {author} {\bibinfo {author} {\bibfnamefont {S.}~\bibnamefont
  {Karlin}}\ and\ \bibinfo {author} {\bibfnamefont {J.}~\bibnamefont
  {McGregor}},\ }\href {http://www.jstor.org/stable/24900526} {\bibfield
  {journal} {\bibinfo  {journal} {Journal of Mathematics and Mechanics}\
  }\textbf {\bibinfo {volume} {7}},\ \bibinfo {pages} {643} (\bibinfo {year}
  {1958})}\BibitemShut {NoStop}%
\bibitem [{\citenamefont {Karlin}\ and\ \citenamefont
  {Tavaré}(1982)}]{linearkilling}%
  \BibitemOpen
  \bibfield  {author} {\bibinfo {author} {\bibfnamefont {S.}~\bibnamefont
  {Karlin}}\ and\ \bibinfo {author} {\bibfnamefont {S.}~\bibnamefont
  {Tavaré}},\ }\href {http://www.jstor.org/stable/3213507} {\bibfield
  {journal} {\bibinfo  {journal} {Journal of Applied Probability}\ }\textbf
  {\bibinfo {volume} {19}},\ \bibinfo {pages} {477} (\bibinfo {year}
  {1982})}\BibitemShut {NoStop}%
\bibitem [{\citenamefont {Askey}\ \emph {et~al.}(1984)\citenamefont {Askey},
  \citenamefont {Ismail},\ and\ \citenamefont {Society}}]{ismail}%
  \BibitemOpen
  \bibfield  {author} {\bibinfo {author} {\bibfnamefont {R.}~\bibnamefont
  {Askey}}, \bibinfo {author} {\bibfnamefont {M.}~\bibnamefont {Ismail}}, \
  and\ \bibinfo {author} {\bibfnamefont {A.~M.}\ \bibnamefont {Society}},\
  }\href {https://books.google.com/books?id=vDLUCQAAQBAJ} {\emph {\bibinfo
  {title} {Recurrence Relations, Continued Fractions and Orthogonal
  Polynomials}}},\ American Mathematical Society: Memoirs of the American
  Mathematical Society\ (\bibinfo  {publisher} {American Mathematical
  Society},\ \bibinfo {year} {1984})\BibitemShut {NoStop}%
\bibitem [{\citenamefont {Morgan}(1979)}]{byron1979}%
  \BibitemOpen
  \bibfield  {author} {\bibinfo {author} {\bibfnamefont {B.~J.}\ \bibnamefont
  {Morgan}},\ }\href {\doibase 10.1080/0020739790100106} {\bibfield  {journal}
  {\bibinfo  {journal} {International Journal of Mathematical Education in
  Science and Technology}\ }\textbf {\bibinfo {volume} {10}},\ \bibinfo {pages}
  {51} (\bibinfo {year} {1979})}\BibitemShut {NoStop}%
\bibitem [{\citenamefont {Gani}\ and\ \citenamefont
  {McNeil}(1971)}]{jointBDintegrals}%
  \BibitemOpen
  \bibfield  {author} {\bibinfo {author} {\bibfnamefont {J.}~\bibnamefont
  {Gani}}\ and\ \bibinfo {author} {\bibfnamefont {D.~R.}\ \bibnamefont
  {McNeil}},\ }\href {http://www.jstor.org/stable/1426175} {\bibfield
  {journal} {\bibinfo  {journal} {Advances in Applied Probability}\ }\textbf
  {\bibinfo {volume} {3}},\ \bibinfo {pages} {339} (\bibinfo {year}
  {1971})}\BibitemShut {NoStop}%
\bibitem [{\citenamefont {Crawford}\ and\ \citenamefont
  {Suchard}(2014)}]{crawford2014birthdeath}%
  \BibitemOpen
  \bibfield  {author} {\bibinfo {author} {\bibfnamefont {F.~W.}\ \bibnamefont
  {Crawford}}\ and\ \bibinfo {author} {\bibfnamefont {M.~A.}\ \bibnamefont
  {Suchard}},\ }\href@noop {} {\enquote {\bibinfo {title} {Birth-death
  processes},}\ } (\bibinfo {year} {2014}),\ \Eprint
  {http://arxiv.org/abs/1301.1305} {arXiv:1301.1305 [stat.ME]} \BibitemShut
  {NoStop}%
\bibitem [{\citenamefont {Crawford}\ \emph {et~al.}(2018)\citenamefont
  {Crawford}, \citenamefont {Ho},\ and\ \citenamefont
  {Suchard}}]{Crawford2018}%
  \BibitemOpen
  \bibfield  {author} {\bibinfo {author} {\bibfnamefont {F.~W.}\ \bibnamefont
  {Crawford}}, \bibinfo {author} {\bibfnamefont {L.~S.~T.}\ \bibnamefont {Ho}},
  \ and\ \bibinfo {author} {\bibfnamefont {M.~A.}\ \bibnamefont {Suchard}},\
  }\href {\doibase 10.1002/wics.1423} {\bibfield  {journal} {\bibinfo
  {journal} {Wiley Interdisciplinary Reviews: Computational Statistics}\
  }\textbf {\bibinfo {volume} {10}},\ \bibinfo {pages} {e1423} (\bibinfo {year}
  {2018})}\BibitemShut {NoStop}%
\bibitem [{\citenamefont {Renshaw}(1972)}]{bd_migration}%
  \BibitemOpen
  \bibfield  {author} {\bibinfo {author} {\bibfnamefont {E.}~\bibnamefont
  {Renshaw}},\ }\href {http://www.jstor.org/stable/2334614} {\bibfield
  {journal} {\bibinfo  {journal} {Biometrika}\ }\textbf {\bibinfo {volume}
  {59}},\ \bibinfo {pages} {49} (\bibinfo {year} {1972})}\BibitemShut {NoStop}%
\bibitem [{\citenamefont {Korbel}\ and\ \citenamefont
  {Wolpert}(2021)}]{korbel2021stochastic}%
  \BibitemOpen
  \bibfield  {author} {\bibinfo {author} {\bibfnamefont {J.}~\bibnamefont
  {Korbel}}\ and\ \bibinfo {author} {\bibfnamefont {D.~H.}\ \bibnamefont
  {Wolpert}},\ }\href@noop {} {\bibfield  {journal} {\bibinfo  {journal} {New
  Journal of Physics}\ }\textbf {\bibinfo {volume} {23}} (\bibinfo {year}
  {2021})}\BibitemShut {NoStop}%
\bibitem [{\citenamefont {Wolpert}\ \emph {et~al.}(2019)\citenamefont
  {Wolpert}, \citenamefont {Kolchinsky},\ and\ \citenamefont
  {Owen}}]{wolpert2019space}%
  \BibitemOpen
  \bibfield  {author} {\bibinfo {author} {\bibfnamefont {D.~H.}\ \bibnamefont
  {Wolpert}}, \bibinfo {author} {\bibfnamefont {A.}~\bibnamefont {Kolchinsky}},
  \ and\ \bibinfo {author} {\bibfnamefont {J.~A.}\ \bibnamefont {Owen}},\
  }\href@noop {} {\bibfield  {journal} {\bibinfo  {journal} {Nature
  communications}\ }\textbf {\bibinfo {volume} {10}},\ \bibinfo {pages} {1}
  (\bibinfo {year} {2019})}\BibitemShut {NoStop}%
\bibitem [{\citenamefont {Ledermann}\ and\ \citenamefont
  {Reuter}(1954)}]{spectral}%
  \BibitemOpen
  \bibfield  {author} {\bibinfo {author} {\bibfnamefont {W.}~\bibnamefont
  {Ledermann}}\ and\ \bibinfo {author} {\bibfnamefont {G.~E.~H.}\ \bibnamefont
  {Reuter}},\ }\href {http://www.jstor.org/stable/91569} {\bibfield  {journal}
  {\bibinfo  {journal} {Philosophical Transactions of the Royal Society of
  London. Series A, Mathematical and Physical Sciences}\ }\textbf {\bibinfo
  {volume} {246}},\ \bibinfo {pages} {321} (\bibinfo {year}
  {1954})}\BibitemShut {NoStop}%
\bibitem [{\citenamefont {Reuter}\ \emph {et~al.}(1953)\citenamefont {Reuter},
  \citenamefont {Ledermann},\ and\ \citenamefont {Bartlett}}]{reuterDiff}%
  \BibitemOpen
  \bibfield  {author} {\bibinfo {author} {\bibfnamefont {G.~E.~H.}\
  \bibnamefont {Reuter}}, \bibinfo {author} {\bibfnamefont {W.}~\bibnamefont
  {Ledermann}}, \ and\ \bibinfo {author} {\bibfnamefont {M.~S.}\ \bibnamefont
  {Bartlett}},\ }\href {\doibase 10.1017/S0305004100028346} {\bibfield
  {journal} {\bibinfo  {journal} {Mathematical Proceedings of the Cambridge
  Philosophical Society}\ }\textbf {\bibinfo {volume} {49}},\ \bibinfo {pages}
  {247–262} (\bibinfo {year} {1953})}\BibitemShut {NoStop}%
\bibitem [{\citenamefont {Hathcock}\ and\ \citenamefont
  {Strogatz}(2022)}]{hathcock2022asymptotic}%
  \BibitemOpen
  \bibfield  {author} {\bibinfo {author} {\bibfnamefont {D.}~\bibnamefont
  {Hathcock}}\ and\ \bibinfo {author} {\bibfnamefont {S.~H.}\ \bibnamefont
  {Strogatz}},\ }\href@noop {} {\bibfield  {journal} {\bibinfo  {journal}
  {Physical Review Letters}\ }\textbf {\bibinfo {volume} {128}},\ \bibinfo
  {pages} {218301} (\bibinfo {year} {2022})}\BibitemShut {NoStop}%
\bibitem [{\citenamefont {Kessler}\ and\ \citenamefont
  {Shnerb}(2023)}]{kessler2023extinction}%
  \BibitemOpen
  \bibfield  {author} {\bibinfo {author} {\bibfnamefont {D.}~\bibnamefont
  {Kessler}}\ and\ \bibinfo {author} {\bibfnamefont {N.~M.}\ \bibnamefont
  {Shnerb}},\ }\href@noop {} {\  (\bibinfo {year} {2023})},\ \Eprint
  {http://arxiv.org/abs/2307.08435} {arXiv:2307.08435 [q-bio.PE]} \BibitemShut
  {NoStop}%
\bibitem [{\citenamefont {Waugh}(1958)}]{Waugh1958}%
  \BibitemOpen
  \bibfield  {author} {\bibinfo {author} {\bibfnamefont {W.~A.~O.}\
  \bibnamefont {Waugh}},\ }\href@noop {} {\bibfield  {journal} {\bibinfo
  {journal} {Biometrika}\ }\textbf {\bibinfo {volume} {45}},\ \bibinfo {pages}
  {241} (\bibinfo {year} {1958})}\BibitemShut {NoStop}%
\bibitem [{\citenamefont {Tavaré}(2018)}]{linearTavare}%
  \BibitemOpen
  \bibfield  {author} {\bibinfo {author} {\bibfnamefont {S.}~\bibnamefont
  {Tavaré}},\ }\href {\doibase 10.1017/apr.2018.84} {\bibfield  {journal}
  {\bibinfo  {journal} {Advances in Applied Probability}\ }\textbf {\bibinfo
  {volume} {50}},\ \bibinfo {pages} {253} (\bibinfo {year} {2018})}\BibitemShut
  {NoStop}%
\bibitem [{\citenamefont {Favard}(1935)}]{favard}%
  \BibitemOpen
  \bibfield  {author} {\bibinfo {author} {\bibfnamefont {J.}~\bibnamefont
  {Favard}},\ }\href@noop {} {\bibfield  {journal} {\bibinfo  {journal} {C. R.
  Acad. Sci. Paris}\ }\textbf {\bibinfo {volume} {200}},\ \bibinfo {pages}
  {2052} (\bibinfo {year} {1935})}\BibitemShut {NoStop}%
\bibitem [{\citenamefont {Koekoek}\ and\ \citenamefont
  {Swarttouw}(1996)}]{askey}%
  \BibitemOpen
  \bibfield  {author} {\bibinfo {author} {\bibfnamefont {R.}~\bibnamefont
  {Koekoek}}\ and\ \bibinfo {author} {\bibfnamefont {R.~F.}\ \bibnamefont
  {Swarttouw}},\ }\href@noop {} {\enquote {\bibinfo {title} {The {Askey}-scheme
  of hypergeometric orthogonal polynomials and its q-analogue},}\ } (\bibinfo
  {year} {1996}),\ \Eprint {http://arxiv.org/abs/math/9602214}
  {arXiv:math/9602214 [math.CA]} \BibitemShut {NoStop}%
\bibitem [{{\relax DLMF}()}]{NIST:DLMF}%
  \BibitemOpen
  {\relax DLMF},\ \href {https://dlmf.nist.gov/} {\enquote {\bibinfo {title}
  {{\it NIST Digital Library of Mathematical Functions}},}\ }\bibinfo
  {howpublished} {\url{https://dlmf.nist.gov/}, Release 1.1.9 of 2023-03-15},\
  \bibinfo {note} {f.~W.~J. Olver, A.~B. {Olde Daalhuis}, D.~W. Lozier, B.~I.
  Schneider, R.~F. Boisvert, C.~W. Clark, B.~R. Miller, B.~V. Saunders, H.~S.
  Cohl, and M.~A. McClain, eds.}\BibitemShut {Stop}%
\bibitem [{\citenamefont {Ross}(1995)}]{ross1995stochastic}%
  \BibitemOpen
  \bibfield  {author} {\bibinfo {author} {\bibfnamefont {S.~M.}\ \bibnamefont
  {Ross}},\ }\href@noop {} {\emph {\bibinfo {title} {Stochastic processes}}}\
  (\bibinfo  {publisher} {John Wiley \& Sons},\ \bibinfo {year} {1995})\
  Chap.~\bibinfo {chapter} {7}\BibitemShut {NoStop}%
\bibitem [{\citenamefont {Karlin}\ and\ \citenamefont
  {McGregor}(1959)}]{randomwalksKarlin}%
  \BibitemOpen
  \bibfield  {author} {\bibinfo {author} {\bibfnamefont {S.}~\bibnamefont
  {Karlin}}\ and\ \bibinfo {author} {\bibfnamefont {J.}~\bibnamefont
  {McGregor}},\ }\href {\doibase 10.1215/ijm/1255454999} {\bibfield  {journal}
  {\bibinfo  {journal} {Illinois Journal of Mathematics}\ }\textbf {\bibinfo
  {volume} {3}},\ \bibinfo {pages} {66 } (\bibinfo {year} {1959})}\BibitemShut
  {NoStop}%
\bibitem [{\citenamefont {Szeg{\H{o}}}(1939)}]{szego}%
  \BibitemOpen
  \bibfield  {author} {\bibinfo {author} {\bibfnamefont {G.}~\bibnamefont
  {Szeg{\H{o}}}},\ }\href {https://books.google.com/books?id=RhsPAAAAIAAJ}
  {\emph {\bibinfo {title} {Orthogonal Polynomials}}}\ (\bibinfo  {publisher}
  {American Mathematical Society},\ \bibinfo {year} {1939})\BibitemShut
  {NoStop}%
\bibitem [{\citenamefont {Chihara}(2011)}]{chihara2011introduction}%
  \BibitemOpen
  \bibfield  {author} {\bibinfo {author} {\bibfnamefont {T.~S.}\ \bibnamefont
  {Chihara}},\ }\href@noop {} {\emph {\bibinfo {title} {An introduction to
  orthogonal polynomials}}}\ (\bibinfo  {publisher} {Courier Corporation},\
  \bibinfo {year} {2011})\BibitemShut {NoStop}%
\bibitem [{\citenamefont {Koornwinder}(1989)}]{koornwinder1989meixner}%
  \BibitemOpen
  \bibfield  {author} {\bibinfo {author} {\bibfnamefont {T.~H.}\ \bibnamefont
  {Koornwinder}},\ }\href@noop {} {\bibfield  {journal} {\bibinfo  {journal}
  {Journal of mathematical physics}\ }\textbf {\bibinfo {volume} {30}},\
  \bibinfo {pages} {767} (\bibinfo {year} {1989})}\BibitemShut {NoStop}%
\bibitem [{\citenamefont {Bank}\ and\ \citenamefont {Ismail}(1985)}]{Bank}%
  \BibitemOpen
  \bibfield  {author} {\bibinfo {author} {\bibfnamefont {E.~D.}\ \bibnamefont
  {Bank}}\ and\ \bibinfo {author} {\bibfnamefont {M.~E.~H.}\ \bibnamefont
  {Ismail}},\ }\href@noop {} {\bibfield  {journal} {\bibinfo  {journal}
  {Constructive Approximation}\ }\textbf {\bibinfo {volume} {1}},\ \bibinfo
  {pages} {103} (\bibinfo {year} {1985})}\BibitemShut {NoStop}%
\bibitem [{\citenamefont {Van~Assche}(1990)}]{van1990pollaczek}%
  \BibitemOpen
  \bibfield  {author} {\bibinfo {author} {\bibfnamefont {W.}~\bibnamefont
  {Van~Assche}},\ }\href@noop {} {\bibfield  {journal} {\bibinfo  {journal}
  {Journal of mathematical analysis and applications}\ }\textbf {\bibinfo
  {volume} {147}},\ \bibinfo {pages} {498} (\bibinfo {year}
  {1990})}\BibitemShut {NoStop}%
\bibitem [{\citenamefont {Rui}\ and\ \citenamefont
  {Wong}(1996)}]{rui1996asymptotic}%
  \BibitemOpen
  \bibfield  {author} {\bibinfo {author} {\bibfnamefont {B.}~\bibnamefont
  {Rui}}\ and\ \bibinfo {author} {\bibfnamefont {R.}~\bibnamefont {Wong}},\
  }\href@noop {} {\bibfield  {journal} {\bibinfo  {journal} {Studies in Applied
  Mathematics}\ }\textbf {\bibinfo {volume} {96}},\ \bibinfo {pages} {307}
  (\bibinfo {year} {1996})}\BibitemShut {NoStop}%
\bibitem [{\citenamefont {Li}\ and\ \citenamefont
  {Wong}(2001)}]{li2001asymptotics}%
  \BibitemOpen
  \bibfield  {author} {\bibinfo {author} {\bibfnamefont {X.}~\bibnamefont
  {Li}}\ and\ \bibinfo {author} {\bibfnamefont {R.}~\bibnamefont {Wong}},\
  }\href@noop {} {\bibfield  {journal} {\bibinfo  {journal} {Constructive
  approximation}\ }\textbf {\bibinfo {volume} {17}},\ \bibinfo {pages} {59}
  (\bibinfo {year} {2001})}\BibitemShut {NoStop}%
\bibitem [{\citenamefont {Yermolayeva}\ and\ \citenamefont
  {Zhedanov}(1999)}]{yermolayeva1999spectral}%
  \BibitemOpen
  \bibfield  {author} {\bibinfo {author} {\bibfnamefont {O.}~\bibnamefont
  {Yermolayeva}}\ and\ \bibinfo {author} {\bibfnamefont {A.}~\bibnamefont
  {Zhedanov}},\ }\href@noop {} {\bibfield  {journal} {\bibinfo  {journal}
  {Methods and Applications of Analysis}\ }\textbf {\bibinfo {volume} {6}},\
  \bibinfo {pages} {261} (\bibinfo {year} {1999})}\BibitemShut {NoStop}%
\bibitem [{\citenamefont {Araaya}(2004)}]{araaya2004meixner}%
  \BibitemOpen
  \bibfield  {author} {\bibinfo {author} {\bibfnamefont {T.~K.}\ \bibnamefont
  {Araaya}},\ }\href@noop {} {\bibfield  {journal} {\bibinfo  {journal}
  {Journal of computational and applied mathematics}\ }\textbf {\bibinfo
  {volume} {170}},\ \bibinfo {pages} {241} (\bibinfo {year}
  {2004})}\BibitemShut {NoStop}%
\bibitem [{\citenamefont {Tavar{\'e}}(1987)}]{Tavar1987TheBP}%
  \BibitemOpen
  \bibfield  {author} {\bibinfo {author} {\bibfnamefont {S.}~\bibnamefont
  {Tavar{\'e}}},\ }\href@noop {} {\bibfield  {journal} {\bibinfo  {journal}
  {Journal of Mathematical Biology}\ }\textbf {\bibinfo {volume} {25}},\
  \bibinfo {pages} {161} (\bibinfo {year} {1987})}\BibitemShut {NoStop}%
\bibitem [{\citenamefont {Finney}(2006)}]{finney2006overview}%
  \BibitemOpen
  \bibfield  {author} {\bibinfo {author} {\bibfnamefont {M.~A.}\ \bibnamefont
  {Finney}},\ }in\ \href@noop {} {\emph {\bibinfo {booktitle} {Fuels
  Management-How to Measure Success: Conference Proceedings. 28-30 March 2006;
  Portland, OR. Proceedings RMRS-P-41. Fort Collins, CO: US Department of
  Agriculture, Forest Service, Rocky Mountain Research Station. p. 213-220}}},\
  Vol.~\bibinfo {volume} {41}\ (\bibinfo {year} {2006})\BibitemShut {NoStop}%
\bibitem [{\citenamefont {Finney}(1998)}]{finney1998farsite}%
  \BibitemOpen
  \bibfield  {author} {\bibinfo {author} {\bibfnamefont {M.~A.}\ \bibnamefont
  {Finney}},\ }\href@noop {} {\emph {\bibinfo {title} {FARSITE, Fire Area
  Simulator--model development and evaluation}}},\ \bibinfo {number} {4}\
  (\bibinfo  {publisher} {US Department of Agriculture, Forest Service, Rocky
  Mountain Research Station},\ \bibinfo {year} {1998})\BibitemShut {NoStop}%
\bibitem [{\citenamefont {Linn}\ \emph {et~al.}(2020)\citenamefont {Linn},
  \citenamefont {Goodrick}, \citenamefont {Brambilla}, \citenamefont {Brown},
  \citenamefont {Middleton}, \citenamefont {O'Brien},\ and\ \citenamefont
  {Hiers}}]{linn2020quic}%
  \BibitemOpen
  \bibfield  {author} {\bibinfo {author} {\bibfnamefont {R.~R.}\ \bibnamefont
  {Linn}}, \bibinfo {author} {\bibfnamefont {S.~L.}\ \bibnamefont {Goodrick}},
  \bibinfo {author} {\bibfnamefont {S.}~\bibnamefont {Brambilla}}, \bibinfo
  {author} {\bibfnamefont {M.~J.}\ \bibnamefont {Brown}}, \bibinfo {author}
  {\bibfnamefont {R.~S.}\ \bibnamefont {Middleton}}, \bibinfo {author}
  {\bibfnamefont {J.~J.}\ \bibnamefont {O'Brien}}, \ and\ \bibinfo {author}
  {\bibfnamefont {J.~K.}\ \bibnamefont {Hiers}},\ }\href@noop {} {\bibfield
  {journal} {\bibinfo  {journal} {Environmental Modelling \& Software}\
  }\textbf {\bibinfo {volume} {125}},\ \bibinfo {pages} {104616} (\bibinfo
  {year} {2020})}\BibitemShut {NoStop}%
\bibitem [{\citenamefont {Richards}(1995)}]{richards1995general}%
  \BibitemOpen
  \bibfield  {author} {\bibinfo {author} {\bibfnamefont {G.~D.}\ \bibnamefont
  {Richards}},\ }\href@noop {} {\bibfield  {journal} {\bibinfo  {journal}
  {International Journal of Wildland Fire}\ }\textbf {\bibinfo {volume} {5}},\
  \bibinfo {pages} {63} (\bibinfo {year} {1995})}\BibitemShut {NoStop}%
\bibitem [{\citenamefont {Pastor}\ \emph {et~al.}(2003)\citenamefont {Pastor},
  \citenamefont {Z{\'a}rate}, \citenamefont {Planas},\ and\ \citenamefont
  {Arnaldos}}]{pastor2003mathematical}%
  \BibitemOpen
  \bibfield  {author} {\bibinfo {author} {\bibfnamefont {E.}~\bibnamefont
  {Pastor}}, \bibinfo {author} {\bibfnamefont {L.}~\bibnamefont {Z{\'a}rate}},
  \bibinfo {author} {\bibfnamefont {E.}~\bibnamefont {Planas}}, \ and\ \bibinfo
  {author} {\bibfnamefont {J.}~\bibnamefont {Arnaldos}},\ }\href@noop {}
  {\bibfield  {journal} {\bibinfo  {journal} {Progress in Energy and Combustion
  Science}\ }\textbf {\bibinfo {volume} {29}},\ \bibinfo {pages} {139}
  (\bibinfo {year} {2003})}\BibitemShut {NoStop}%
\bibitem [{\citenamefont {Linn}\ \emph {et~al.}(2002)\citenamefont {Linn},
  \citenamefont {Reisner}, \citenamefont {Colman},\ and\ \citenamefont
  {Winterkamp}}]{Linn2002StudyingWB}%
  \BibitemOpen
  \bibfield  {author} {\bibinfo {author} {\bibfnamefont {R.~R.}\ \bibnamefont
  {Linn}}, \bibinfo {author} {\bibfnamefont {J.~M.}\ \bibnamefont {Reisner}},
  \bibinfo {author} {\bibfnamefont {J.}~\bibnamefont {Colman}}, \ and\ \bibinfo
  {author} {\bibfnamefont {J.}~\bibnamefont {Winterkamp}},\ }\href@noop {}
  {\bibfield  {journal} {\bibinfo  {journal} {International Journal of Wildland
  Fire}\ }\textbf {\bibinfo {volume} {11}},\ \bibinfo {pages} {233} (\bibinfo
  {year} {2002})}\BibitemShut {NoStop}%
\bibitem [{\citenamefont {Peterson}\ \emph {et~al.}(2009)\citenamefont
  {Peterson}, \citenamefont {Morais}, \citenamefont {Carlson}, \citenamefont
  {Dennison}, \citenamefont {Roberts}, \citenamefont {Moritz}, \citenamefont
  {Weise} \emph {et~al.}}]{peterson2009using}%
  \BibitemOpen
  \bibfield  {author} {\bibinfo {author} {\bibfnamefont {S.~H.}\ \bibnamefont
  {Peterson}}, \bibinfo {author} {\bibfnamefont {M.~E.}\ \bibnamefont
  {Morais}}, \bibinfo {author} {\bibfnamefont {J.~M.}\ \bibnamefont {Carlson}},
  \bibinfo {author} {\bibfnamefont {P.~E.}\ \bibnamefont {Dennison}}, \bibinfo
  {author} {\bibfnamefont {D.~A.}\ \bibnamefont {Roberts}}, \bibinfo {author}
  {\bibfnamefont {M.~A.}\ \bibnamefont {Moritz}}, \bibinfo {author}
  {\bibfnamefont {D.~R.}\ \bibnamefont {Weise}},  \emph {et~al.},\ }\href@noop
  {} {\emph {\bibinfo {title} {Using HFire for spatial modeling of fire in
  shrublands}}}\ (\bibinfo  {publisher} {Pacific Southwest Research Station,
  Forest Service, United States Department of Agriculture},\ \bibinfo {year}
  {2009})\BibitemShut {NoStop}%
\bibitem [{\citenamefont {Papadopoulos}\ and\ \citenamefont
  {Pavlidou}(2011)}]{papadopoulos2011comparative}%
  \BibitemOpen
  \bibfield  {author} {\bibinfo {author} {\bibfnamefont {G.~D.}\ \bibnamefont
  {Papadopoulos}}\ and\ \bibinfo {author} {\bibfnamefont {F.-N.}\ \bibnamefont
  {Pavlidou}},\ }\href@noop {} {\bibfield  {journal} {\bibinfo  {journal} {IEEE
  systems Journal}\ }\textbf {\bibinfo {volume} {5}},\ \bibinfo {pages} {233}
  (\bibinfo {year} {2011})}\BibitemShut {NoStop}%
\bibitem [{\citenamefont {Flajolet}\ and\ \citenamefont
  {Sedgewick}(2009)}]{analyticcombs}%
  \BibitemOpen
  \bibfield  {author} {\bibinfo {author} {\bibfnamefont {P.}~\bibnamefont
  {Flajolet}}\ and\ \bibinfo {author} {\bibfnamefont {R.}~\bibnamefont
  {Sedgewick}},\ }\href@noop {} {\emph {\bibinfo {title} {Analytic
  Combinatorics}}}\ (\bibinfo  {publisher} {Cambridge University Press},\
  \bibinfo {year} {2009})\BibitemShut {NoStop}%
\bibitem [{\citenamefont {Flajolet}\ \emph {et~al.}(2006)\citenamefont
  {Flajolet}, \citenamefont {Fusy}, \citenamefont {Gourdon}, \citenamefont
  {Panario},\ and\ \citenamefont {Pouyanne}}]{flajolet2006hybrid}%
  \BibitemOpen
  \bibfield  {author} {\bibinfo {author} {\bibfnamefont {P.}~\bibnamefont
  {Flajolet}}, \bibinfo {author} {\bibfnamefont {E.}~\bibnamefont {Fusy}},
  \bibinfo {author} {\bibfnamefont {X.}~\bibnamefont {Gourdon}}, \bibinfo
  {author} {\bibfnamefont {D.}~\bibnamefont {Panario}}, \ and\ \bibinfo
  {author} {\bibfnamefont {N.}~\bibnamefont {Pouyanne}},\ }\href@noop {} {\
  (\bibinfo {year} {2006})},\ \Eprint {http://arxiv.org/abs/math/0606370}
  {arXiv:math/0606370 [math.CO]} \BibitemShut {NoStop}%
\bibitem [{\citenamefont {Wong}(2001)}]{wong2001asymptotic}%
  \BibitemOpen
  \bibfield  {author} {\bibinfo {author} {\bibfnamefont {R.}~\bibnamefont
  {Wong}},\ }\href@noop {} {\emph {\bibinfo {title} {Asymptotic approximations
  of integrals}}}\ (\bibinfo  {publisher} {SIAM},\ \bibinfo {year}
  {2001})\BibitemShut {NoStop}%
\end{thebibliography}%

\end{document}